\documentclass[epj]{svjour}
\usepackage{graphics,graphicx,color,soul}
\usepackage[normalem]{ulem}  
\def\simge{\mathrel{\rlap{\raise 0.511ex
    \hbox{$>$}}{\lower 0.511ex \hbox{$\sim$}}}}
\def\simle{\mathrel{\rlap{\raise 0.511ex
      \hbox{$<$}}{\lower 0.511ex \hbox{$\sim$}}}}
\begin{document}
\title{Constraints on the Symmetry Energy Using the Mass-Radius Relation of Neutron Stars}
\author{James M. Lattimer\inst{1} \and Andrew W. Steiner\inst{2}
}                     
%
%
\institute{Dept. of Physics \& Astronomy, Stony Brook University, Stony Brook, NY 11794-3800, USA \and Institute for Nuclear Theory, Seattle, WA 98195, USA}
\date{Received: date / Revised version: date}
%
\abstract{ The nuclear symmetry energy is intimately connected with
  nuclear astrophysics. This contribution focuses on the estimation of
  the symmetry energy from experiment and how it is related to the
  structure of neutron stars. The most important connection is between
  the radii of neutron stars and the pressure of neutron star matter
  in the vicinity of the nuclear saturation density $n_s$. This
  pressure is essentially controlled by the nuclear symmetry energy
  parameters $S_v$ and $L$, the first two coefficients of a Taylor
  expansion of the symmetry energy around $n_s$. We discuss
  constraints on these parameters that can be found from nuclear
  experiments. We demonstrate that these constraints are largely
  model-independent by deriving them qualitatively from a simple
  nuclear model. We also summarize how recent theoretical studies of
  pure neutron matter can reinforce these constraints. To date,
  several different astrophysical measurements of neutron star radii
  have been attempted. Attention is focused on photospheric radius
  expansion bursts and on thermal emissions from quiescent low-mass
  X-ray binaries. While none of these observations can, at the present
  time, determine individual neutron star radii to better than 20\%
  accuracy, the body of observations can be used with Bayesian
  techniques to effectively constrain them to higher precision.  These
  techniques invert the structure equations and obtain estimates of
  the pressure-density relation of neutron star matter, not only near
  $n_s$, but up to the highest densities found in neutron star
  interiors. The estimates we derive for neutron star radii are in
  concordance with predictions from nuclear experiment and theory.}


\authorrunning{Lattimer \& Steiner}
\titlerunning{Neutron Star Constraints}
\maketitle

\section{Introduction}
\label{intro}
Neutron stars are laboratories for the study of dense nuclear matter
under conditions that are beyond those that can be achieved in
experiments.  The equation of state and internal compositions of large
portions of neutron stars are poorly understood at present.  However,
there has been substantial recent progress in unraveling these
mysteries.  This progress has come from theoretical studies of nuclear
and neutron matter, nuclear experiments, and astrophysical
observations.  The most profound aspect of the nuclear interaction for
neutron stars, in many respects, concerns the nuclear symmetry energy
which largely controls the composition and pressure of neutron-rich matter, and
therefore, many aspects of neutron star structure such as the radius,
moment of inertia and crustal properties.  

For most practical purposes, the interior of a neutron star can be
divided into a dense core and a less-dense crust. The density of the
core-crust boundary is believed to be near $n_s/2$, where
$n_s\simeq0.16$ fm$^{-3}$ is the nuclear saturation density, with a
weak dependence on the incompressibility and symmetry properties of
bulk nuclear matter. While matter just below the neutron star crust is
likely a uniform liquid of hadrons, electrons and muons, the crust
itself is composed of an equilibrium mixture of dense nuclei and a
neutron gas together with electrons. This division into two coexisting
phases is a natural consequence of the fact that uniform nuclear
matter at subnuclear densities, for large proton fractions, has
negative pressure. For these densities, phase coexistence involves
pressure and neutron and proton chemical potential equality in both
phases, which together determine the relative concentrations of nuclei
(the dense phase) and neutron gas (the less dense phase). It is important
to point out that in the crust the pressure mostly originates from the
degenerate relativistic electrons, for which the pressure is
$p_e=\hbar cnx(3\pi^2nx)^{1/3}$, where $n$ is the baryon density and
$x$ is the proton fraction (charge neutrality dictates that the number
of electrons per baryon is also $x$). Baryon pressure originates from
both nuclei and the neutron gas. However, the overall pressure of a
nucleus must equal the neutron gas pressure, or it would expand or
contract. This pressure remains very small until densities approach
$n_s$. In fact, the dominant baryonic pressure results from the
attractive Coulomb energy stemming from the Coulomb lattice, leading
to a net negative pressure that, like $p_e$, scales as $n^{4/3}$. The
ratio of the magnitudes of the lattice and electron pressures is only
a few percent, however, so in spite of uncertainties regarding the
nuclear force, the equation of state (at least, the pressure-density
relation) in the crust is very well-understood.

Matter in the interior of a neutron star, unlike that in laboratory
nuclei, is very neutron rich.  On timescales long compared to $\beta$
decay timescales of seconds, neutron star matter evolves into weak
interaction equilibrium, or $\beta$-equilibrium, in which the total
energy is at a minimum with respect to composition:
\begin{equation}\label{beta}
{\partial (E+E_e)\over\partial x}=\mu_p-\mu_n+\mu_e-(m_n-m_p)c^2=0,
\end{equation}
where $E$ is the baryon energy per baryon and $E_e$ is the electron
energy per baryon.  The $\mu$s are chemical potentials, which, for
baryons, are measured with respect to their rest masses. Electrons are
relativistic, and $\mu_e=\hbar c(3\pi^2nx)^{1/3}$.
The energy of uniform hadronic matter, in its ground state, is
essentially a function of baryon density ($n$), temperature ($T$), and
composition, which is usually parameterized in terms of its charge
fraction $x$.  For baryonic matter composed solely of neutrons and
protons, $x=n_p/(n_n+n_p)$.  It is convenient to define the symmetry
energy $S(n)$ as the difference between the energy per baryon of pure neutron
matter ($x=0$) and symmetric nuclear matter ($x=1/2$).  Since matter
in neutron stars under nearly all conditions of interest here is
highly degenerate, we only consider the case $T=0$.

In most theoretical models of cold uniform nuclear matter, the energy at a
given density can be well approximated by keeping only the first term
of a quadratic expansion:
\begin{equation}\label{energy}
E(n, x) \simeq E(n, 1/2) + S_2(n)(1-2x)^2+\dots
\end{equation}
so that the symmetry energy $S(n)\simeq S_2(n)$. However, it has not
been experimentally verified that quartic and higher-order terms are
negligible. We will indicate where this neglect might have an
appreciable effect. The symmetry energy is experimentally accessible
from nuclear masses and other experiments such as dipole resonances
and neutron skin thicknesses which sample matter near the nuclear
saturation density $n_s$. It is therefore convenient to consider a
Taylor expansion of $S_2$ near $n_s$:
\begin{equation}\label{symmetry}
S_2(n)\simeq S_v+{L\over3}(n-n_s)+{K_{\rm sym}\over18}(n-n_s)^2+\cdots
\end{equation}
which defines the symmetry parameters $S_v, L$ and $K_{\rm sym}$.

From Equation (\ref{energy}), we now find that
\begin{equation}\label{beta1}
\mu_p-\mu_n={\partial E\over\partial x}=4S_2(n)(1-2x).
\end{equation}
The solution of Equation (\ref{beta}) at $n_s$ yields
\begin{equation}\label{beta2}
x\simeq{1\over3\pi^2n_s}\left({4S_v\over\hbar c}\right)^3\simeq0.04,
\end{equation} 
i.e., neutron star matter is very nearly pure neutron matter.  At
higher densities, $x$ follows the behavior of $S_2(n)$. Below $n_s$,
where nuclei exist, Equation (\ref{beta1}) shows that the neutron
excess of the system, and individual nuclei, increases with density.
The minimum value of $x$ in beta equilibrium generally occurs at the
core-crust boundary just below $n_s$.

For pure neutron matter at $n_s$ and in the quadratic approximation,
the energy and pressure are given by
\begin{eqnarray}\label{sat}
E_N(n_s)&=&E(n,0)\simeq S_v+B,\cr 
p_N(n_s)&=&p(n_s,0)=n_s^2\left({\partial E\over\partial n}
\right)_{n_s,x=0}\!\!\simeq{L\over3}n_s,
\end{eqnarray}
where $B=-E(n_s,1/2)\simeq16$ MeV is the binding energy of symmetric
matter at the saturation density. For matter in $\beta$-equilibrium,
it follows that
\begin{equation}\label{sat1}
p_\beta(n_s)\simeq{L\over3}n_s\left[1-\left({4S_v\over\hbar c}
\right)^3{4-3S_v/L\over3\pi^2n_s}+\dots\right].
\end{equation}
This important result shows that the pressure of matter at the
nuclear saturation density can be expressed solely in
terms of the standard symmetry parameters $S_v$ and $L$, in the
quadratic approximation.

The symmetry energy is not only important in determining the
composition and pressure of matter in the interior, but also plays an
important role in determining the overall structure of the star.
Lattimer \& Prakash \cite{LP01} found that the neutron star radius
$R$, for a given stellar mass $M$, is highly correlated with the
neutron star matter pressure $p_\beta$ at densities in the vicinity of
$n_s$. This relation can be expressed as
\begin{equation}\label{radius}
R_M=C(n,M)[p_\beta(n)/{\rm MeV~fm^{-3}}]^{1/4},
\end{equation}
where $R_M$ is the radius of a star of mass $M$ and $C$ are
coefficients that depend on the density and mass. The upper set in
table~\ref{tab:1} shows the coefficients $C(n,1.4M_\odot)$ compiled
from about 3 dozen equations of state for three densities, $n_s,
1.5n_s$ and $2n_s$. Lattimer \& Lim \cite{Lattimer13} re-analyzed this
relation restricted to EOSs which could satisfy the constraint $\hat
M=2.0M_\odot$ where $\hat M$ is the minimum value for the maximum
neutron star mass given by the largest precisely measured neutron star
mass. Currently, this determined from measurements of PSR J1614+2230
\cite{Demorest:2010}, with $M=1.97\pm0.04M_\odot$, and PSR J0348+0432
\cite{Antoniadis13}, with $M=2.01\pm0.04 M_\odot$. It is observed that
the coefficients $C(n,M)$ become more accurate at higher densities,
but since $p_\beta$ can be expressed relatively model-independently in
terms of $S_v$ and $L$ at $n=n_s$, we can only usefully employ
$C(n_s,1.4M_\odot)$ to relate neutron star radii to symmetry energy
parameters.

\begin{table}
\begin{center}
\caption{Coefficients $C(n,1.4M_\odot)$, in km, for the
  pressure-radius correlation. $\hat M$ is the minimum value for the
  maximum neutron star mass.\label{tab:1}}
\begin{tabular}{l|ccc}
\hline\noalign{\smallskip}
$\hat M/M_\odot$&$n_s$ & $1.5n_s$ & $2n_s$\\
\noalign{\smallskip}\hline\noalign{\smallskip}
1.3&$9.30\pm0.58$&$6.99\pm0.30$&$5.72\pm0.25$\\
\noalign{\smallskip}\hline\noalign{\smallskip}
2.0&$9.52\pm0.49$ &$7.06\pm0.24$ &$5.68\pm0.14$\\
\noalign{\smallskip}\hline
\end{tabular}
\end{center}
\end{table}

\section{Symmetry Parameters From Nuclear Experiments}
\subsection{Correlations from the liquid drop model}

The distribution of neutrons and protons within nuclei differ, and,
furthermore, these distributions vary with $Z$ and $A$.  Therefore,
measurements of nuclear properties, especially for neutron-rich
nuclei, offer hope of constraining nuclear symmetry energy
parameters.  The most obvious manisfestation of the effects of
symmetry is visible in the liquid drop expression of the nuclear
energy
\begin{eqnarray} E(Z,A)&=&A(-B+S_vI^2)+A^{2/3}(E_s-S_sI^2)~+\cr
+~{3\over5}{e^2Z^2\over
r_0A^{1/3}}&+&E_{\mathrm{shell}}(Z,A-Z)+E_{\mathrm{pairing}}(A),\label{ld} 
\end{eqnarray} 
where $E_s\simeq19$ MeV is the symmetric matter surface energy
parameter, $I=(A-2Z)/A$ is the neutron excess, $S_s$ is the surface
symmetry energy parameter, and $r_0=(4\pi n_s/3)^{-1/3}$. The last
three terms in Eq. (\ref{ld}) represent the Coulomb, shell and pairing
energies, respectively. We will ignore shell and pairing effects for
the present discussion.

The net symmetry energy of an isolated nucleus is then 
\begin{equation}
E_{{\rm DM},i}=I_i^2(S_vA_i-S_sA_i^{2/3}).\label{ldsym} 
\end{equation} 
The parameters of the liquid drop model are typically determined by a
least-squares fit to measured masses, so a linear correlation between
$S_v$ and $S_s$ is therefore expected from minimizing the differences
between model predictions and experimentally measured symmetry
energies, i.e., minimizing
\begin{equation}
{\chi}^2=\sum_i(E_{{\rm exp},i}-E_{{\rm DM},i})^2/\sigma_{\rm DM}^2,\label{chi}
\qquad {\bar \chi}^2 \equiv \chi^2/{\cal N}
\end{equation}
where ${\cal N}$ is the total number of nuclei and $\sigma_{\rm DM}$ is a
nominal error. A ${\bar \chi}^2$ contour one unit above the minimum value
represents the $1-\sigma$ confidence interval which is an ellipse in
this linear example.

The properties of the confidence ellipse are determined by the second
derivatives of ${\bar \chi}^2$ at the minimum,
\begin{eqnarray}
[{\bar \chi}_{vv},~{\bar \chi}_{vs},~{\bar \chi}_{ss}]&=&
{2\over{\cal N}\sigma_{\rm DM}^2}\sum_iI_i^4[A_i^2,~-A_i^{5/3},~A_i^{4/3}]\cr
&\simeq&[61.6,~-10.7,~1.87]/\sigma_{\rm DM}^{2},\label{chider}
\end{eqnarray}
where ${\bar \chi}_{vs}=\partial^2{\bar \chi}^2/\partial S_v\partial
S_s$, etc. The specific values quoted follow from the set of 2336
nuclei with $N$ and $Z$ greater than 40 from Ref. \cite{Audi03}. The
confidence ellipse in $S_s-S_v$ space has orientation $\alpha_{\rm
  DM}=(1/2)\tan^{-1}|2{\bar \chi}_{vs}/({\bar \chi}_{vv}-{\bar
  \chi}_{ss})|\simeq9.8^\circ$ with respect to the $S_s$ axis, with
error widths $\sigma_{v,{\rm DM}}=\sqrt{({\bar
    \chi}^{-1})_{vv}}\simeq2.3\sigma_{\rm DM}$ and $\sigma_{s,{\rm
    DM}}=\sqrt{({\bar \chi}^{-1})_{ss}}\simeq13.2\sigma_{\rm DM}$
where $({\bar \chi}^{-1})$ is the matrix inverse. The correlation
coefficient is $r_{\rm DM}={\bar \chi}_{vs}/\sqrt{{\bar
    \chi}_{vv}{\bar \chi}_{ss}}\simeq0.997$. In this simple example,
the shape and orientation of the confidence interval depend only on
$A_i$ and $I_i$ and not on the binding energies themselves or the
location of the ${\bar \chi}^2$ minimum or the other drop parameters.
This correlation is therefore largely model-independent and the most
valuable of constraints from nuclear experiment.

In practice, the liquid droplet model~\cite{Myers69}, which differs
from the liquid drop model by accounting for varying neutron/proton
ratios within the nucleus that produce neutron skins in neutron-rich
nuclei, is an improved treatment.  Its symmetry energy is
\begin{equation}
E_{{\rm LD},i}=S_vI_i^2A_i(1+S_sA_i^{-1/3}/S_v)^{-1},
\label{ld1}
\end{equation} 
and therefore predicts a
linear correlation between $S_s/S_v$ and $S_v$ rather than between
$S_s$ and $S_v$ as in the drop model. 
 The same methodology as for the liquid
drop model can be used to determine the confidence interval in
$S_s/S_v - S_v$ space.  In contrast to the drop model, the properties
of the confidence interval now also depend, but only slightly, on the measured
masses, $E_{{\rm exp},i}$, as well as the parameters of the droplet model.

The liquid droplet model also predicts correlations from other
observational properties of nuclei. These include the dipole
polarizability, which is the linear response of a nucleus excited from
its ground state to an excited state due to the action of an external
isovector oscillating (dipole) electric field,
\begin{equation}\label{dp}
\alpha_D={AR^2\over20S_v}\left[1+{5\over3}{S_s\over S_v}A^{-1/3}\right],
\end{equation}
where $R=r_oA^{1/3}$ is the nuclear radius, and the neutron skin
thickness
\begin{equation}\label{skin} 
R_n-R_p={2r_o\over3}{S_sI\over S_v+S_sA^{-1/3}}, 
\end{equation}
where $R_{n,p}$ are the mean neutron and proton radii.

Comparing the functional forms of Equations (\ref{ld1}), (\ref{dp})
and (\ref{skin}), we observe that, respectively keeping $E_{\rm
    LD}, \alpha_D$ and $R_n-R_p$ fixed, different slopes are
predicted for variations of $S_s/S_v$ relative to $S_v$:
\begin{eqnarray}
{d(S_s/S_v)\over dS_v}&=&{1\over S_v}\left[A^{1/3}+
{S_s\over S_v},~{3\over5}A^{1/3}+{S_s\over S_v},~0\right]\cr
&\simeq&\left[0.25,~ 0.17,~0\right] {\rm MeV}^{-1},
\end{eqnarray}
where we assumed $S_v\simeq30$ MeV, $S_s/S_v\simeq3/2$ and
$A^{1/3}\simeq6$. As a result, comparison of experimental results for
these nuclear properties can potentially tightly constrain the liquid
droplet symmetry parameters, $S_v$ and $S_s$.

However, we have seen that astrophysical constraints on the neutron
star radius restrict the value of $p_\beta(n_s)$, which can be
expressed in terms of $S_v$ and $L$ via Equation (\ref{sat1}). To make
nuclear experimental constraints astrophysically relevant therefore
requires them to be expressed in $S_v-L$ space rather than $S_V-S_s$
space. In the context of the liquid droplet model, the surface energy
term represents the correction to the volume energy which accounts for
the fact that the density within the nucleus is not constant.
Therefore, we can expect that the density dependence of $S$ (or $S_2$)
is critical to this transformation and we can write $S_s(S_2)$ in the
quadratic approximation.

One can analytically predict these liquid droplet correlations as well
as the relation $S_s(S_2)$ by means of the so-called {\it
  hydrodynamical} model proposed by Lipparini \& Stringari
\cite{LS89}. If $S_2$ could be adequately described in the density
range $0<n<n_s$ with only the parameters $S_v$ and $L$, this
automatically would lead to an expression for $S_s(S_v, L)$. The
original model \cite{LS89} assumed $S_2(n)=S_v+L(n-n_s)/3$, which is
strictly valid only as long as $n/|n-n_s| \ll 1$. As a result, the
relation for $S_s(S_v,L)$ and the experimental correlations between
$S_v$ and $L$ estimated in Ref. \cite{LS89} are model-dependent. Here,
we generalize the model to allow for an arbitrary dependence of $S_2$
on the density so that the model-dependence of higher-order terms in
the density expansion of $S_2$, such as $K_{sym}$, can be determined.

We emphasize that the hydrodynamical model is not a substitute for
more sophisticated microscopic treatments of nuclei, including
Thomas-Fermi and Hartree-Fock analyses.  However, it does allow one to
straightforwardly observe the origins of correlations between $S_v$
and $L$ that have been proposed to result from nuclear experiments.

\subsection{The hydrodynamical model}

Following Lipparini \& Stringari,we assume a simplified nuclear Hamiltonian energy density
\begin{eqnarray}\label{ham}
{\cal H}&=&{\cal H}_B(n,\alpha)+{\cal Q}(n)\left(n^{\prime}\right)^2, \cr 
{\cal H}_B(n,\alpha)&=&{\cal H}_B(n,0)+v_{sym}(n)\alpha^2
\end{eqnarray}
where the uniform matter contribution is ${\cal H}_B(n,\alpha)$,
${\cal Q}(n)$ controls the gradient contributions, $v_{sym}=S_2/n$,
$n=n_n+n_p$ is the isoscalar density, and $\alpha=n_n-n_p$ is the
isovector density.  We will optimize the total nuclear energy
subject to the constraints
\begin{equation}\label{const}
A=\int\rho d^3r,\qquad N-Z=\int\alpha d^3r,
\end{equation}
 producing the chemical potentials $\mu$ and $\bar\mu$:
\begin{equation}\label{ham1}
{\delta\over\delta n}[{\cal H}-\mu n]=0,\qquad {\delta\over\delta\alpha}[{\cal H}-\bar\mu\alpha]=0.
\end{equation}
These lead to
\begin{eqnarray}\label{ham2}
2{d\over dr}\left[{\cal Q}n^\prime\right]-{\partial{\cal Q}\over\partial n}\left(n^\prime\right)^2&=&{\partial{\cal H}_B\over\partial n}-\mu,\cr
0={\partial{\cal H}_B\over\partial\alpha}-\bar\mu&=&2v_{sym}\alpha-\bar\mu.
\end{eqnarray}

Using the second of the constraints (Equation \ref{const}) and the
second of Equation (\ref{ham2}), we obtain
\begin{equation}\label{ham4}
N-Z={\bar\mu\over2}\int {1\over v_{sym}}d^3r\equiv{\bar\mu\over2}H,
\end{equation}
which defines $H$.  It then follows that
\begin{equation}\label{ham5}
\alpha={\bar\mu\over2v_{sym}}={N-Z\over v_{sym}H}.
\end{equation}

Separating out the total symmetry energy, and neglecting the Coulomb
energy for the moment, we obtain
\begin{equation}\label{sym}
E_{sym0}=\int v_{sym}\alpha^2 d^3r={(N-Z)^2\over H}.
\end{equation}
The dipole static polarizability, $\alpha_D$, hereafter referred
  to simply as the dipole polarizability, is found by performing the
constrained variation \cite{LS89}
\begin{equation}\label{dipole}
{\delta\over\delta\alpha}\left(\int{\cal H}d^3r-\epsilon\int z\alpha d^3r\right)=0,
\end{equation}
with $\epsilon$ a small parameter. Defining $\alpha_d$ as the function
$\alpha(r)$ which solves Eq.~\ref{dipole}, the dipole polarizability
is
\begin{equation}
\alpha_D={1\over2\epsilon}\int z\alpha_d d^3r.
\end{equation}
The solutions for $\alpha_d$ and the dipole
polarizability are
\begin{equation}\label{dipole2}
\alpha_d={\epsilon z\over2v_{sym}},\qquad \alpha_D={1\over12}\int{r^2\over v_{sym}}d^3r,
\end{equation}
where $z^2=r^2/3$ within the integral. It is also possible to show
\cite{LS89} that the mean excitation energy $\omega_D$ of the dipole
resonance is
\begin{equation}\label{dipole3}
\hbar^2\omega^2_D={\hbar^2\over3mA}\int v_{sym}n^{\prime2}d^3r
\end{equation}

The neutron skin thickness $R_n-R_p$, the difference between the mean
radii of neutrons and protons, is defined by
\begin{equation}\label{skin3}
{4\pi\over3}\left(R_n^3-R_p^3\right)=\int\left({n_n\over n_{no}}-{n_p\over n_{po}}\right)d^3r
\end{equation}
where $n_{no}$ and $n_{po}$ are the central values of the neutron and
proton densities. $R_n$ and $R_p$ represent the 'squared-off' radii.
From Equation (\ref{ham5}), the central isovector density is
\begin{equation}\label{skin2}
\alpha_o={N-Z\over v_{sym,o}H}={(N-Z)n_o\over S_vH},
\end{equation}
where $n_o$ is the central density, which for symmetric matter would
be $n_s$. $v_{sym,o}=S_v/n_o$ is the central value of $v_{sym}$. One
can then show, treating $R_n-R_p<<R$ and keeping the lowest-order
term:
\begin{equation}\label{skin4}
{R_n-R_p\over R}\simeq{2I\over3}\left(1-{A\over S_vH}\right)\left(1-{(N-Z)^2\over S_v^2H^2}\right)^{-1},
\end{equation}
where $I=(N-Z)/A$.

Experimentally, however, it is preferable to measure the differences
of the mean-square neutron and proton radii,
\begin{equation}\label{skin7}
r_{np}=r_n-r_p\equiv\sqrt{\int n_nr^2d^3r\over\int n_nd^3r}-\sqrt{\int n_pr^2d^3r\over\int n_pd^3r},
\end{equation}
which, if the densities are uniform, is $\sqrt{3/5}(R_n-R_p)$. In the
hydrodynamical model, the mean-square radii are
\begin{eqnarray}\label{skin8}
r^2_{n,p}&=&{1\over (N,Z)}\int n_{n,p}r^2d^3r={1\over2(N,Z)}\int(n\pm\alpha)r^2d^3r\cr
&=&{1\over1\pm I}\left[{3\over5}R^2\pm12{I\alpha_D\over H}\right],
\end{eqnarray}
where the upper (lower) sign refers to $n(p)$.  We then find
\begin{equation}\label{skin10}
{r_{np}\over R}\simeq I\sqrt{3\over5(1-I^2)}\left[{20\alpha_D\over HR^2}-1\right].
\end{equation}
This relation shows a clear connection between the neutron skin
thickness and the dipole polarizability.

Now we focus on the behavior of the total density as a function of
radius. Multiplying the first of Equation (\ref{ham2}) by $n^\prime$
and the second by $\alpha^\prime$, their sum can be integrated:
\begin{equation}\label{ham3}
{\cal Q}(n)n^{\prime2}={\cal H}_B(n,\alpha)-\mu n-\bar\mu\alpha,
\end{equation}
for which the boundary condition $\mu n_o+\bar\mu\alpha_o={\cal
  H}_B(n_o,\alpha_o)$ at the center. To make further progress, it is
necessary to have specific functional forms for ${\cal H}_B(n,0)$ and
${\cal Q}(n)$.

We make the common quadratic approximation for the uniform symmetric
matter energy density:
\begin{equation}\label{bulk}
{\cal H}_B(n,0)=n\left[-B+{K\over18}(1-u)^2\right]
\end{equation}
with compressibility parameter $K$, bulk binding energy $B$, and
$u=\rho/\rho_s$, where $\rho_s$ is the saturation density. In the case
that $\alpha_o\simeq0$, one has $\rho_o=\rho_s=0.16$ MeV fm$^{-3}$.
When $\alpha_o>0$, we can redefine $n_s$ to be the new central density
and $B$ to the new bulk binding for that neutron excess, and still
keep the quadratic behavior for ${\cal H}_B(n,0)$. With the choice
${\cal Q}(n)=Q/n$, we now find the equation for the isoscalar density
as a function of position:
\begin{equation}\label{bulk1}
{du\over dz}=-u(1-u),\qquad a=3\sqrt{2Q\over K},
\end{equation}
where $u=n/n_s$ and $z=r/a$, which defines the surface thickness
parameter $a$. This has the solution of a Fermi function, or
Woods-Saxon distribution,
\begin{equation}\label{fermi}
u={1\over1+e^{z-y}}
\end{equation}
where $y$ is a constant of integration, determined from the first
constraint:
\begin{eqnarray}
A&=&\int nd^3r=4\pi n_o a^3F_2(y),\cr
\hspace*{-.5cm}F_i(y)&=&\int_0^\infty\!\!\!\!{z^idz\over1+e^{z-y}}\simeq{y^{i+1}\over i+1}\left[1+{i(i+1)\over6}\left({\pi\over y}\right)^2\right].
\end{eqnarray}
Here $F_i$ is the usual Fermi integral, and the right-most
approximation holds for $y>>1$ and $i\ne-1$, ignoring an exponentially
small term. This is justified, since one finds that $y\simeq
r_oA^{1/3}/a\simeq13$ for $^{208}$Pb (the value for $a$ is determined
below). The choice of ${\cal Q}(n)$ results in the Woods-Saxon density
distribution assumed by Ref. \cite{LS89}.

The parameter $K\simeq240$ MeV from experiment, and the value of $Q$
follows from the observed value of the 90-10 surface thickness:
\begin{equation}\label{bulk2}
t_{90-10}=a\int_{0.1}^{0.9}{du\over u^\prime}=4a\ln(3)\simeq2.3{\rm~fm},
\end{equation}
giving $a=0.523$ fm and
\begin{equation}\label{bulk3}
Q={K\over18}\left({t_{90-10}\over4\ln(3)}\right)^2\simeq3.65{\rm~MeV~fm}^2.
\end{equation}
As a check, the liquid droplet surface tension parameter is the
semi-infinite, symmetric matter, surface thermodynamic potential per
unit area:
\begin{eqnarray}\label{bulk4}
\hspace*{-.5cm}\sigma_o&=&\int[{\cal H}-\mu n]dz=2Q\int_0^\infty{n^{\prime2}\over n}dz\cr
&=&{2Qn_o\over a}\int_0^1(1-u)du
={Qn_o\over a}\simeq1.17{\rm~MeV~fm}^{-2}.
\end{eqnarray}
This gives a value $E_s=4\pi r_o^2\sigma_o\simeq19.2$ MeV for the
symmetric matter surface energy parameter in the liquid droplet model,
which is very close to the accepted value \cite{Myers69,MS66}.
Therefore, the simple energy density functional we assume fits the
most important observed properties of the symmetric matter nuclear
interface, its tension and thickness, as well as the observed nuclear
incompressibility.

Although Lipparini and Stringari assumed a simple form for
$v_{sym}(n)$, this is not necessary to find analytic solutions. We
note that the function $S_2(n)$ can be represented by the series
expansion $S_2(u)=(\sum_ib_iu^i)^{-1}$ in the domain $0<u<1$, and that
integrals of the form $\int r^jv_{sym}^{-1}d^3r$ are analytically
expressible in terms of a series expansion of integer Fermi integrals:
\begin{equation}\label{sym13}
\int {r^j\over v_{sym}}d^3r=4\pi\rho_oa^{3+j}\left[{F_{2+j}(y)\over S_v}-(2+j){\cal T}F_{1+j}(y)+\cdots\right],
\end{equation}
where ${\cal T}$ is given by the series expansion
\begin{equation}\label{texp}
{\cal T}=b_1+3b_2/2+11b_3/6+25b_4/12+137b_5/60\cdots
\end{equation} 
for any $j$.  Note that $\sum_ib_i=S_v^{-1}$.
For example, the conventional density expansion
\begin{equation}\label{sym6}
S_2(u)\simeq S_v+{L\over3}(u-1)+{K_{sym}\over18}(u-1)^2+\cdots,
\end{equation}
keeping just the first three terms in the expansion of $S_2^{-1}$, leads to
\begin{eqnarray}\label{sym1}
b_0&=&{1\over S_v}\left[1+{L\over3S_v}+\left({L\over3S_v}
\right)^2-{K_{sym}\over18S_v}\right],\cr
b_1&=&{1\over3S_v^2}\left({K_{sym}\over3}-L-{2L^2\over3S_v}\right),\cr
b_2&=&{1\over18S_v^2}\left({2L^2\over S_v}-K_{sym}\right),\cr
{\cal T}&=&{1\over3S_v^2}\left[{K_{sym}\over12}-L-{L^2\over6S_v}\right].
\end{eqnarray}

The total symmetry energy of a nucleus, neglecting Coulomb effects,
now becomes, after expanding the Fermi integrals in powers of $y$ and
keeping the first two terms,
\begin{equation}\label{sym3}
E_{sym0}={(N-Z)^2\over H}\simeq AI^2S_v\left[1-{3S_v{\cal T}\over y}\right]^{-1}.
\end{equation}
This expression is identical to the liquid droplet model symmetry
energy in the same approximation,
$E_{sym0}=AI^2S_v/[1+S_sA^{-1/3}/S_v]$, if we identify
\begin{equation}\label{sym4}
S_s=-{3aS_v^2{\cal T}\over r_o}.
\end{equation}
This important result is the generalized hydrodynamical model
prediction for $S_s(S_2)$. It is essentially the same as the result
for $S_s$ established by Steiner et al.\cite{Steiner05}, which is
\begin{equation}\label{sple}
{S_s\over S_v}\propto\int_0^1u\left[{S_v\over S_2(u)}-1\right]{\cal Q}^{-1/2}{du\over u^\prime}
\end{equation}
We observe that the simple linear approximation $S_2\simeq
S_v+(L/3)(u-1)$ adopted by Ref. \cite{LS89} implies that $S_s\approx
aL/r_o$.

We note that keeping higher-order terms in $y$ in the expansions of the Fermi
integrals would allow determination of curvature and constant
contributions to the symmetry energy. The contributions of
these terms has not yet been carefully studied.

Other important results stemming from Eq. (\ref{sym13}), keeping the
lowest-order terms in $y$, are
\begin{eqnarray}\label{sym5}
\int{r^j\over v_{sym}}d^3r&\simeq&{A\over S_v}\left({3\over3+j}+{S_s\over S_v}A^{-1/3}\right),\cr
H&\simeq&{A\over S_v}\left(1+{S_s\over S_v}A^{-1/3}\right),\cr
\alpha_D&\simeq&{AR^2\over20S_v}\left(1+{5\over3}{S_s\over S_v}A^{-1/3}\right),
\end{eqnarray}
exactly as anticipated by Equations (\ref{ld1}) and (\ref{dp}).

For the neutron skin thicknesses, we use Equations (\ref{skin4}),
(\ref{sym5}) and (\ref{skin10}) to find
\begin{eqnarray}\label{skin5}
R_n-R_p&=&{2Ir_o\over3}{S_s\over Sv}\left(1+{S_s\over S_v}A^{-1/3}\right)^{-1}\times\cr
&\times&\left[1-I^2\left(1+{S_s\over S_v}A^{-1/3}\right)^{-2}\right]^{-1},
\end{eqnarray}
which is also very similar to the liquid droplet result when Coulomb effects are neglected.   Similarly, we find
\begin{equation}\label{rskin}
r_{np}\simeq\sqrt{3\over5}{2Ir_o\over3\sqrt{1-I^2}}{S_s\over S_v}\left(1+{S_s\over S_v}A^{-1/3}\right)^{-1}.
\end{equation}

\subsection{Inclusion of Coulomb effects}
In nuclei, the charge repulsion among the protons redistributes
neutrons and protons and reduces the neutron skin thickness. To take
this into account, and to extend the model of \cite{LS89}, we now
include a Coulomb contribution ${\cal H}_C=n_p V_C/2$ in the energy
density ${\cal H}$, where, in spherical symmetry, the Coulomb
potential is
\begin{equation}\label{coul}
V_C(r)={e^2\over r}\int_0^r n_p(r^\prime)d^3r^\prime+\int_r^\infty {e^2\over r^\prime} n_p(r^\prime)d^3r^\prime.
\end{equation}
If the protons are uniformly distributed for $r<R$,
\begin{equation}\label{coul1}
V_C={Ze^2\over R}\left({3\over2}-{r^2\over2R^2}\right)
\end{equation}
for $r<R$ and $V_C=Ze^2/r$ for $r>R$. We have found that a reasonable
approximation for a Woods-Saxon proton distribution, and one that
keeps the model analytic, is provided by assuming Equation
(\ref{coul1}) to apply for all $r$. Furthermore, the Coulomb potential
and the total Coulomb energy when the Coulomb potential is
self-consistently determined are adequately described by the same
approximation. Where the discrepancy between this approximation and
the real potential is large, the proton density is small. In addition,
we will assume that $V_C$ does not significantly alter the total
density $n(r)$
so that the relations derived in Equation (\ref{sym5}) remain valid
whether or not one considers the effects of the Coulomb potential.

For the moment, consider the limit in which  the effects of
the Coulomb potential on the asymmetry density $\alpha$ are negligible,
so that $\alpha=(N-Z)/(v_{sym}H)$.  Then, the total Coulomb energy is
\begin{equation}\label{coul2}
E_{C0}={1\over2}\int n_pV_Cd^3r={1\over4}\int nV_Cd^3r-{G\over4}{N-Z\over H},
\end{equation}
where
\begin{eqnarray}\label{newham1}
G&=&\int{V_C\over v_{sym}}d^3r={3\over2}{Ze^2\over R}\left(H-{4\alpha_D\over R^2}\right)\cr
&=&{6\over5}{Ze^2A\over RS_v}\left(1+{5\over6}{S_s\over S_v}A^{-1/3}\right).
\end{eqnarray}
Assuming that the
overall nucleon density $n$ retains the Fermi profile of Equation
(\ref{fermi}), we find
\begin{eqnarray}\label{coul3}
E_{C0}&=&{Ze^2\over R}\Biggl({3\pi\over2}n_oa^3\left[F_2(y)-{a^2\over3R^2}F_4(y)\right]-\cr
&-&{3\over8}(N-Z)\left[1-{4\alpha_D\over HR^2}\right]\Biggr)\cr
&\simeq&{3Z^2e^2\over5R}\left[1+{N-Z\over12Z}{S_sA^{-1/3}\over S_v+S_sA^{-1/3}}\right],
\end{eqnarray}
where we keep only the lowest order terms in $y$ and so ignore
diffuseness corrections.  The second term in the brackets of the last
line is generally of order 1\% of the first term for heavy nuclei, and we therefore
find, as expected, that $E_{C0}$ is essentially the same as in the case of a uniform proton
distribution.  This indicates that the adopted shape of $V_C$ is
relatively unimportant, justifying our approach.

To include the effects of the Coulomb potential on the asymmetry density, we
perform the second variation in Equation (\ref{ham1}):
\begin{equation}\label{newham}
\alpha={\bar\mu+fV_C/4\over2v_{sym}}={N-Z-f(G-V_CH)/8\over v_{sym}H}.
\end{equation}
The factor $f=1+\partial\ln V_C/\partial\ln n_p$ depends on
whether $V_C$ is assumed to depend on $n_p$ for the
purposes of the variational operator $\delta\alpha$.  Since the
variation of the total energy is done at fixed $N$ and $Z$, it could
be argued that $f=1$.  On the other hand, the definition of $V_C$
(Equation \ref{coul}) shows it is proportional to $n_p$, in which
case $f=2$.  We will assume $f=2$ in our subsequent calculations.

We now observe that
\begin{eqnarray}\label{newham3}
\alpha_o&=&{n_o\over S_vH}\left[N-Z-{f\over8}\left(G-{3\over2}H{Ze^2\over R}\right)\right]\cr
&=&{n_oA\over S_vH}\left[I+{3f\over4}{Ze^2\over R}{\alpha_D\over AR^2}\right],
\end{eqnarray}
showing that polarization effects result in an increase in asymmetry
near the center of a nucleus, as previously noted by Danielewicz \cite{Danielewicz03}.

The total symmetry and Coulomb energy is
\begin{eqnarray}\label{newham2}
E_{sym}&+&E_C=\int v_{sym}\alpha^2 d^3r+{1\over2}\int n_pV_C d^3r\cr
&\simeq&E_{sym0}+E_{C0}+{3f(2-f)\over11200}{Z^2e^4A\over R^2S_v}\times\cr
&\times&\left(1+{10S_s\over3S_v}A^{-1/3}\right)\left(1+{S_s\over S_v}A^{-1/3}\right)^{-1}.
\end{eqnarray}
When $f=2$, the last term vanishes; even if
$f=1$, its magnitude is negligible in comparison to the other terms,
being at most 1 MeV, or .1\% of the leading terms, for
$^{208}$Pb.  Thus, the total symmetry and Coulomb energy is barely
affected by the inclusion of Coulomb potential effects on the
distributions of neutrons and protons.

The major role of the Coulomb energy on the symmetry energy
correlations arises from the term proportional to $N-Z$ in $E_{C0}$ in Equation
(\ref{coul3}).  
We note this term is the same as in the liquid droplet model \cite{Steiner05} but is
1/2 the value derived by Danielewicz \cite{Danielewicz03}.
This term affects the derived slope of the correlation between $L$ and
$S_v$, which is therefore different from that in Ref. \cite{Danielewicz03}.
Given that $I\simeq0.2$ for typical heavy nuclei, the $N-Z$ term
increases the correlation slope $dL/dS_v$ by about 11\% in our
approach compared to about 25\% in Danielewicz's.  Steiner et
al. \cite{Steiner05} noted that Danielewicz's approach (labelled the
``$\mu_\alpha$'' model) led to an offset and an appreciably steeper correlation than
the liquid droplet approach (labelled the ``$\mu_n$'' model) which are
apparently incompatible with Thomas-Fermi and Hartree-Fock fits to
nuclear binding energies (such as those found by Refs. \cite{Kortelainen10,MS90,Moller12}).
 In spite of the symmetric treatment of neutrons and
protons in the ``$\mu_\alpha$'' approach, which physically seems more
justified than the asymmetric teatment of the liquid droplet approach \cite{MS66}, 
Ref. \cite{Steiner05} was unable to satisfactorily explain this discrepancy.  We conclude
that the ``$\mu_\alpha$'' model overestimates the Coulomb modifications to the
total energy.
 
Interestingly, including Coulomb potential effects on the asymmetry in nuclei
does not change the dipole polarizability.  Applying the dipole
constraint, as in Equation (\ref{dipole}), one finds
\begin{eqnarray}\label{dipcoul}
\alpha_d&=&{\epsilon z+fV_C/2\over2v_{sym}},\cr 
\alpha_D&=&{1\over4\epsilon}\int z{\epsilon z+fV_C/2\over v_{sym}}d^3r={1\over12}\int{r^2\over v_{sym}}d^3r,
\end{eqnarray}
where the second term in the middle expression for $\alpha_D$ vanishes
because of symmetry.  

To examine the role of Coulomb effects on the neutron skin thickness, we now note that
\begin{equation}\label{newham4}
{n_n\over n_{no}}-{n_p\over n_{po}}={2n_o\over n_o^2-\alpha_o^2}\left[\left({S_v\over v_{sym}}-n\right){\alpha_o\over n_o}-{f\over16v_{sym}}{Ze^2\over R}{r^2\over R^2}\right].
\end{equation}
The neutron skin thickness when Coulomb effects are included becomes
\begin{eqnarray}\label{skincoul}
{4\pi\over3}(R_n^3&-&R_p^3)={2n_o\over n_o^2-\alpha_o^2}\times\cr
&\times&\left[(S_vH-A){\alpha_o\over n_o}-{3Ze^2f\over4R}{d\over R^2}\right],
\end{eqnarray}
or
\begin{eqnarray}\label{skincoul1}
R_n-R_p&\simeq&{2r_o\over3}\left[I{S_s\over S_v}-{3Ze^2f\over80r_oS_v}\left(1+{5S_s\over3S_v}A^{-1/3}\right)\right]\times\cr
& &\left(1+{S_s\over S_v}A^{-1/3}\right)^{-1}\left[1-\left({\alpha_o\over\rho_o}\right)^2\right]^{-1}.
\end{eqnarray}
A related result was found in
  Ref.~\cite{Danielewicz03}:
\begin{eqnarray}\label{dan}
R_n-R_p&\simeq&{2r_o\over3(1-I^2)}{S_s\over S_v}\left[I-{Ze^2\over20RS_v}\right]\times\cr
& &\left(1+{S_s\over S_v}A^{-1/3}+{e^2A^{2/3}\over80RS_v}{S_s\over S_v}\right)^{-1}.
\end{eqnarray}
In both models Coulomb effects reduce the neutron skin thickness for
neutron-rich nuclei, and they induce a proton skin in symmetric
nuclei. However, the Coulomb effects are much larger within the
hydrodynamical model. It is straightforward to see how this leads to a
negative correlation between $L$ and $S_v$. If we approximate
$S_s\simeq(a/r_o)L$, and ignore the last term in the denominator of
Equation (\ref{skincoul1}), one finds by variation at fixed $R_n-R_p$
\begin{eqnarray}\label{skincoul2}
&&{dL\over L}\left(I-{2\over3}{P\over S_v}A^{-1/3}\right)\simeq{dS_v\over S_v}\times\cr
&\times&\left[I-{P\over S_s}\left(1+{10S_s\over3S_v}A^{-1/3}+{5S_s^2\over3S_v^2}A^{-2/3}\right)\right],
\end{eqnarray}
where $P=3Ze^2f/(80r_o)\simeq7.8$ for $f=2$ and for $^{208}$Pb. As
long as $S_s/S_v>0$, the bracket on the second line is negative and
$dL/dS_v<0$. According to Equation (\ref{dan}),
the equivalent expression for the treatment of Ref. \cite{Danielewicz03} is
\begin{equation}\label{skincoul3}
{dL\over L}\left(I-{q\over S_v}\right)\simeq{dS_v\over S_v}\left(I-{2q\over S_v}\right),
\end{equation}
where $q=Ze^2/(20R)\simeq0.875$ for $^{208}$Pb.  Since $I>2q/S_v$,
however, the correlation $dL/dS_v$ is always positive in this approach.

Nevertheless, Coulomb
effects contribute to an even greater reduction in the mean-square neutron skin thickness.  In the hydrodynamical
model with Coulomb effects included, the mean-square radii are
\begin{eqnarray}\label{skin12}
r^2_{n,p}&=&{1\over2(N,Z)}\int(n\pm\alpha)r^2d^3r\cr
&\simeq&{3\over5}{R^2\over1\pm I}\Biggl[1\pm I{1+{5\over3}{S_s\over S_v}A^{-1/3}\over1+{S_s\over S_v}A^{-1/3}}\cr
&\mp&{Ze^2f\over140RS_v}{1+{10\over3}{S_s\over S_v}A^{-1/3}\over1+{S_s\over S_v}A^{-1/3}}\Biggr],
\end{eqnarray}
where the upper (lower) sign refers to $n(p)$.  We then find
\begin{eqnarray}\label{skin13}
r_{np}&\simeq&\sqrt{3\over5(1-I^2)}{2r_o\over3}\left(1+{S_s\over S_v}A^{-1/3}\right)^{-1}\times\cr
&\times&\left[I{S_s\over S_v}-{3Ze^2f\over280r_oS_v}\left(1+{10\over3}{S_s\over S_v}A^{-1/3}\right)\right].
\end{eqnarray}
Note that the coefficient of the last term in the square brackets is 2/7 of that in Equation (\ref{skincoul1}).  It is interesting
that Ref. \cite{Danielewicz03} obtained a nearly identical result for $r_{np}$ (if
$f=2$ is assumed) in spite of the fact that their results for $R_n-R_p$ and $E_{sym0}$ 
differ.  

If we were to approximate
$S_s\simeq aL/r_o$ and ignore the $1-I^2$ denominator term in
Equation (\ref{skin10}), then one finds by variation at fixed $r_{np}$:
\begin{eqnarray}\label{skin11}
&&{dL\over L}\left(I-{2\over3}{P\over S_v}A^{-1/3}\right)\simeq{dS_v\over S_v}\times\cr
&\times&\left[I-{2P\over7S_s}\left(1+{20S_s\over3S_v}A^{-1/3}+{10S_s^2\over3S_v^2}A^{-2/3}\right)\right],
\end{eqnarray}
The correlation $dL/dS_v$ is negative only as long as\break
$S_s/S_v\simle0.6$.  However, this result is particularly sensitive to
the dependence of $S_s$ on $S_v$ and $L$, as we  now show.

Suppose
\begin{eqnarray}\label{dep}
d\ln S_s&=&a~d\ln L + b~d\ln S_v;\cr 
d\ln r_{np}&=&\alpha~d\ln S_s-\beta~d\ln S_v.
\end{eqnarray}  
Then, holding $r_{np}$ fixed implies that
\begin{equation}\label{dep1}
{d\ln L\over d\ln S_v} = {\beta-b\alpha\over a\alpha}.
\end{equation}
The skin thickness correlation will be negative if $b>\beta/\alpha$.
Lattimer \& Lim \cite{Lattimer13} found that Thomas-Fermi
semi-infinite surface calculations of $S_s$ could be approximated by
\begin{equation}\label{dep2}
{S_s\over S_v}\simeq0.6461+{S_v\over97.85{\rm~MeV}}+0.4364{L\over S_v}+0.0873{L^2\over S_v^2}.
\end{equation}
Assuming $S_v\simeq31$ and $L\simeq45$ MeV, one finds $S_s/S_v\simeq1.76$,
$a\simeq0.56$, $b\simeq0.62$, $\alpha\simeq0.18$ and
$\beta\simeq0.07$, so that $d\ln L/d\ln S_v\simeq-0.4$.  It appears that the
hydrodynamical model formula, Equation (\ref{sym1}), 
underpredicts both $S_s$ and its dependence on $S_v$, possibly due to
the neglect of cross terms involving gradients of $\alpha$ in Equation (\ref{ham}).

Despite these shortcomings of the model, it allows for
a qualitative understanding of the role of the symmetry energy in
nuclear properties and for the correlations between $S_v$ and $L$ that
have been observed.  In particular, it demonstrates that a precision
measurement of the neutron skin thickness is extremely important
because its implied correlation between $S_v$ and $L$ is essentially
orthogonal to those provided by fitting nuclear masses and measuring dipole excitation energies.

\subsection{Comparison with experimental data\label{sec:comp}}

\begin{figure}
\vspace*{-0.9cm}\hspace*{-.5cm}\includegraphics[width=10cm]{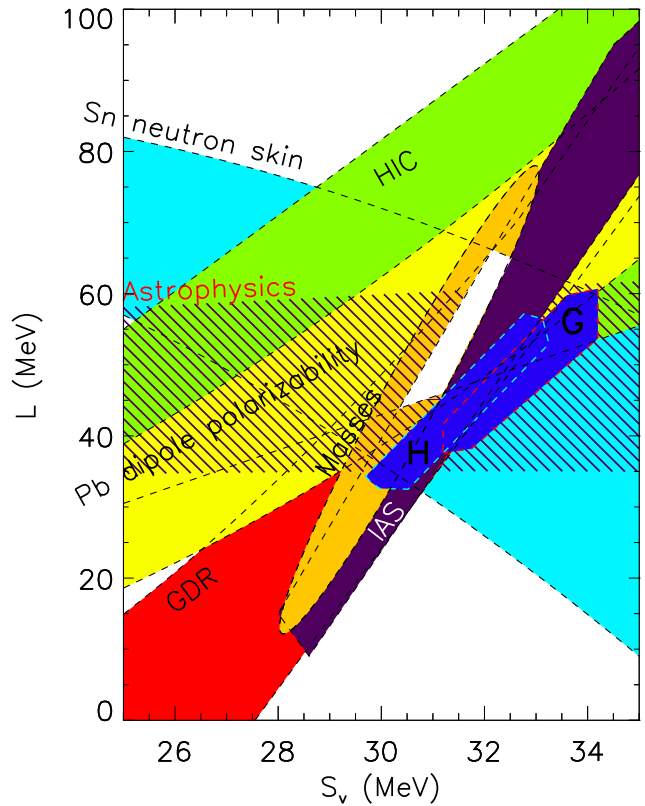}
\vspace*{-1.2cm}\caption{Experimental constraints for symmetry energy parameters, adapted and revised from \cite{Lattimer13}.  See the text for further discussion.  G and H refer to the neutron matter studies of Gandolfi et al. \cite{Gandolfi12} and Hebeler et al. \cite{Hebeler10}, respectively.\label{symcor}}
\end{figure}

It is now useful to compare experimental results for binding energies,
neutron skin thicknesses, dipole polarizabilities, and centroids of
giant dipole resonances.  These were recently reviewed in
Ref. \cite{Lattimer13} and we summarize the results in
Fig. \ref{symcor}.

The correlation between $L$ and $S_v$ for measured nuclear masses is taken from
Hartree-Fock calulations with the UNEDF0 density
functional\cite{Kortelainen10}, in which the nominal fitting error was arbitrarily
chosen to be $\sigma=2$ MeV.  In all likelihood, this value is overestimated, as negative values for $L$, which give
negative neutron matter pressures, are reached.  As a result, the
confidence ellipse is probably smaller than given in Ref. \cite{Kortelainen10} and we have instead chosen the value
$\sigma=1$ MeV.  Importantly, the shape and
orientation of the ellipse are the same as predicted by the liquid droplet model, Eq. (\ref{ld1}), once the dependence of $S_s$ on $S_v$ and $L$ is taken into account. 

The constraints for the neutron skin thickness of $^{208}$Pb are taken
from a study by Chen et al. \cite{Chen10}, who
converted the experimental results\cite{Ray79,Krasznahorkay94,Krasznahorkay99,Trzeinska01,Klimkiewicz07,Terashima08} for Sn isotopes into an equivalent value for $^{208}$Pb:
$r_{np}\simeq(0.175\pm0.020)$ fm.  Performing a series of Skyrme
Hartree-Fock calculations of $^{208}$Pb, in which values of $S_v$ and
$L$ were systematically varied, they also established that
\begin{eqnarray}\label{skinpb}
{r_{np}\over{\rm fm}}&\simeq&-0.094669+{7.2028S_v\over{\rm GeV}}+{2.3107L\over{\rm GeV}}\cr
&-&{8.8453S_v^2\over{\rm GeV}^2}-{47.837S_vL\over{\rm GeV}^2}+{4.003L^2\over{\rm GeV}^2}.
\end{eqnarray}
This formula, with the aforementioned value for $r_{np}$, establishes the
correlation slope $d\ln L/d\ln S_v\simeq-3.75$, assuming $S_v=31$ MeV and $L=45$
MeV. This result is somewhat steeper than the slope predicted by the
hydrodynamical model (which is essentially flat), and it should be
explored with a greater variety of effective interactions.

Similarly, the constraint for the electric dipole polarizability
$\alpha_D$ of $^{208}$Pb is taken from data produced by Tamii et
al. \cite{Tamii11}: $\alpha_D\simeq(20.1\pm0.6)$ fm$^3$.  Roca-Maza et
al. \cite{Roca-Maza13} showed, from studies with a series of
relativistic and non-relativistic interactions, that the dipole
polarizability, bulk symmetry parameter, and the neutron skin
thickness for $^{208}$Pb can be constrained by
\begin{equation}\label{ajr}
\alpha_DS_v\simeq(325\pm14)+(1799\pm70)(r_{np}/{\rm fm}){\rm~Mev~fm}^3.
\end{equation}
By use of Equation (\ref{skinpb}), this is converted onto the $S_v-L$
correlation shown in Figure \ref{symcor}.  

Equation (\ref{ajr}) is not functionally the same as that established
from the hydrodynamical model, whose dependence on $S_v$ and $S_s$ can
be gleaned from Equations (\ref{dipcoul}) and (\ref{skin13}). Although
the predicted slopes of both correlations are similar, being positive
and somewhat less steep than that from nuclear masses, the functional
difference suggests that the correlation of Ref. \cite{Roca-Maza13}
might retain some model-dependence and should be further explored.
Note that the slope of this correlation is significantly different
than shown in \cite{Lattimer13}, which relied on the analysis in
Ref.~\cite{Reinhard10} that erroneously concluded $\alpha_D\propto
r_{np}$.

The constraint for the centroid energy of the giant dipole resonance
for $^{208}$Pb is taken from Trippa, Col\'o and Vigezzi
\cite{Trippa08}. They concluded that the measured energy was best fit
by those forces having a bulk symmetry energy $S_2(0.1)$, evaluated at
the density $n=0.1$ fm$^{-3}$, in the range
$S_2(0.1)\simeq(24.1\pm0.9)$ MeV. This symmetry energy value can be
converted into a correlation between $S_v$ and $L$ for a given nuclear
force model. Lattimer \& Lim \cite{Lattimer13} deduced the band shown
in Figure \ref{symcor} by studying a wide range of plausible density
functionals. Unfortunately, the hydrodynamical model does not yield an
analytic prescription for this correlation, but one should expect it
to be similar to that of the dipole polarizability, which is borne out
by the results shown in Figure \ref{symcor}.

Additional correlations depicted in Figure \ref{symcor} are due to
studies of isospin diffusion in heavy-ion collisions \cite{Tsang:2009}
and energies of excitations to isobaric analog states
\cite{Danielewicz13}. The model-dependence of the former analysis has
not been fully explored, although the depicted results are consistent
with multifragmentation studies in intermediate-energy heavy ion
collisions \cite{Shetty07} which imply 40 MeV $<L<125$ MeV. Excitation
energies are sensitive to shell effects, but are closely related to
the energies of nuclear ground states. Thus, the latter correlation
bears a great deal of resemblance to that determined from nuclear
binding energies. Danielewicz \& Lee also utilized
measurements\cite{Ray79,Friedman12,Clark03,Zenhiro10,Starodubsky94} of
the $^{208}$Pb skin thickness to further restrict the allowed $S_v-L$
parameter space. The weighted average of neutron skin thicknesses they
employed was $r_{np}=(0.179\pm0.023)$ fm, 0.004 fm larger than
determined by Ref. \cite{Chen10} and used in Fig. \ref{symcor}. Their
combined analysis of isobaric analog states and skin thickness
measurements result in the restriction 33 MeV $<L<72$ MeV (not shown
in Fig. \ref{symcor}).

In contrast, the white region displayed in Fig. \ref{symcor}
represents the consensus agreement of the six experimental constraints
we have discussed, giving a somewhat smaller range 44 MeV $<L<66$ MeV.
(In comparison to the consensus region found in Ref.
\cite{Lattimer13}, the region displayed in Fig. \ref{symcor} is
slightly smaller because of the incorporation of the additional
constraint from isobaric analog states.) Since the model dependencies
of these constraints have not been thoroughly explored, the size of
this consensus region may well be underestimated. If we treat the
white region as a 68\% confidence interval for the experimental
determination of $S_v$ and $L$, it can be used with Monte Carlo
sampling to determine a distribution of neutron star matter pressures
by means of Eq. (\ref{sat1}). This can then be combined with further
Monte Carlo sampling of Eq. (\ref{radius}), whose uncertainty reflects
one standard deviation, to determine the confidence interval for radii
of $1.4M_\odot$ stars: $R_{1.4}\simeq(12.1\pm1.1)$ km to 90\%
confidence. (Employing the $r_{np}$ constraint of Ref.
\cite{Danielewicz13} instead of Ref. \cite{Chen10}, the upper limit to
$L$ increases to about 69 MeV and the lower and upper limits to
$r_{np}$ increase by about 0.1 km.) As we will see, this range is
quite compatible with several astrophysical observations.

\subsection{Neutron matter studies}

Two recent studies of pure neutron matter using realistic two- and
three-nucleon interactions coupled with low-energy scattering phase
shift data, the first employing chiral Lagrangian methods
\cite{Hebeler10} and the second using quantum Monte Carlo techniques
\cite{Gandolfi12}, can also render constraints on symmetry energy
coefficients and neutron star radii \cite{Steiner:2012,Hebeler13}.  With the
important assumption that higher-than-quadratic terms in Eq.
(\ref{energy}) are ignored, the values of the neutron matter energy
and pressure at $n_s$ provide direct estimates of $S_v$ and $L$.  The
estimated error ranges for the symmetry parameters determined
\cite{Hebeler10} from neutron matter studies are also displayed in
Fig. \ref{symcor}.  These estimates are very consistent with those
determined from nuclear experiments, but their small displacement may represent the effects of neglecting quartic or
higher-order terms in the symmetry energy expansion.

\section{Estimates of Neutron Star Radii}
Although nearly three dozen neutron star masses have been determined
very accurately \cite{Lattimer12}, there are no precise simultaneous
measurements of a star's mass and radius. To date, several different
astrophysical measurements of neutron star radii have been attempted.
We will focus attention on radius estimates inferred from photospheric
radius expansion (PRE) bursts and thermal emissions from quiescent
low-mass X-ray binaries (QLMXBs) and isolated neutron stars.
Unfortunately, no single observation, at the present time, can
reliably determine a neutron star radius to better than 20\% accuracy.
This translates into nearly a 100\% error for the determination of
$L$, since $L\propto R^4$ using Equations (\ref{sat1}) and
(\ref{radius}), which is substantially larger than the accuracy
afforded by nuclear experiments and neutron matter
theory~\cite{Lattimer13}. Moreover, with such large errors in $M$ and
$R$ for an individual source, the direct inversion of the neutron star
structure equations cannot credibly limit the pressure-density
relation.

Even taking the ensemble of measurements and attempting to invert the
neutron star structure equations to infer the $M-R$ relation is
problematic without physical guidance. How does one choose the
weighting for a particular EOS and $M-R$ curve: does one place more
emphasis on its passing close to the central values of the
measurements, or does one integrate the effective weight along the
entire $M-R$ curve? Fortunately, Bayesian techniques make it clear how
to add integrate the weight along the curve. The body of observations
can be coupled to the structure equations, as shown by \cite{SLB10},
to effectively determine the $M-R$ relation and, further, to obtain
estimates of the pressure-density relation of neutron star matter, not
only near $\rho_s$, but up to the highest densities found in neutron
star interiors. Relative estimated errors of the pressure of
high-density matter can be as large as 150\% but are generally much
smaller. The final results and model comparisons are given in \S
\ref{sec:bayes} and discussed in \S \ref{sec:disc}.

The nuclear symmetry energy is also connected to other outstanding
problems in nuclear astrophysics, including the crustal properties of
neutron stars, the possible onset of the direct Urca process at high
densities in the neutron star interior, and the quark-hadron
transition and the appearance of other exotica such as hyperons and
meson condensates.  
  
\subsection{Photospheric Radius Expansion Bursts}
\label{sec:PRE}
Accretion onto neutron stars in binaries often leads to X-ray bursts
from the unstable burning of the accreted material from its companion. The nuclear burning spreads across the stellar surface and gives
rise to a sudden increase in X-ray luminosity and temperature. Some
of these X-ray bursts are energetic enough to reach the so-called
Eddington limit at which radiation pressure is sufficiently large
to overcome gravity, leading to expansion of the star's photosphere.
These PRE bursts can constrain $M$ and $R$
because the largest flux during the burst must be near
the Eddington flux
\begin{equation}\label{edd}
F_{\rm Edd}={cGM\over\kappa R^2}(1+z)
\end{equation}
where $z=(1-2\beta)^{-1/2}-1$ is the redshift of the source and
$\beta=GM/Rc^2$ is the dimensionless compactness parameter.  The
observed Eddington flux is diluted by distance and is twice redshifted
(once for energy, once for time):
\begin{equation}\label{edd1}
F_{{\rm Edd},\infty}={cGM\over\kappa D^2}(1+z)^{-1}.
\end{equation}
It has been usually assumed that the flux measured when the effective
temperature is a maximum corresponds to the Eddington flux and that
``touchdown'' has occurred, i.e., the photosphere is coincident with
$R$~\cite{Ozel06}. If this is the case, the appropriate redshift to be
applied is as given above.  If the maximum temperature point is
reached while the photosphere is above the stellar surface, the
effective redshift might actually be negligible. In any case,
measurement of the peak flux (or Eddington flux) constitutes an
observable which is a function of $M, \kappa, D$, and possibly $R$.
The reproducability of the maximum flux from repeated energetic bursts
from the same source supports the identification of this flux with the
Eddington limit.

A second observable in these systems is the nearly constant angular
emitting area during the cooling tail several seconds after the
burst's peak. The emission is nearly thermal, and simultaneous
measurement of the observed (redshifted) flux and (redshifted)
temperature can yield an angular diameter if the effects of the
atmosphere, through the color correction factor
$f_c=T_c/T_{\mathrm{eff}}$ between effective temperature and color
temperature, are known:
\begin{equation}\label{flux}
A\equiv{F_\infty\over\sigma T_{{\rm eff},\infty}^4}=f_c^{-4}\left({R_\infty\over D}\right)^2,
\end{equation}
where $R_\infty=R(1+z)$ is the apparent radiation radius and $T_{{\rm
    eff},\infty}=T/(1+z)$ is the observed effective temperature.
Repeated bursts from the same source show the same emitting areas,
suggesting strongly that the entire neutron star surface is emitting
during the cooling tail and that non-spherically symmetric effects
during this phase are small. The observable $A$ is thus a function of
$M, R, f_c$ and $D$. With knowledge of $D, f_c$ and $\kappa$, the mass
and radius can be deduced from $F_{{\rm Edd},\infty}$ and $A$.

The observables $F_{{\rm Edd},\infty}$ and $A$ can be combined into
two parameters,
\begin{eqnarray}\label{ag}
\alpha&=&{F_{{\rm Edd},\infty}\over\sqrt{A}}{\kappa D\over f_c^2c^3}=\beta(1-2\beta),\cr
\gamma&=&{A\over F_{{\rm Edd},\infty}}{f_c^4c^3\over\kappa}={R\over\beta(1-2\beta)^{3/2}},
\end{eqnarray}
where we used Equation (\ref{edd1}) to establish the second set of
equalities.  These, in turn, can be solved for $M$ and $R$:
\begin{eqnarray}\label{sol}
\beta&=&{1\over4}\pm{1\over4}\sqrt{1-8\alpha},\cr
R&=&\alpha\gamma\sqrt{1-2\beta},\qquad
M={\alpha^{3/2}\gamma\beta^{1/2}c^2/G}.
\end{eqnarray}
Note that $\gamma$ is independent of $D$ and $R_\infty=\alpha\gamma$ is
independent of $\kappa$ and $F_{{\rm Edd},\infty}$.  For real
solutions to exist, $\alpha$ must be less than or equal to $1/8$.
However, we shall see that observations imply this condition is
usually not met.

\begin{table*}[t]
\begin{center}
  \caption{PRE X-ray bursters and estimated Eddington fluxes, angular areas and distances taken from the indicated references.  Values and uncertainties for $\alpha, \gamma$ and $R_\infty$ reflect assumptions about $f_c$ and $X$ as discussed in the text.\label{tab:2}}
\begin{tabular}{l|cccccc}
\hline\noalign{\smallskip}
PRE Source&$D$&$F_{{\rm Edd},\infty}$&$A$&$\alpha$&$\gamma$&$R_\infty$\\
&kpc&$10^{-8}$ erg cm$^{-3}$s$^{-1}$&km$^2$kpc$^{-2}$&&km&km\\
\noalign{\smallskip}\hline\noalign{\smallskip}
EXO 1745-248\cite{Ozel09}&$6.3\pm0.6$&$6.25\pm0.2$&$1.17\pm0.13$&$0.188\pm0.035$&$76.86\pm17.33$&$14.57\pm1.64$\\
4U 1608-522\cite{Guver10a}&$5.8\pm1.0$&$15.41\pm0.65$&$3.246\pm0.024$&$0.247\pm0.058$&$90.22\pm17.09$&$20.36\pm3.68$\\
4U 1820-30\cite{Guver10b}&$8.2\pm0.7$&$5.39\pm0.12$&$0.9198\pm0.0186$&$0.235\pm0.041$&$69.16\pm13.62$&$15.82\pm1.58$\\
KS 1731-260\cite{Ozel12}&$8.0\pm0.4$&$4.45\pm0.12$&$0.884\pm0.051$&$0.199\pm0.032$&$82.79\pm16.57$&$15.63\pm1.18$\\
SAX J1748.9-2021\cite{Guver13}&$8.2\pm0.6$&$4.03\pm0.44$&$0.897\pm0.096$&$0.177\pm0.036$&$97.64\pm23.08$&$15.74\pm1.61$\\
\noalign{\smallskip}\hline
\end{tabular}
\end{center}
\end{table*}

\begin{table*}[t]
\begin{center}
\caption{PRE X-ray burster solutions resulting from Monte Carlo trials with parameters taken from their uncertainty intervals.  Only solutions with real values of $R$ are accepted; the fraction of Monte Carlo acceptances is shown in the last column.\label{tab:3}}
\begin{tabular}{l|cccccc}
\hline\noalign{\smallskip}
PRE Source&$\alpha$&$\gamma$&$R_\infty$&R&M&acceptance\\
&&km&km&km&$M_\odot$&\%\\
\noalign{\smallskip}\hline\hline\noalign{\smallskip}
&\multicolumn{6}{c}{$z_{\rm ph}=z$}\\
\noalign{\smallskip}\hline\noalign{\smallskip}
EXO 1745-248&$0.117\pm0.006$&$109.0\pm14.2$&$12.77\pm1.62$&$9.11\pm1.55$&$1.45\pm0.28$&4.87\\
4U 1608-522&$0.115\pm0.010$&$110.8\pm16.4$&$12.73\pm2.22$&$9.21\pm1.74$&$1.41\pm0.38$&0.861\\
4U 1820-30&$0.121\pm0.004$&$103.4\pm7.5$&$12.48\pm0.96$&$8.81\pm1.04$&$1.46\pm0.19$&0.0311\\
KS 1731-260&$0.121\pm0.004$&$124.5\pm9.0$&$15.01\pm1.03$&$10.58\pm1.23$&$1.76\pm0.21$&1.01\\
SAX J1748.9-2021&$0.116\pm0.008$&$132.9\pm17.4$&$15.27\pm1.65$&$11.05\pm1.86$&$1.69\pm0.33$&9.67\\
\noalign{\smallskip}\hline\noalign{\smallskip}
&\multicolumn{6}{c}{$z_{\rm ph}=0$}\\
\noalign{\smallskip}\hline\noalign{\smallskip}
EXO 1745-248&$0.158\pm0.021$&$85.35\pm15.55$&$13.25\pm1.67$&$10.00\pm1.45$&$1.42\pm0.27$&66.3\\
4U 1608-522&$0.167\pm0.020$&$103.5\pm16.2$&$17.20\pm3.08$&$12.41\pm1.98$&$1.96\pm0.49$&20.7\\
4U 1820-30&$0.173\pm0.014$&$87.04\pm10.39$&$15.03\pm1.58$&$10.63\pm1.25$&$1.77\pm0.25$&24.5\\
KS 1731-260&$0.163\pm0.018$&$92.29\pm13.68$&$14.87\pm1.21$&$11.01\pm1.28$&$1.64\pm0.22$&59.2\\
SAX J1748.9-2021&$0.154\pm0.023$&$102.1\pm20.7$&$15.26\pm1.64$&$11.70\pm1.61$&$1.58\pm0.30$&72.9\\
\noalign{\smallskip}\hline
\end{tabular}
\end{center}
\end{table*}

\begin{figure*}
\hspace*{-1cm}\includegraphics[width=8.5cm,angle=180]{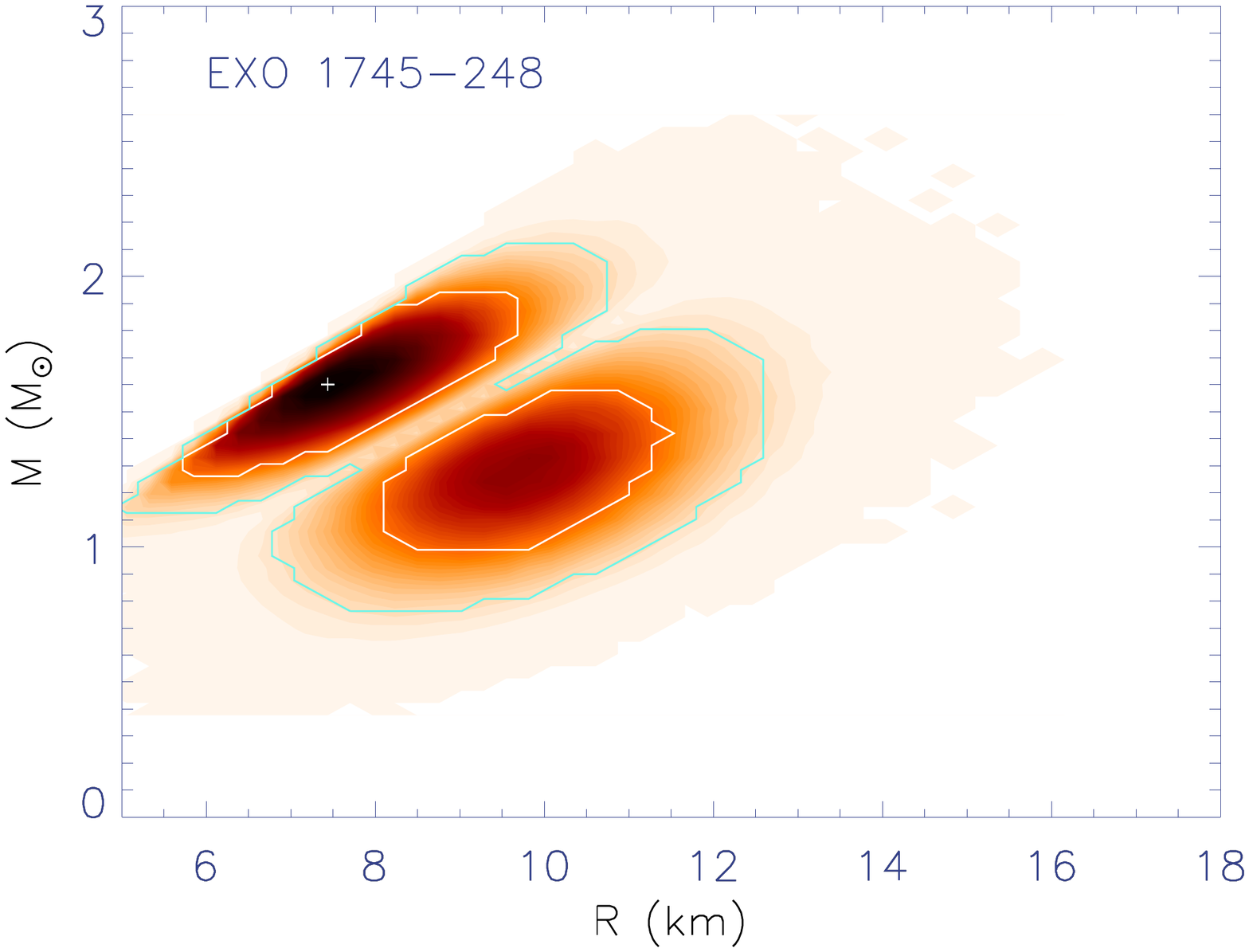}
\hspace*{-3.0cm}\includegraphics[width=8.5cm,angle=180]{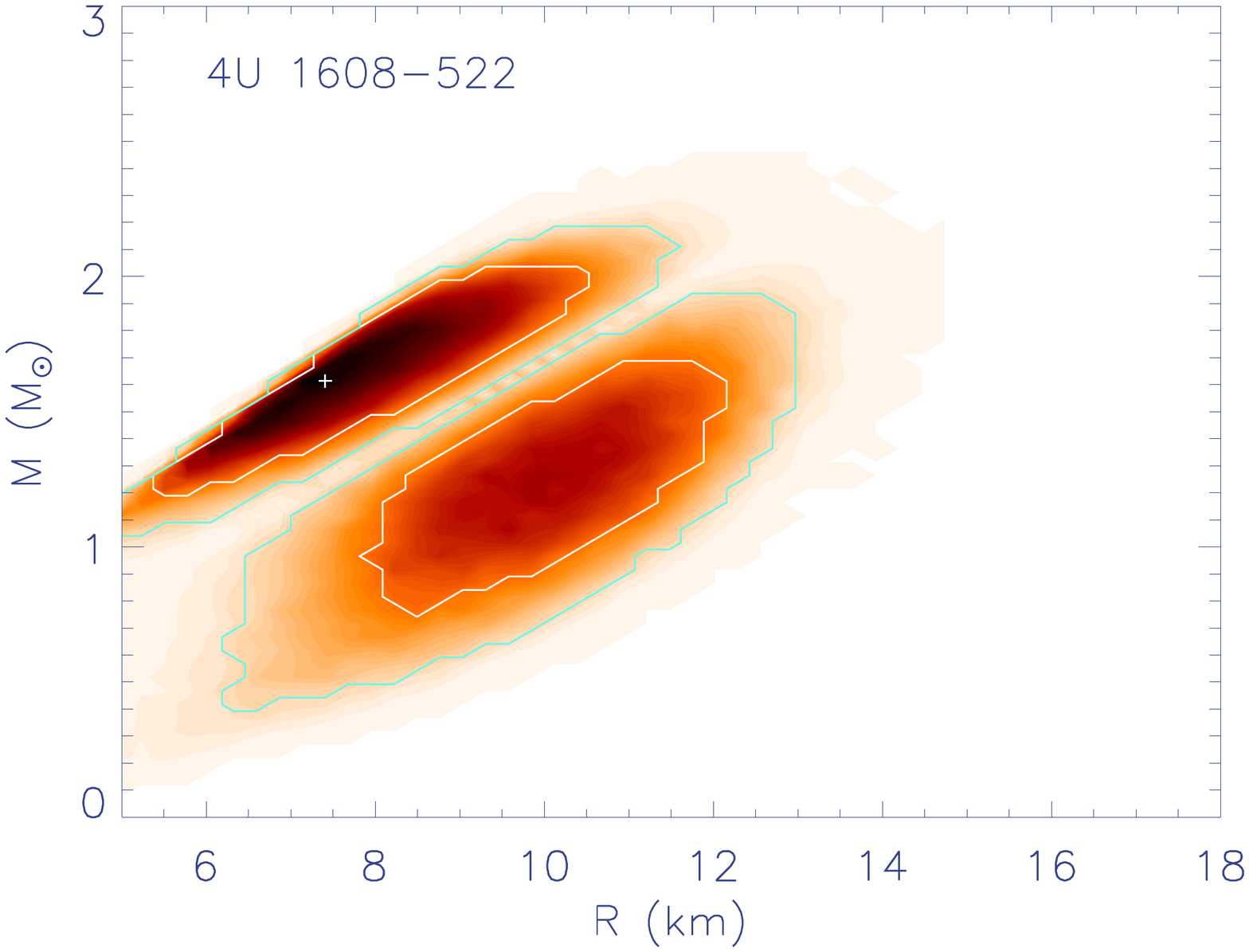}
\hspace*{-3.0cm}\includegraphics[width=8.5cm,angle=180]{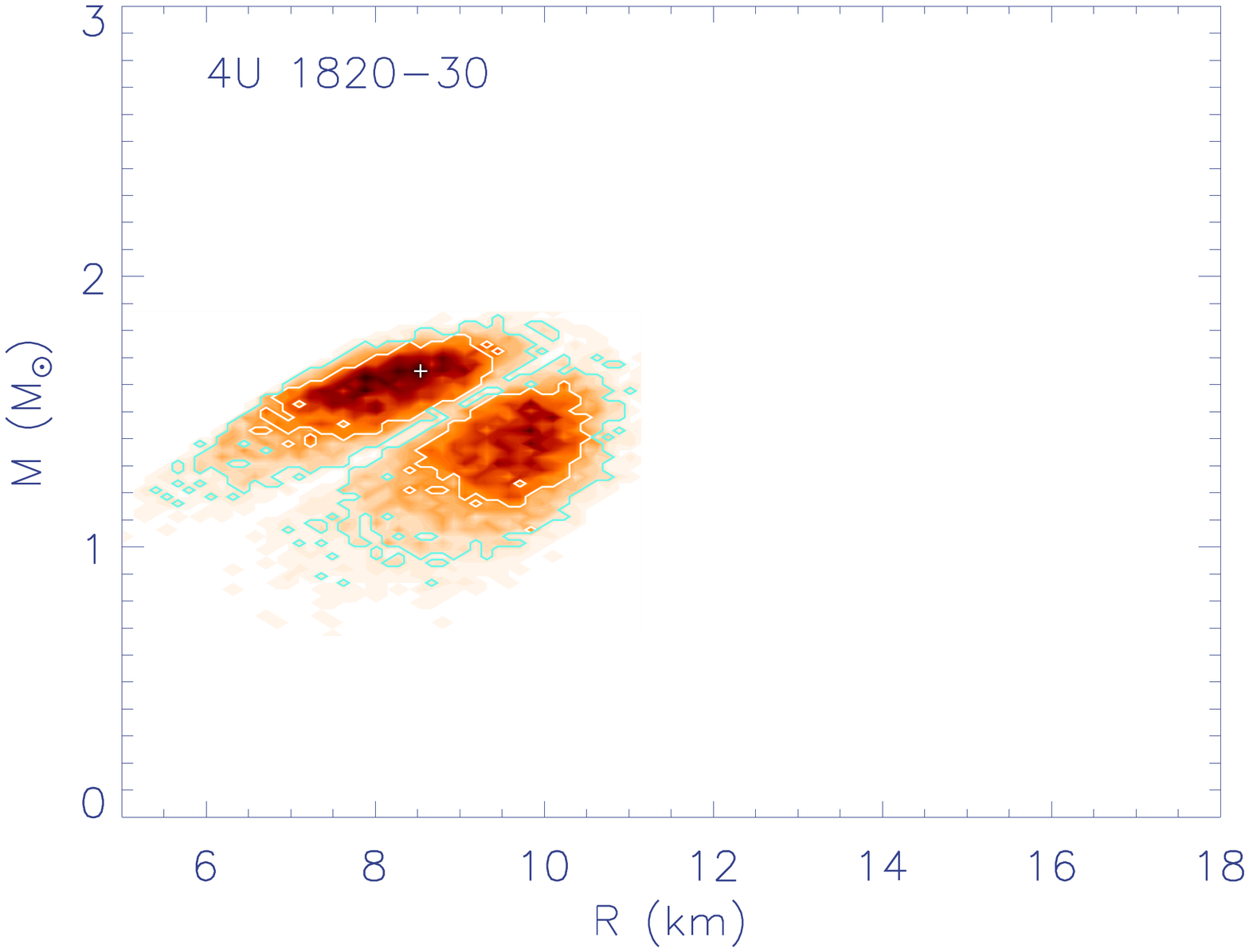}

\vspace*{-1.7cm}\hspace*{-1cm}\includegraphics[width=8.5cm,angle=180]{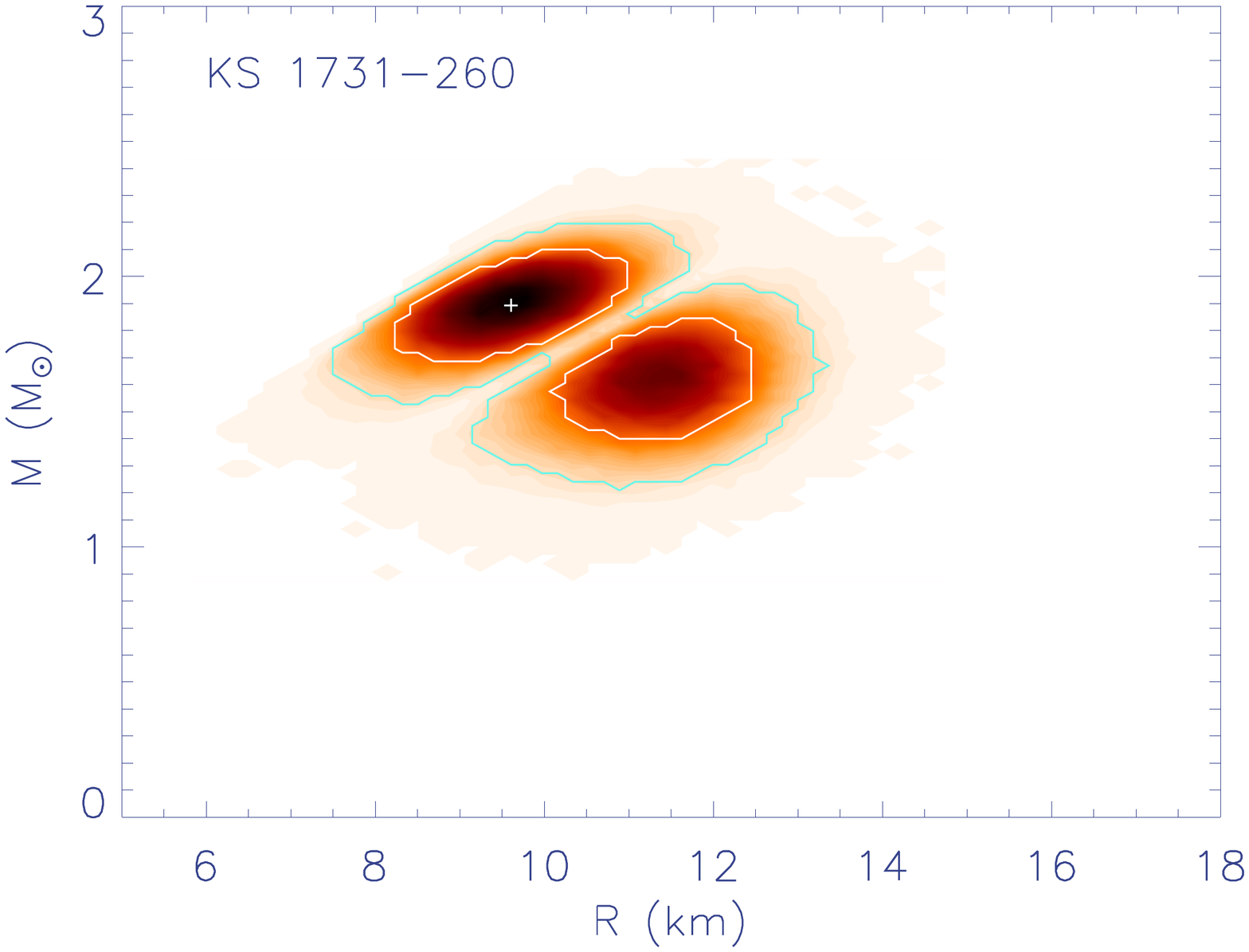}
\hspace*{-3.0cm}\includegraphics[width=8.5cm,angle=180]{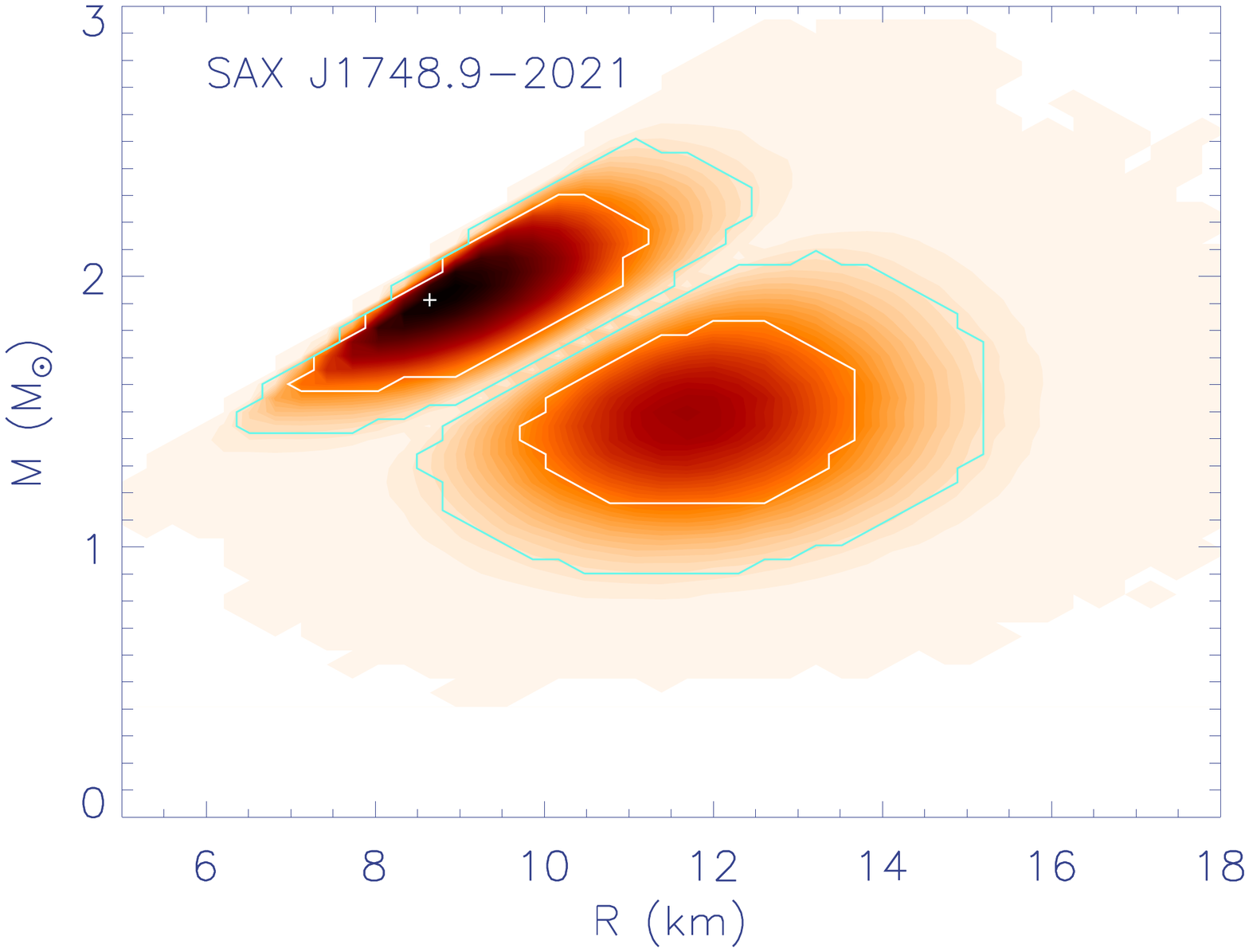}

\vspace*{-.5cm}\caption{$M-R$ probability contours for PRE X-ray burst sources assuming that $z_{\rm ph}=z$.  Crosses indicate maximum probabilities and white (green) contours show $1\sigma$ ($2\sigma$) uncertainty contours.\label{fig:1}}
\end{figure*}

\begin{figure*}
\hspace*{-1cm}\includegraphics[width=8.5cm,angle=180]{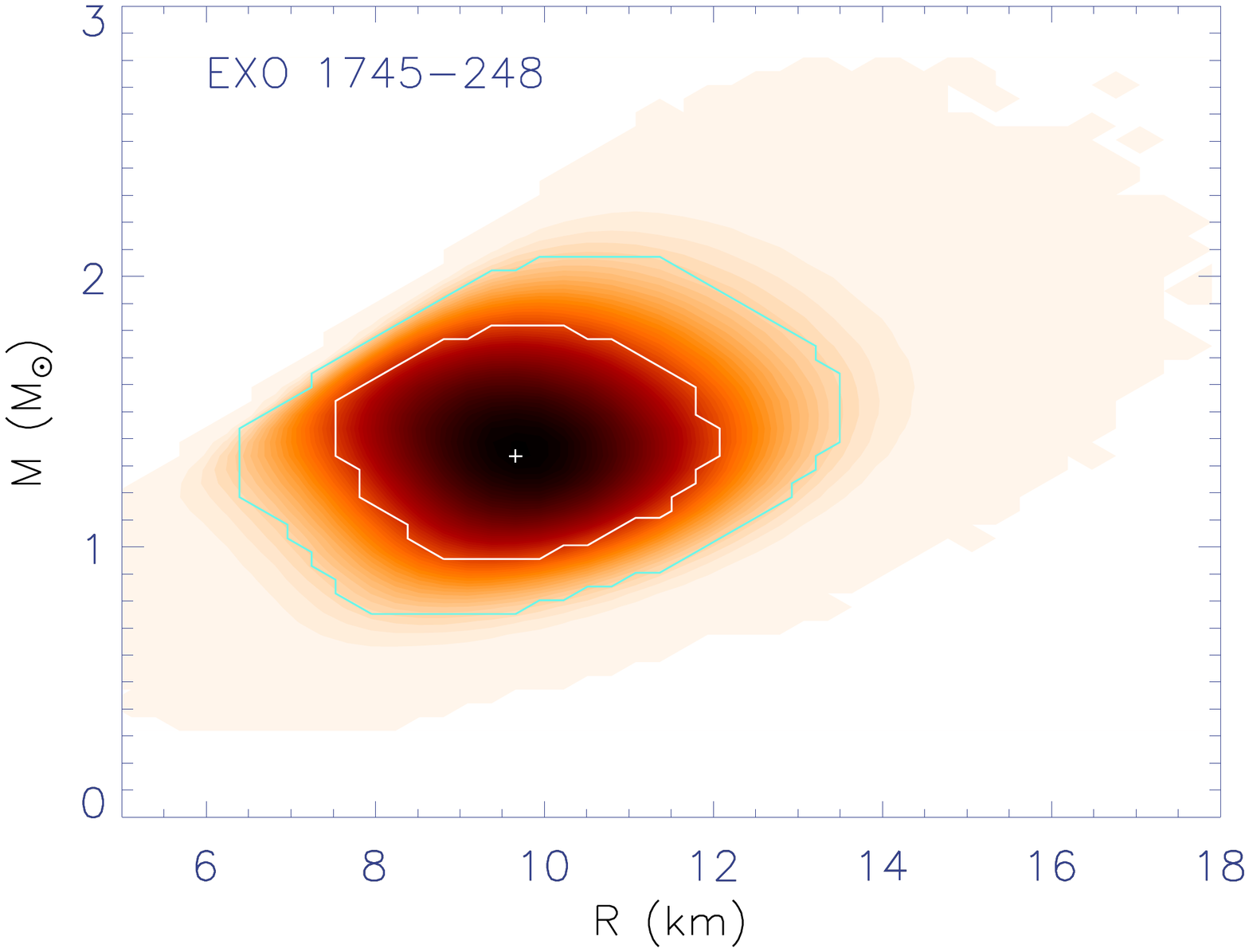}
\hspace*{-3.0cm}\includegraphics[width=8.5cm,angle=180]{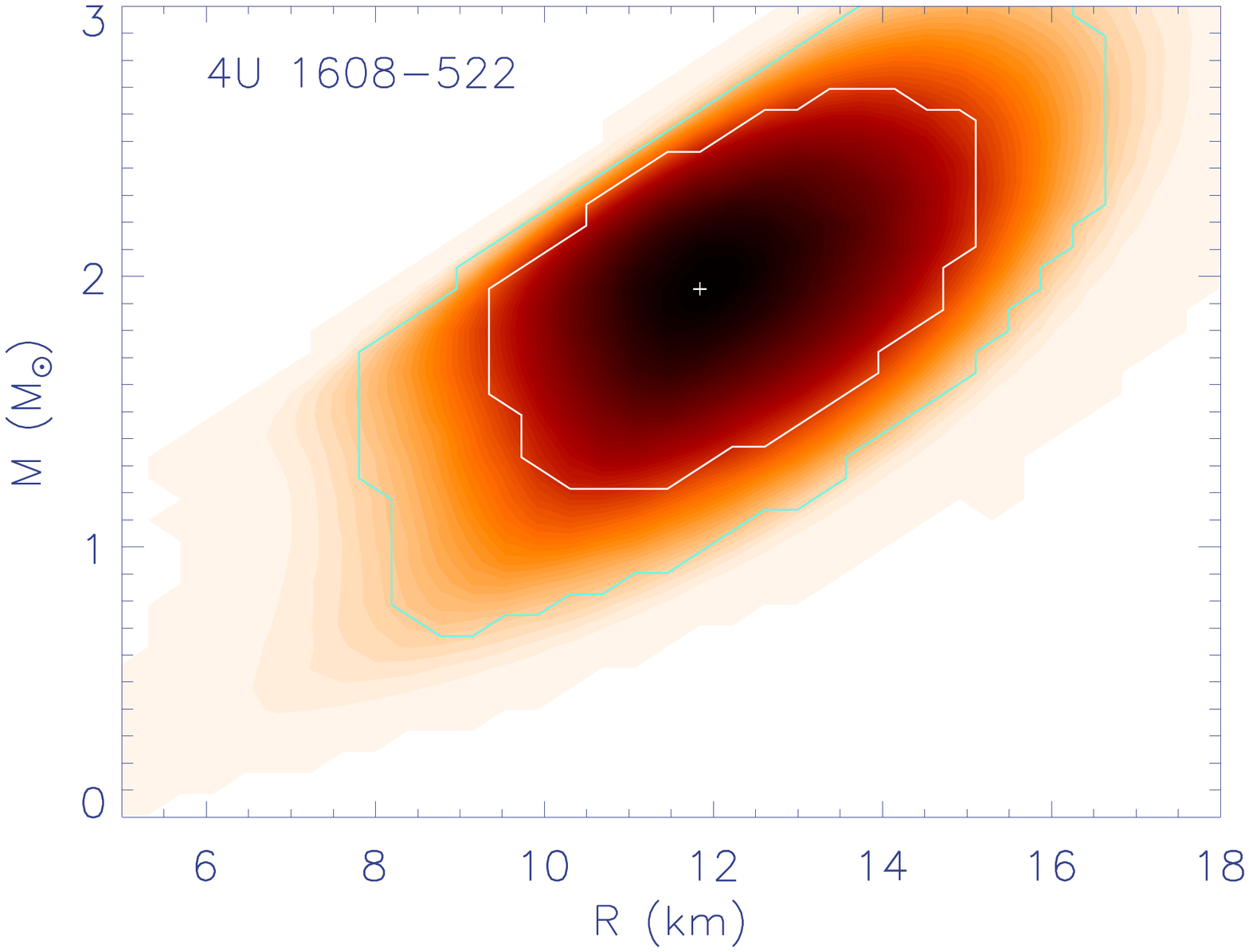}
\hspace*{-3.0cm}\includegraphics[width=8.5cm,angle=180]{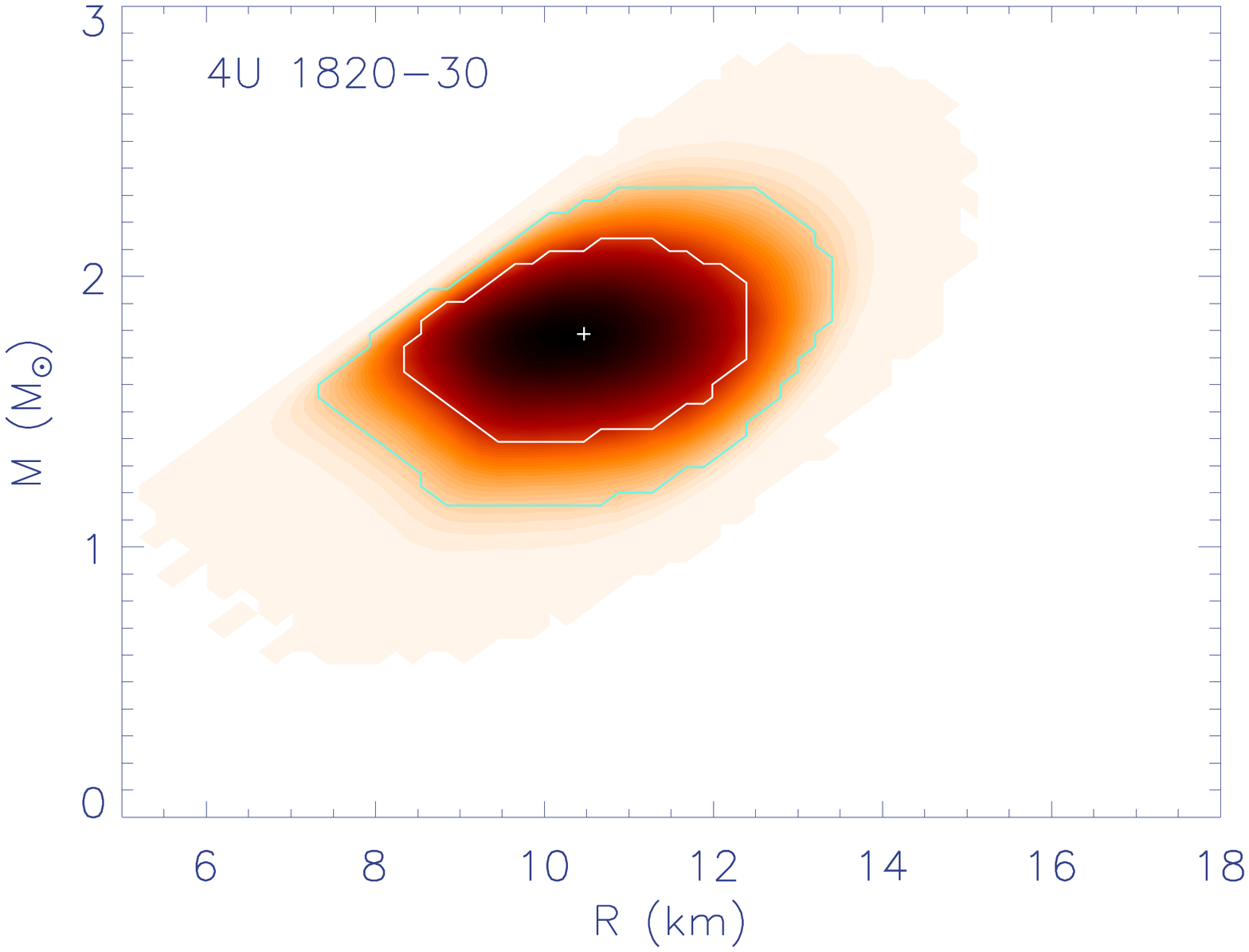}

\vspace*{-1.7cm}\hspace*{-1cm}\includegraphics[width=8.5cm,angle=180]{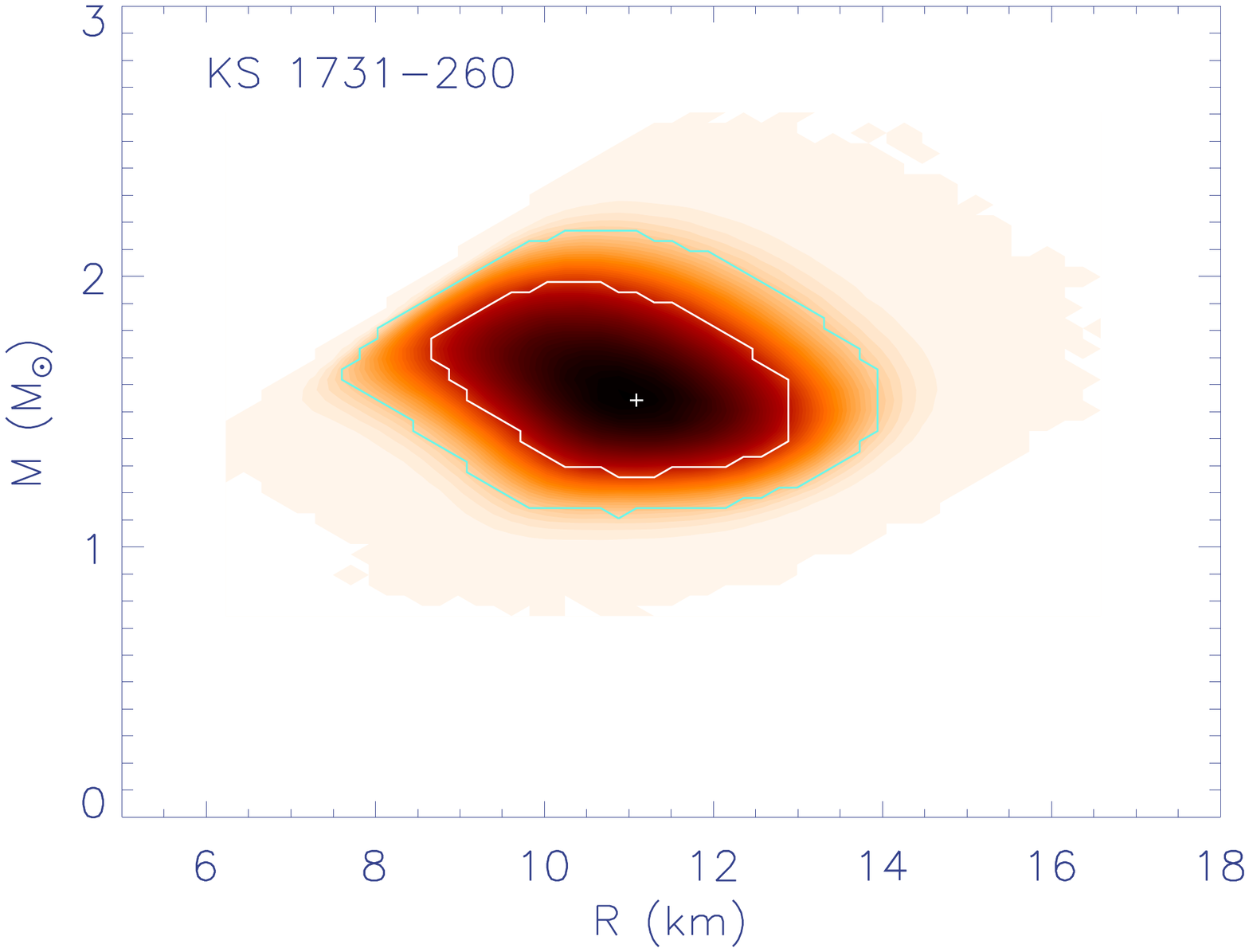}
\hspace*{-3.0cm}\includegraphics[width=8.5cm,angle=180]{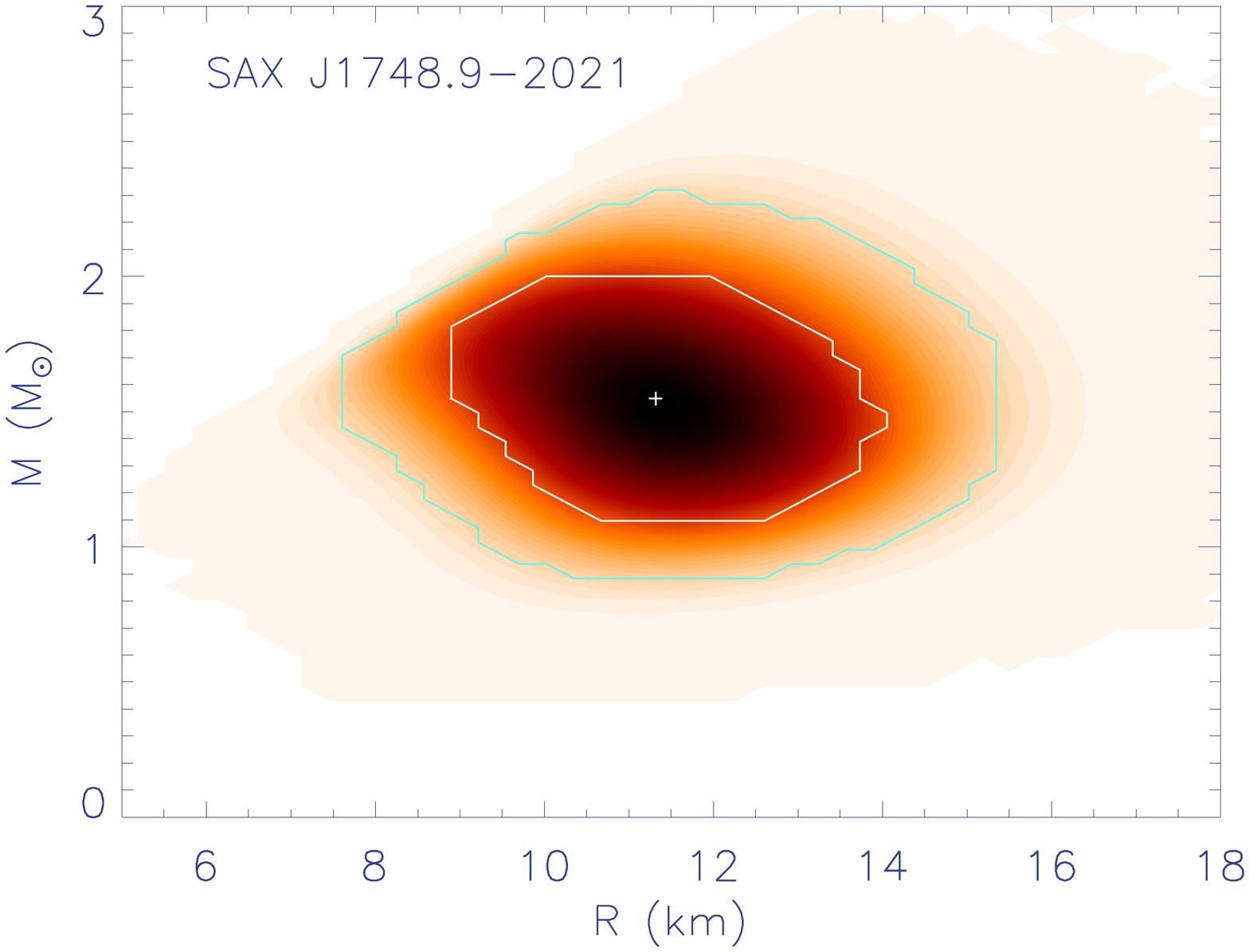}

\vspace*{-.5cm}\caption{The same as Figure \ref{fig:1} except assuming that $z_{\rm ph}=0$.\label{fig:2}}
\end{figure*}

To date, five PRE X-ray bursters have been studied by \"Ozel and
collaborators. We collect results for $D, F_{{\rm Edd},\infty}$ and
$A$ for each source in table~\ref{tab:2}. Uncertainties for these
quantities are assumed to be Gaussian with the indicated $1\sigma$
error bars. Values of $\alpha$ and $\gamma$ can be found with
additional assumptions for $\kappa$ and $f_c$. The opacity in these
high-temperature sources is dominated by electron scattering, and we
employ the Thomson opacity $\kappa=0.200m_B^{-1}(1+X)$ cm$^2$
g$^{-1}$, where $m_B$ is the baryon mass and $0<X<0.7$ is the
mass fraction of hydrogen. This range for $X$ reflects the uncertainty
in compositions which could range from pure He at one extreme to solar
at the other. Also, following \cite{SLB10}, we assume that
$f_c=1.40\pm0.07$, a slightly larger range than \"Ozel et al. have
assumed. The uncertainty distributions for $\kappa$ and $f_c$ are
taken to be boxcar shaped. The computed values for $\alpha$ and
$\gamma$ reflect the combined uncertainties of observables and
physical parameters using Monte Carlo sampling.

It is clear from table~\ref{tab:2} that none of the sources satisfy
$\alpha\le1/8$ to within $1\sigma$. If the observed quantities
$F_{{\rm Edd},\infty}, D$ and $A$ are Monte Carlo sampled within their
probability distributions, nearly all the resulting values of $\alpha$
will be greater than 1/8 and those trials must be rejected. Those
trials that will be accepted will cluster near $\alpha=1/8$, which is
more easily accomplished if $X\simeq0$ and $f_c\simeq1.47$. The
fraction of accepted trials are shown in the last column of
table~\ref{tab:3}. In turn, accepted $\gamma$ values will be near the
upper end of the ranges given in table~\ref{tab:2}. As a consequence,
in this model, $\alpha\simeq1/8$ to within $1\sigma$, irrespective of
the values of the observables, as can be seen in table~\ref{tab:3}.
Also, the error bars of the derived quantities are seen to be
artificially reduced in this case, which does not seem justified. The
probability distributions for inferred values of $R$ and $M$ are shown
in Figure \ref{fig:1}, which shows that the distributions are bimodal
with each lobe carrying equal total weight. The average values listed
in table~\ref{tab:2} are intermediate between the two probability
clumps. Note that the distributions computed by Refs.
\cite{Ozel09,Guver10a,Guver10b} are incorrect, but were corrected by
Ref. \cite{Ozel:2010}.

Sources of systematic uncertainties not accounted for in this simple
model are possible asymmetries in emission as well as the assumption
that the photosphere is at the neutron star surface when $F_{{\rm
    Edd},\infty}$ is measured.  Steiner et al. \cite{SLB10} attempted
to include these uncertainties by allowing the effective redshift of
the photosphere, $z_{\rm ph}$, to be randomly chosen from a
distribution uniformly populated with $1/(1+z_{\rm ph})^2$ between 0
and $1/(1+z)^2$.  In the extreme case that $z_{\rm ph}=0$, Equation
\ref{ag} becomes
\begin{eqnarray}\label{ag1}
\alpha&=&{F_{{\rm Edd},\infty}\over\sqrt{A}}{\kappa D\over f_c^2c^3}=\beta\sqrt{1-2\beta},\cr
\gamma&=&{A\over F_{{\rm Edd},\infty}}{f_c^4c^3\over\kappa}={R\over\beta(1-2\beta)}.
\end{eqnarray}
Solving these for $M$ and $R$ yields a new set of relations:
\begin{eqnarray}\label{sol1}
\beta&=&\left[1+\sqrt{3}\sin\left({\theta/3}\right)-\cos\left({\theta/3}\right)\right]/6,\cr
R&=&\alpha\gamma\sqrt{1-2\beta},\qquad
M={\alpha^2\gamma c^2/G},
\end{eqnarray}
where $\theta=\cos^{-1}(1-54\alpha^2)$. When
$\alpha<3^{-3/2}\simeq0.192$, $\theta$ is real and there are 3 real
roots for $\beta$. One of these is negative, and another is greater
than 1/3 which nearly violates the causality constraint for neutron
stars\cite{LP01}. The remaining real root is the one given in Equation
(\ref{sol1}). When $\alpha>3^{-3/2}$, $\theta$ and all roots for
$\beta$ are imaginary. Table~\ref{tab:2} indicates that, to within
$1\sigma$, all five sources have $\alpha<0.192$, so that Monte Carlo
sampling of the observables within their probability distributions
should yield a much larger fraction of physically acceptable
solutions. Indeed, this is borne out by the acceptance fraction shown
in table~\ref{tab:3}. Accepting only those trials for which
$\alpha<0.192$, table \ref{tab:3} shows averages and standard
deviations for $M$ and $R$ under the assumption that $z_{\rm ph}=0$.
Interestingly, values of $\alpha$ are larger and values of $\gamma$
are smaller than in the previous case, with the consequence that in
three of the five cases the values of $R_\infty=\alpha\gamma$ are
largely unchanged. The uncertainty ranges are much less compressed
compared to the case $z_{\rm ph}=z$, and derived values of $R$ are on
average 1.2 km larger. While the high percentage of accepted trials
with $z_{\rm ph}=0$ is encouraging, this model remains oversimplified
and the possibility that $z_{\rm ph}=z$ cannot be ruled out. The
average neutron star mass and radius implied by these results are
$\bar R=10.77\pm0.65$ km and $\bar M=1.65\pm0.12 M_\odot$, and the
probability distributions are displayed in Figure \ref{fig:2}. Given
expectations that neutron star radii don't change much with neutron
star mass, these relatively small standard deviations are interesting.

\subsection{Quiescent Low-Mass X-ray Binaries}
\label{qlmxb}
Certain neutron stars in binary systems may intermittently accrete
matter from an evolving companion star, with episodes of accretion
separated by long periods of quiescence.  While the neutron star
accretes, compression of matter in the crust induces nuclear reactions
that release heat in sufficient amounts to warm the star to
temperatures not seen since its birth, these neutron stars cool via
neutrino radiation from their interiors and X-rays from their
surfaces.  It is generally believed that accretion suppresses surface
magnetic fields, which is an advantage of using these systems for
radius measurements compared to isolated neutron stars for which
strong, uncertain, magnetic fields may exist.  Strong magnetic fields
can significantly affect a star's atmosphere and introduce large
uncertainties in radius measurements.  In addition, due to the rapid
gravitational settling timescales (of order seconds), only the
lightest element in accreted matter remains in its atmosphere.  Thus,
these transient X-ray sources, also known as QLMXBs, are believed to have low-magnetic field H or He
atmospheres.  The emitted X-ray spectra, for a given composition, will
depend largely on $R$ and $T_{\rm eff}$, and, to a lesser extent, on
gravity $g=GM(1+z)/R^2$.

In contrast, the observed spectrum will depend on the distance $D$ and
on the amount of interstellar absorption between the source and the
observer, usually parameterized by $N_H$, the column density of H.
The absorption is important, as it has an energy dependence of
$E^{-8/3}$ and can significantly reduces the observed flux near the
peak and at lower energies. It is often difficult
to determine distances to field sources, while distance determinations
of globular clusters are relatively accurate.  For this reason,
attention has been focused on systems in globular clusters.

Fitting the observed spectrum in principle can provide estimates for
$R_\infty$, $T_{{\rm eff},\infty}$, $g$ and $N_H$, but due to lack of
resolution and poor statistics, the deduced $N_H$ is often at odds
with the amount of absorption deduced from HI radio surveys.  Although
it is obvious that an underestimate of absorption will lead to an
underestimate of mass and radius, because decreasing the absorption
has a similar effect to decreasing the distance, it is possible
through analytic considerations to predict the magnitude of the
effect.  For simplification, we first consider the case of a blackbody
emitter. The observed energy dependence of the flux from an absorbed
blackbody with an effective temperature $T$ obeys
\begin{equation}\label{fbb}
F(E, T, N_H)=\alpha E^3{e^{-bN_{H21}/E^{8/3}}\over e^{E/kT}-1},
\end{equation}
where $\alpha$ is a constant and $b\simeq0.16$ keV$^{8/3}$ represents the approximate effects of absorption
\cite{Wilms00}.  $N_{H21}$
is the hydrogen column density in units of $10^{21}$ cm$^{-2}$.

For a given $T$, the maximum flux occurs at $E_0$ where $dF/dE=0$, or
\begin{equation}\label{ebb}
E_0=\left[3+(8/3)bN_{H21}E_0^{-8/3}\right]\left(1-e^{-E_0/kT}\right)kT.
\end{equation}
Therefore $E_0>3kT$ in general, and the exponential term is small.  The observed
flux, neglecting gravity and redshift, is
\begin{equation}\label{fbb1}
\left({R\over D}\right)^2\int_{E_L}^{E_U}F(E,T,N_H)dE,
\end{equation}
where $E_L\sim0.3$ keV and $E_H\sim10$ keV represent the low- and
high-energy cutoffs of the X-ray detector response. To compare the
effect of changing the amount of absorption on the inferred radius, we
assume that both the total observed flux and the peak energy $E_0$ are
held fixed as $N_H$ is varied. Changing the H column density from
$N_1=N_{1,H21}$ to $N_2=N_{2,H21}$ will alter the effective
temperature from $T_1$ to $T_2$:
\begin{equation}\label{tbb}
{T_2\over T_1}\simeq{9E_0^{8/3}+8bN_1\over9E_0^{8/3}+8BN_2},
\end{equation}
neglecting the exponential term in Equation \ref{ebb}. Thus, the
effective temperature will decrease with an increase in absorption.
The ratio of deduced radii follows from Equation \ref{fbb1}:
\begin{equation}\label{rbb}
\left({R_2\over R_{1}}\right)^2=
{\int_{E_L}^{E_U}F(E,T_1,N_1)dE\over\int_{E_L}^{E_U}F(E,T_2,N_2)dE}.
\end{equation}
These integrals can be approximated by the method of steepest descents
to high accuracy: the integration limits are extended to $-\infty$ and
$\infty$ and the integrand is replaced by a Gaussian centered at
$E_0$. These approximations yield
\begin{equation}\label{rbb1}
\left({R_2\over R_{1}}\right)^2\simeq{F_1\over F_2}\sqrt{F_2^{\prime\prime}F_1\over F_2F_2^{\prime\prime}},
\end{equation}
where $F_1=F(E_0,T_1,N_1)$ and $\prime\prime$ indicates a second
derivative evaluated at $E_0$. One has
\begin{eqnarray}\label{fbb2}
{F_1\over F_2}&\simeq&\exp\left[{11b(N_2-N_1)\over3E_0^{8/3}}\right],\cr
{F_2^{\prime\prime}F_1\over F_2F_1^{\prime\prime}}&\simeq&{27E_0^{8/3}+88bN_2\over27E_0^{8/3}+88bN_1}.
\end{eqnarray}
An increase in $N_H$ necessarily leads to an increase in $R$ since
both factors in Equation (\ref{fbb2}) are greater than unity. For
example, for $T_1=0.10$ keV, $N_1=0.9$ and $N_2=1.8$, one finds
$E_0\simeq0.52$ keV, $T_2\simeq0.07$ keV, and $R_2/R_1\simeq5.35$. The
analytic expressions in Equations (\ref{rbb1}) and (\ref{fbb2}) are
accurate in this case to better than 1\%, compared to the exact
integrations of Equation (\ref{rbb}) and differentiations of Equation
(\ref{fbb}). For comparision, Guillot et al. \cite{Guillot13} find a
ratio $R_2/R_1\simeq2$ for similar conditions using an H atmosphere in
the case of a source in $\omega$ Cen.

\begin{figure}
\includegraphics[width=9.5cm]{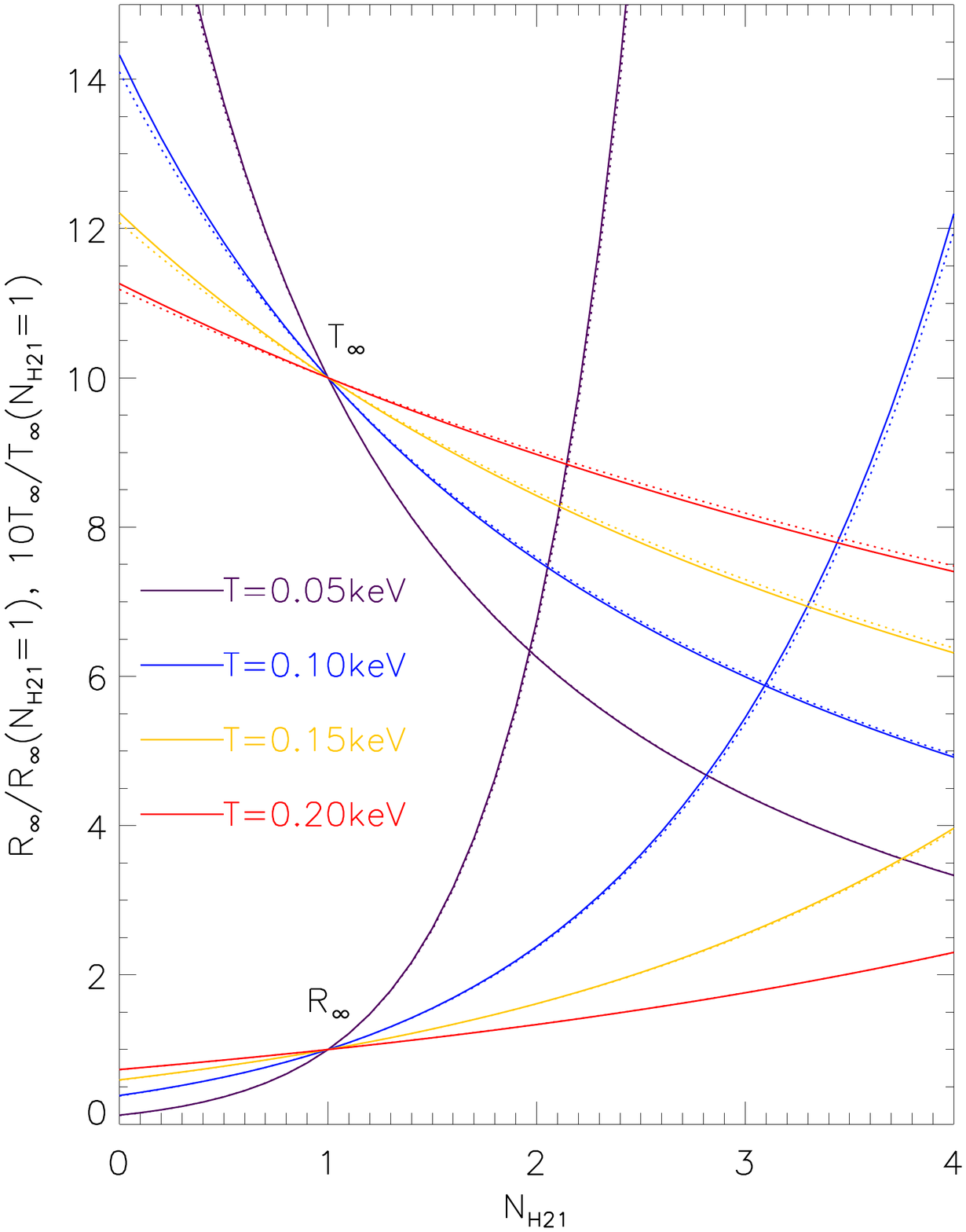}
\caption{Relative radii (lower curves) and effective temperatures
  (upper curves) for H atmospheres as functions of H column densities.
  Results are shown as ratios relative to the values obtained for the
  indicated base effective temperature and column density
  ($N_{H21}=1.0$). Solid lines show full integrations while dotted
  lines are the analytic expressions derived from steepest descent
  integration described in the text. \label{fig:3} }
\end{figure}

The radius change overestimate in the case of a blackbody can be
understood using a simple approximation to the shape of the spectrum
from an H atmosphere. Following Lattimer \& Steiner \cite{LS13}, a
hydrogen atmosphere can be approximated by
\begin{equation}\label{fh}
F(E,T,N_H)\simeq\alpha^\prime E^3{e^{-bN_{H21}/E^{8/3}}\over e^{\beta(E/kT)^p}-1}
\end{equation}
where $\alpha^\prime$ depends weakly on $T$, approximately as
$T^{0.2}$, $\beta\simeq1.35$ and $p\simeq5/7$.  The value of $p$ is a
consequence of the dominance of electron scattering in H atmospheres,
for which the cross section varies as $E^{-3}$.  Including the effects
of absorption, the peak in the observed spectrum occurs when
\begin{equation}\label{eh}
E_0\simeq\left({9+8bN_{H21}E_0^{-8/3}\over3\beta p}\right)^{1/p} kT,
\end{equation}
again ignoring the small exponential term. Unabsorbed spectra have an
energy peak at $E_0/(kT)\simeq(3/\beta p)^{1/p}\simeq4.9$, and absorption only 
makes this factor larger, justifying
this approximation. Keeping the peak energy fixed, two different
column densities lead to temperatures
\begin{equation}\label{th}
{T_2\over T_1}\simeq\left({9E_0^{8/3}+8bN_1\over9E_0^{8/3}+8bN_2}\right)^{1/p}.
\end{equation}
The ratio of inferred radii can be found using
\begin{eqnarray}\label{fh1}
{F_1\over F_2}&\simeq&\left({T_1\over T_2}\right)^{0.2}\exp\left[\left({8\over3p}+1\right)b{N_2-N_1\over E_0^{8/3}}\right],\cr
{F_2^{\prime\prime}F_1\over F_2F_1^{\prime\prime}}&\simeq&{27pE_0^{8/3}+8bN_2(8+3p)\over27pE_0^{8/3}+8bN_1(8+3p)}.
\end{eqnarray}
For the same conditions as previously, one finds
$R_2/R_1\simeq2.24$, in closer agreement with realistic 
atmospheres. The relative increase in radius with increasing
absorption is very temperature sensitive: it is more pronounced for
smaller temperatures, as can be seen in Figure \ref{fig:3}.

It is also interesting to explore helium atmospheres, which could be
relevant in the case of ultracompact binaries in which the companion is
a white dwarf.  As in H atmospheres, one expects electron scattering
to dominate, especially at higher temperatures.  Ref. \cite{LS13}
found that the predicted spectrum of He atmospheres is similar to that
of H atmospheres, with $p\simeq5/7$, but with the value of
$\beta\simeq1.24$ instead of 1.35.  Thus, the inferred temperature of
a He atmosphere is about 13\% less than that of an H atmosphere, and
the inferred radius is about 28\% larger, assuming that $E_0$ is
unchanged.

In order to estimate the gravity and/or redshift of the atmosphere
from the observed spectrum, it is required that an additional aspect
of the atmosphere that is sensitive to $g$ and/or $z$ as well as
$T_{\mathrm{eff}}$ and $R$, in addition to the peak energy and the
overall flux, be measured.

\begin{table*}
\begin{center}
\caption{Inferred properties of QLMXBs\label{tab:4}}
\begin{tabular}{lc|ccc|ccc}
\hline\noalign{\smallskip}
Source&$D$ (kpc)&$N_{H21}$&$R_\infty$ (km)&$z$&$N_{H21}$&$R_\infty$ (km)&$z$\\
&&\multicolumn{3}{|c|}{Guillot et al. (2013)}&\multicolumn{3}{|c}{Lattimer \& Steiner (2013)}\\
\noalign{\smallskip}\hline\hline\noalign{\smallskip}
M28&$5.5\pm0.3$&2.52&$12.84^{+1.50}_{-1.48}$&$0.198^{+0.485}_{-0.120}$&1.89&$10.65^{+1.27}_{-1.14}$&$0.212^{+0.456}_{-0.123}$\\
NGC 6397&$2.02\pm0.18$&0.96&$8.42^{+1.32}_{-1.36}$&$0.242^{+0.278}_{-0.106}$&1.4&$11.66^{+1.94}_{-1.72}$&$0.241^{+0.279}_{-0.102}$\\
M13&$6.5\pm0.6$&0.08&$11.48^{+2.54}_{-2.29}$&$0.308^{+0.376}_{-0.212}$&0.145&$12.93^{+2.91}_{-2.53}$&$0.286^{+0.392}_{-0.190}$\\
$\omega$ Cen&$4.8\pm0.3$&1.82&$23.03^{+4.48}_{-3.86}$&$0.187^{+0.492}_{-0.144}$&1.04&$13.25^{+2.57}_{-2.08}$&$0.200^{+0.456}_{-0.134}$\\
NGC 6304&$6.22\pm0.026$&3.46&$11.52^{+2.73}_{-2.10}$&$0.212^{+0.467}_{-0.120}$&2.66&$9.39^{+2.09}_{-1.75}$&$0.212^{+0.407}_{-0.108}$\\
\noalign{\smallskip}\hline
\end{tabular}
\end{center}
\end{table*}

\begin{figure}
\includegraphics[width=9.0cm]{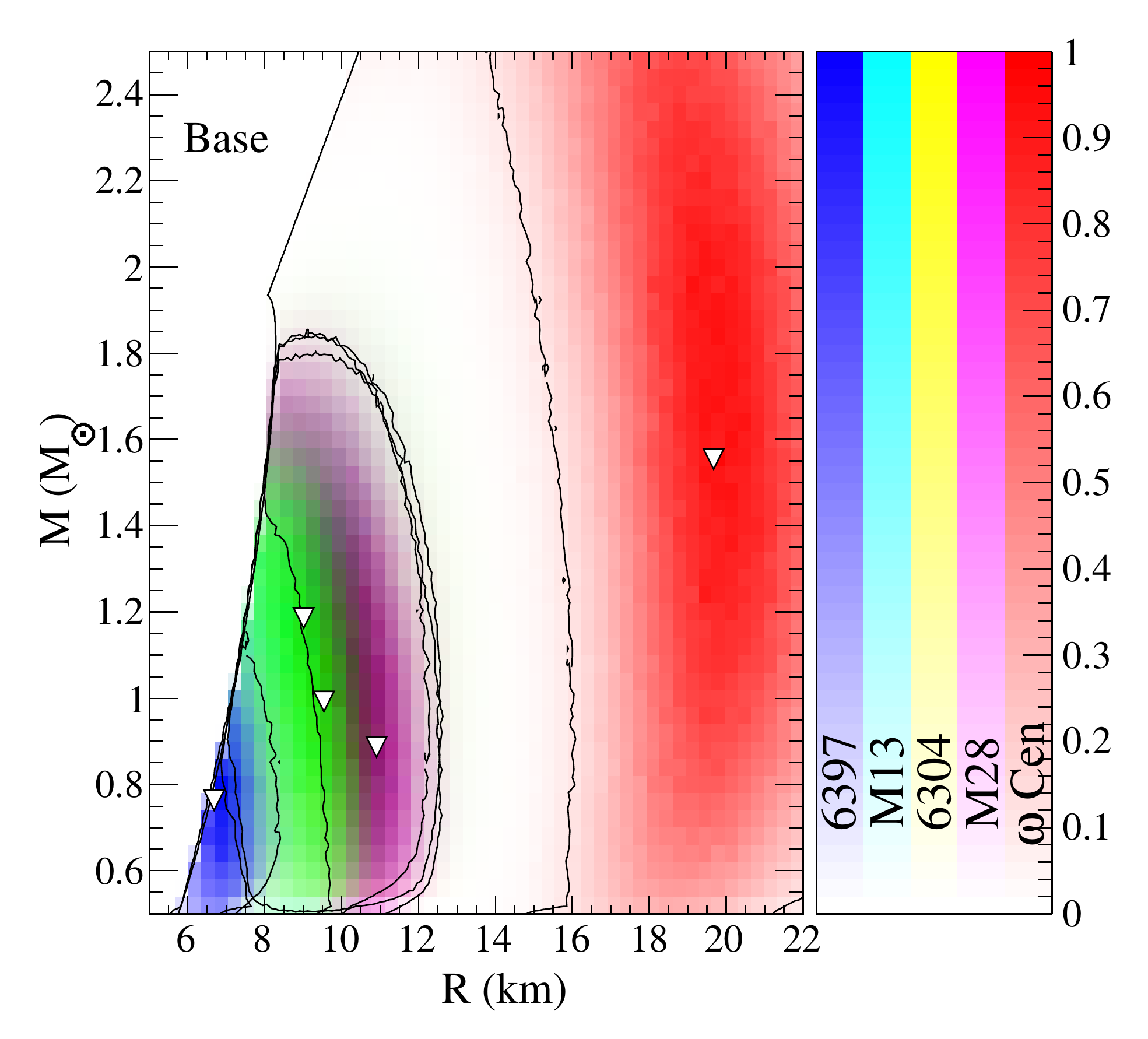}
\caption{Probability distributions in $M$ and $R$ for the 5 QLMXBs
  studied by Guillot et al. \cite{Guillot13} in which $N_H$ values are
  fit to the observed spectra. Color coding for the relative
  probabilities are indicated by the bar graphs on the right, which
  are ordered according to their most probable radii (marked by
  triangles). Solid curves denote 90\% confidence boundaries. The
  left-most curve bounds the region permitted by general relativity,
  causality, and the observation of a $2M_\odot$ neutron star
  \cite{LS13}.\label{base}}
\end{figure}

Guillot et al. \cite{Guillot13} recently summarized the observed
properties of 5 QLMXBs with known distances and modeled them with
H atmospheres to derive masses, radii and H column densities.  Their
results are shown in table~\ref{tab:4} and in Figure \ref{base}.  The
most striking feature in these results is that the optimum inferred
values of $R_\infty$ range from 8.4 to 23.0 km, of $R$ from 6.4 to
19.4 km, and of $M$ from 1.25 to 2.69 $M_\odot$.  Such large
variations are not expected from evolutionary considerations for a
relatively uniform class of sources.

\begin{figure}
\includegraphics[width=9.0cm]{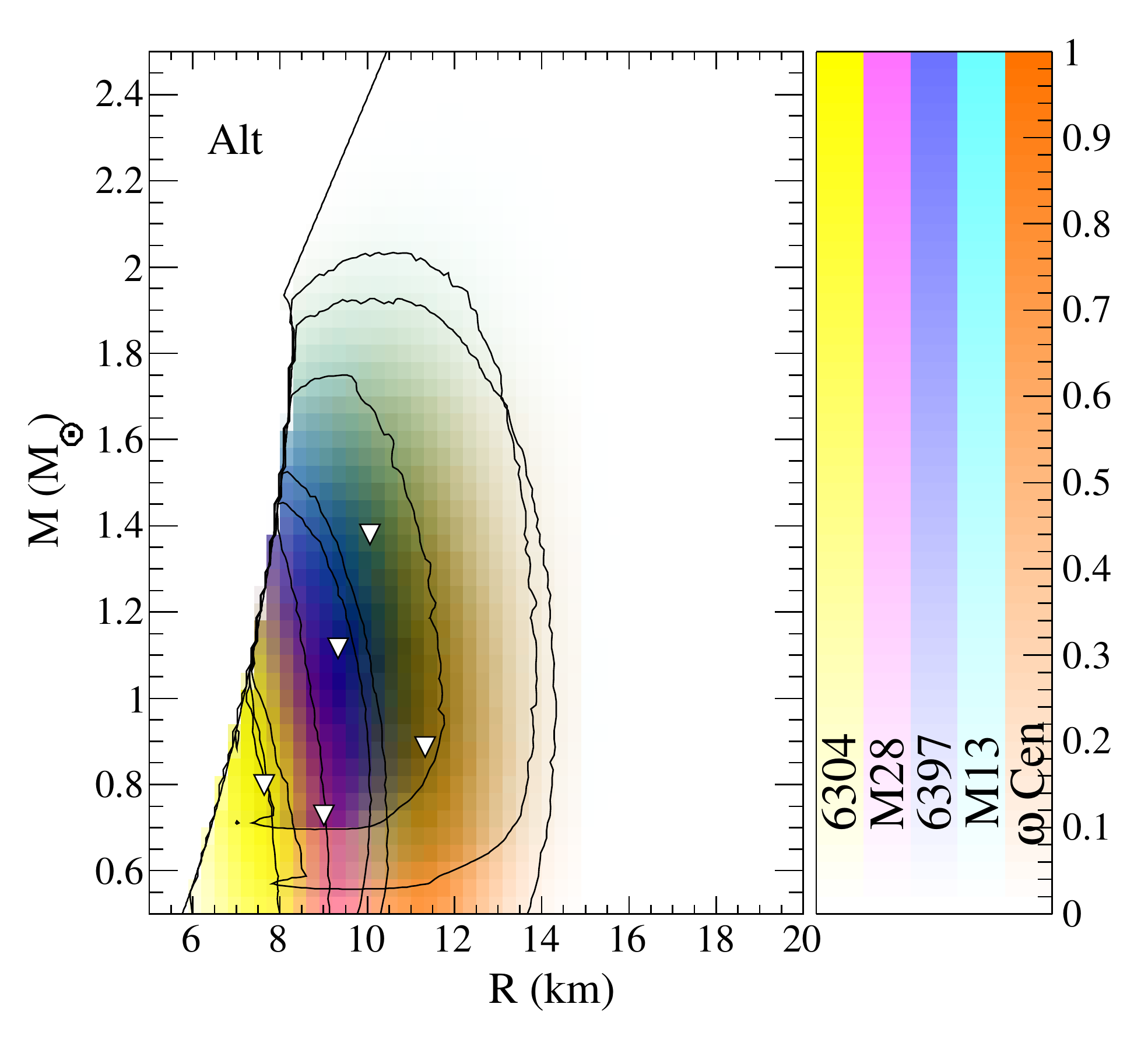}
\caption{The same as Figure \ref{base}, but $M$ and $R$ probabilities
  are derived using fixed $N_H$ values from Ref.
  \cite{Dickey90} using the procedure described in the text
  \cite{LS13}.\label{alt}}
\end{figure}

Guillot et al. \cite{Guillot13} noted that in the most extreme cases
of large and small radii (the sources in $\omega$ Cen and NGC 6397,
respectively), the values of $N_H$ they inferred were markedly
different from those independently determined \cite{Dickey90} by
observations of HI column densities in the directions toward the
respective globular clusters. Lattimer \& Steiner \cite{LS13}
observed, furthermore, that these differences in $N_H$ values acted in
such a way as to enhance the disparity of inferred radii found by
\cite{Guillot13}. Using an analytic procedure as described above,
Ref. \cite{LS13} estimated new values of $R_\infty$ using the alternate
$N_H$ values from Ref. \cite{Dickey90} for each source. The new values
for neutron star properties are summarized in table~\ref{tab:4} and
displayed in Figure \ref{alt}. In this analysis, inferred values for
$z$ were barely affected. Values of $R_\infty$, $R$ and $M$, which now
range from 9.4 to 13.2 km, 7.8 to 11 km, and 1.23 to 1.64 $M_\odot$,
respectively, have much smaller variations. The mean value of the
inferred radii is about 9.5 km, which is at the lower extremity of
values inferred from PRE bursts (\S \ref{sec:PRE}). The source in
$\omega$ Cen has been confirmed to have an H atmosphere
\cite{Haggard04}. However, it is possible that one or more of the
remaining four QLMXB sources has a He atmosphere rather than an H
atmosphere; if so, the inferred radii of those sources having He
atmospheres would be increased by approximately 30\% as discussed in
Ref. \cite{LS13}.

\section{Bayesian Analysis of Mass and Radius Observations}
\label{sec:bayes}

The basic problem we want to solve is how to compute the M-R curve
from a set of neutron star mass and radius observations. We also want
the EOS, taking advantage of the well-known bijection between the M-R
curves and EOSs provided by the TOV equations. The first critical
point is that observations are never perfectly accurate, and thus this
problem has an inherently statistical nature: what we really want is
the probability distribution of M-R curves and EOSs. The second
critical point is that this is a highly underconstrained problem: a
curve has an infinite number of degrees of freedom, and we will always
have only a finite set of observations. Bayesian statistics is thus
already a natural choice, since its application to underconstrained
problems is a bit simpler. In an overconstrained least-squares
problem, there can be one unique best fit represented by a point in
the model space. In the underconstrained system, there is no
unique best-fit, but rather an entire subspace inside our model space
which consists of ``best-fits''. As we will see below, Bayes theorem
provides for us a recipe for characterizing that subspace.

The joint probability of event $a$ in event space $A$ and event $b$ in
event space $B$ can be denoted $P[A,B]$ (sometimes to be more explicit
we write $P[A=a,B=b]$), and can be thought of as a two-dimensional
function of $a$ and $b$. The reader is forewarned that there are a
plethora of notations for the same quantity including $P_{A,B}(a,b)$,
$P[a,b]$, $P[A \cap B]$, $P[A~\mathrm{and}~B]$, and several other
options using parentheses instead of or in addition to square
brackets. The appropriate units for the joint probability,
$[a^{-1}b^{-1}]$, are clear from the normalization condition
\begin{equation}
1 = \int_A~\int_B~P[A,B]~da~db
\end{equation}
The marginal probability $P[A]$ (with units $[a^{-1}]$) is then given by
\begin{equation}
P[A] = \int_B~P[A,B]~db
\end{equation}
and similarly for $P[B]$. The conditional probability is then defined
by $P[A|B]\equiv P[A,B]/P[B]$, i.e. the probability of A given B. From this
definition, the celebrated ``Bayes theorem'' can be be directly
written 
\begin{equation}
P[A|B] P[B] = P[B|A] P[A] \, .
\end{equation}
In our context, we replace $A$
with the model space ${\cal M}$ and $B$ with the ``data space'' ${\cal
  D}$ (the space of all possible data sets), and
\begin{equation}
P[{\cal M}|{\cal D}] P[{\cal D}] = P[{\cal D}|{\cal M}] P[{\cal M}]
\end{equation}
where $P[{\cal M}|{\cal D}]$ is the conditional probability of the
model given the data, $P[{\cal D}|{\cal M}]$ is the conditional
probability of the data given the model, and $P[{\cal M}]$ and
$P[{\cal D}]$ are the prior probabilities for the model and the data.
By analogy to the definitions above,
\begin{equation}
P[{\cal D}] = \int_{\cal M}P[{\cal D},{\cal M}]~dm = 
\int_{\cal M}P[{\cal D}|{\cal M}] P[{\cal M}]~dm
\end{equation}
(this is sometimes referred to as the ``law of total probability'')
and thus 
\begin{equation}
P[{\cal M}|{\cal D}] = \frac{P[{\cal D}|{\cal M}] P[{\cal M}]}
{\int_{\cal M}P[{\cal D}|{\cal M}=m] P[{\cal M}]~dm}
\end{equation}
What we want to compute is the conditional probability of all models
in our model space, given the data actually observed $d$ inside the
space of all possible data sets ${\cal D}$, i.e. $P[{\cal M}|{\cal
    D}=d]$. In some cases we only require relative probabilities, i.e.
$P[{\cal M}=m_1|{\cal D}=d]/P[{\cal M}=m_2|{\cal D}=d]$ for models
$m_1$ and $m_2$ inside our model space ${\cal M}$, so we do not need
to compute the integral in the denominator. The function $P[{\cal
    D}|{\cal M}]$ is analogous to the likelihood function familiar
from frequentist statistics (as will be described below), and $P[{\cal
    M}]$ is referred to as the prior distribution, reflecting the
prior probability of a given model $m$.

In a typical data set of several one-dimensional data points, the
likelihood function is just a multi-dimensional Gaussian, 
$P[{\cal D}|{\cal M}]=\exp(-\chi^2/2)$ where
\begin{equation}
\chi^2 = \sum_i \left( \frac{x_{\mathrm{pred},i}-x_{\mathrm{obs},i} }
{\sigma_i} \right)^2
\end{equation}
Our neutron star data set is inherently two-dimensional, and this is
sometimes referred to as a ``Type II regression''. A typical
frequentist approach is to minimize the distance from the the observed
data point $(R,M)$ and the model $M-R$ curve. (This is not entirely
unambiguous because one must still choose the relevant mass and radius
scales to measure a distance.) One way to proceed in the Bayesian
formalism is to treat the mass of each neutron star as a new model
parameter. Our model space, ${\cal M}$, now includes the neutron star
masses, $M_i$ in addition to the EOS parameters $p_i$. In the case
that the observations are of the form of two-dimensional Gaussians
centered at $(R_{\mathrm{obs},i},M_{\mathrm{obs},i})$ with width
$(\sigma_{R,i}, \sigma_{M,i})$, the conditional probability is
\begin{eqnarray}
&P[{\cal D}|{\cal M}] = & \prod_{i=1}^{N_O}
\left(2 \pi \sigma_{M,i} \sigma_{R,i}\right)^{-1} \nonumber \\
&& \times \exp \left[ - 
\frac{1}{2} \left( \frac{M_{i}-M_{\mathrm{obs},i} }
{\sigma_{M,i}} \right)^2 \right. \nonumber \\
&& \left. - \frac{1}{2} \left( \frac{R_i(M_i,\{p_j\})-R_{\mathrm{obs},i} }
{\sigma_{R,i}} \right)^2 \right] 
\label{eq:pdm}
\end{eqnarray}
for $N_{O}$ neutron star observations. It is the evaluation of the
function $R_i(M_i,{p_j})$ here which requires a solution of the TOV
equations for each point in the model space. In general, the
observations are not two-dimensional Gaussians, and the conditional
probability for each observation is a general distribution ${\cal
  D}_i(R_i,M_i)$ normalized so that
\begin{equation}
1=\int~d R_i~dM_i~{\cal D}_i(R_i,M_i) \, .
\end{equation}

The full prior probability, $P[{\cal M}]$, in this context is now an
$(N_P+N_O)$-dimensional function reflecting the prior probability
given a set of $N_P$ EOS parameters and $N_O$ neutron star masses. It
is reasonable to assume that the prior can be factorized into separate
prior distributions for the EOS and the masses. A simple uniform prior
distribution for the EOS parmameters is not unreasonable. A physical
interpretation for the prior on the neutron star masses is that it is
equal to the neutron star initial mass function, which we will vary
below.

Bayes theorem itself is a result which can be obtained from
basic axioms of probability theory. The frequentist and Bayesian
approaches diverge in how the theorem ought to be applied. The
standard Bayesian approach is to compute the desired results by
integrating (marginalizing) over the parameters not currently being
considered. Explicitly, for the posterior probability distribution of
one of the EOS parameters, $p_i$, one computes the integral
\begin{eqnarray}
P[p_i] &=& \int P[{\cal D}|{\cal M}] P[{\cal M}]
dp_1~dp_2~\ldots~dp_{i-1}\nonumber \\
&& dp_{i+1}~\ldots~dp_{N_p}
dM_1~\ldots~dM_{N_O}
\label{eq:post1}
\end{eqnarray}
After normalizing the posterior distribution, 
one can compute the ``Bayesian confidence region''. 
When the posterior is sufficiently unimodal, the 68\% 
confidence region is the range $(p_{iL},p_{iR})$ 
surrounding the maximum value of $P[p_i]$ for which 
\begin{equation}
0.68 = \int_{p_{iL}}^{p_{iR}} P[p_i]~d p_i 
\end{equation}
and $P[p_{iL}]=P[p_{iR}]$. For a multimodal distribution, the 68\% confidence region is the region ${\cal S}_i(\eta)$, defined as the union of all intervals over $p_i$ for which
$P[p_i]>\eta$,
which is obtained by solving $\int_{{\cal S}_i(\eta)}P[p_i]~d
p_i=0.68$ for $\eta$.

Alternatively, one can write $P[p_i]$ using a $\delta$-function, and
a helpful simplification comes from the fact that we can
write almost all the quantities of interest using the same
kernel, $P[{\cal D}|{\cal M}] P[{\cal M}]$. The expressions
\begin{eqnarray}
P[p_i=\hat{p_i}] &=& \int \delta(p_i - \hat{p_i}) 
\times \nonumber \\
&& P[{\cal D}|{\cal M}] P[{\cal M}]
d \{p\}d \{M\} \nonumber \\
P[R(M)=\hat{R},M=\hat{M}] &=& \int \delta[R(\hat{M}) - \hat{R}]
\times \nonumber \\
&& P[{\cal D}|{\cal M}] P[{\cal M}]
d \{p\}d \{M\} \nonumber \\
P[\epsilon=\hat{\varepsilon},P(\varepsilon)=\hat{P}]
&=& \int \delta[P(\hat{\varepsilon}) - \hat{P}] \times \nonumber \\ 
&& P[{\cal D}|{\cal M}] P[{\cal M}] d \{p\}d \{M\} \nonumber \\
\label{eq:margs}
\end{eqnarray}
give the posterior distributions for the parameters, the probability
distribution for the radius given a fixed mass (the M-R curve), and
the probability distribution for the pressure given a fixed energy
density (the EOS), respectively. Unless we are computing the Bayes
factor (defined below), we only need to determine these integrals up
to a scale factor and one can replace the $\delta$-functions by pairs
of step-functions, i.e.
\begin{equation}
\delta(P-\hat{P}) \rightarrow \theta[P-(\hat{P}-\Delta P)]
\theta[(\hat{P}+\Delta P)-P]
\end{equation}
One can perform a single Markov chain Monte Carlo simulation of the
integration kernel $P[{\cal D}|{\cal M}] P[{\cal M}]$, and construct a
histograms to select only those points in the chain which satisfy the
conditions given by the step-functions in the various integrals.

\subsection{Simplified Models Without Equations of State}

A simple model, suggested by Ref.~\cite{Guillot13} partly based on 
results from Ref.~\cite{SLB13}, is that all neutron stars have the
same radius. This model is beneficial because it
contains no assumptions about the nature of the compact objects being
observed. In this case, the model space has only one parameter, and
the only remaining integrals are those over the individual neutron
star masses. We also assume uniform prior distributions for all of the
neutron star masses corresponding to flat neutron star initial mass
functions. In the case that the observations are two-dimensional
Gaussians, the integrals over the masses are trivial Gaussian
integrals and Equation~(\ref{eq:post1}) can be written
\begin{equation}
P(R) = \prod_{i=1}^{N_O} \left(2 \pi \right)^{-1/2} \sigma_{R,i}^{-1}
\exp \left[ -\frac{1}{2} \left( \frac{R-R_{\mathrm{obs},i}}
{\sigma_{R,i}} \right)^2 \right] \, .
\end{equation}
This is propotional to the frequentist likelihood function for
$N_O$ neutron star radius measurements and the peak of this
distribution is exactly the radius which minimizes the 
corresponding $\chi^2$. 

We use this simple "common radius" model to analyze both the PRE
observations (with $z_{\mathrm{ph}}=0$ as in Fig.~\ref{fig:2}) and the
QLMXB observations. 
One can enforce causality with the additional
restriction $R_i<2.94~G M_i$ for each object. It is clear from
Fig.~\ref{base} that there are very few radii which intersect the
regions for the neutron star in $\omega$ Cen and the other four
neutron stars. If all neutron stars indeed have the same radius, the
probability of actually observing the data given in Fig.~\ref{base}
would be extremely small. After the adjustment for the hydrogen column
densities described in section \ref{qlmxb} above and obtaining the
results in Fig.~\ref{alt}, there are several vertical lines which go
through all the data sets, and the Bayesian 95\% confidence interval
for the neutron star radius is $10.8 \pm 1.2$ km. Our result is not
significantly different from the range of radii earlier predicted in
Ref.~\cite{SLB13}, in part because the average PRE data with $z=0$
discussed in \S \ref{sec:PRE} also has $\bar R=10.8$ km and partly
because the QLMXB data was analyzed with similar assumptions regarding
the hydrogen column density in both cases. Our predicted radius is
larger than in G13 for this common radius model by about 1.5 km to
2.8 km, depending on whether $N_H$ values are frozen at their
"best-fit" values or not. This difference cannot be attributed to
different assumptions for $N_H$, because the average radius found by
Ref. \cite{LS13} are actually 1.6 km smaller than found by Ref.
\cite{Guillot13}. Partly, the difference is due to the unequal
weighting assigned to individual sources in Ref. \cite{Guillot13} on
the basis of the relative quality of the observational data: the
source with the smallest individual radius, in NGC 6397, has 35\% of
the statistical weight and the source with the largest individual
radius, in $\omega$ Cen, has 7.8\% of the statistical weight
\cite{Guillot13}. This unequal weighting only partially explains G13's
small common radius. With unequal weighting, their average radius is
reduced to 9.8 km, still larger than their common radius.

There is a straightforward explanation why the common radius found by
Ref.~\cite{Guillot13} is as much as 1.7 km smaller than their average
value~\cite{LS13}. The key is that values of $R_\infty$ found by Ref.
\cite{Guillot13} in their joint analysis of QLMXBs are nearly the same
as those found in their independent determinations. Because $R_\infty$
is determined with greater accuracy than $z$, the joint analysis can
find a common radius by shifting the value of $z$ for each source. We
can thus estimate the common radius $R$ by minimizing the function
\begin{equation}\label{chicom}
\chi^2=\prod_i\left\{\exp\left(-w_i\left[{z(R,R_{\infty,i})-
z(R_i,R_{\infty,i})\over\Delta_i}\right]^2\right)\right\}
\end{equation}
with respect to $R$. $R_i$, $R_{\infty,i}$, $w_i$ and $\Delta_i$ are
the radius, $R_\infty$, weight and $z$ uncertainty, respectively,
associated with source $i$. Since $z(R,R_\infty)+1=R_\infty/R$,
minimization leads to
\begin{equation}\label{chicom1}
R=\sum_i{w_iR_{\infty,i}^2\over\Delta_i^2}\bigg/
\sum_i{w_iR_{\infty,i}^2\over R_i\Delta_i^2}\simeq8.1{\rm~km},
\end{equation}
nearly the value ($8\pm1$ km) Ref. \cite{Guillot13} found when $N_H$
values were assumed frozen.

\begin{figure}
\includegraphics[width=9.0cm]{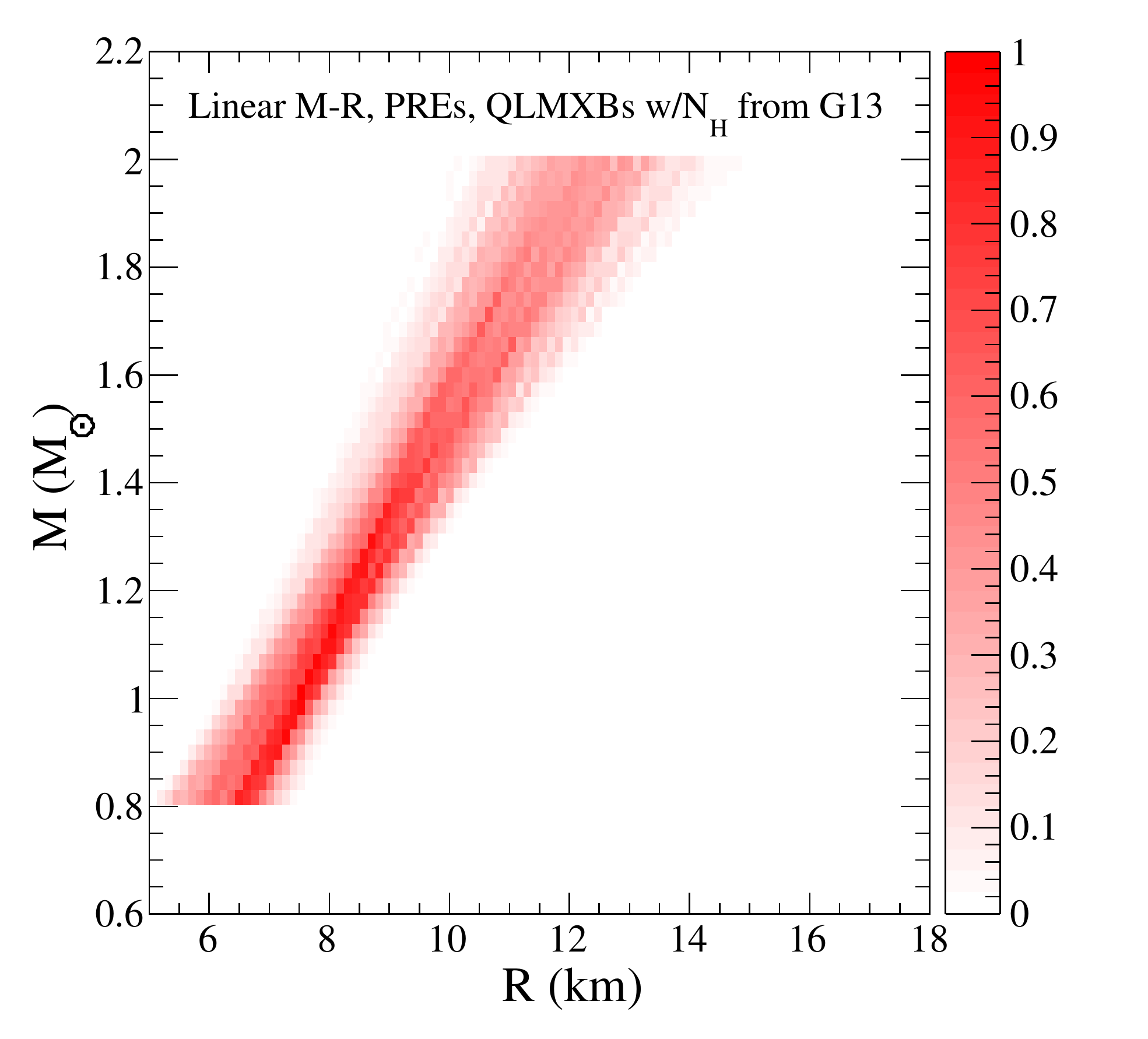}
\includegraphics[width=9.0cm]{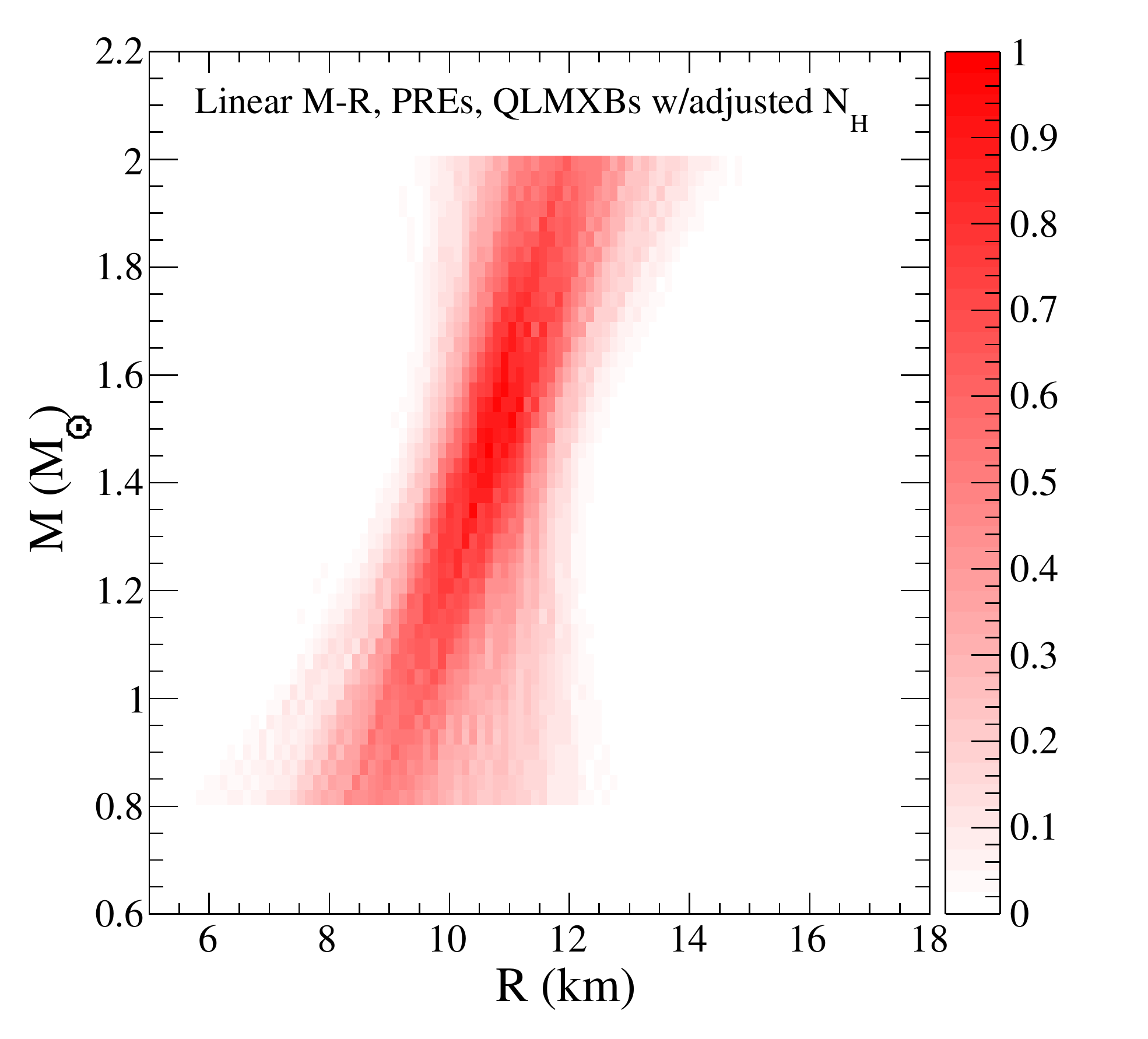}
\caption{$M-R$ curves obtained from the PRE and QLMXB data under the
  assumption that the $M-R$ curve is a line with constant slope. The
  top panel gives results assuming that for the QLMXBs, $N_H$ is given
  by the values in G13. The bottom panel uses the adjusted values of
  $N_H$ from Ref. \cite{Dickey90}. The PRE $(R,M)$ distributions are
  the same for both panels.\label{line_mvsr}}
\end{figure}

An alternative model with two parameters assumes that the $M-R$ curve
is a line with arbitrary slope. The top panel in Figure
\ref{line_mvsr} gives the results assuming the hydrogen column
densities from Ref.~\cite{Guillot13} and the bottom panel gives the
results assuming the adjusted hydrogen column densities from
Ref.~\cite{Dickey90} and allowing for the presence of helium
atmospheres. The PRE sources are treated the same in both panels,
under the assumption that $z_{\mathrm{ph}}=0$ at touchdown. The
results are cut off for masses above $\hat M=2M_{\odot}$ and we use
0.8$M_{\odot}$ as a lower limit. Both sets of $M-R$ line
distributions, independent of the assumptions about the hydrogen
column densities, show a clear preference for very small radii for
low-mass stars. In the context of current models of neutron stars with
crusts based on modern nucleon-nucleon interactions, such small radii
are very difficult without very small values of $L$ and strong phase
transitions just above the nuclear saturation density.

Note that by using a different prior distribution for two model
parameters (slope and y-intercept), this alternative model would give
exactly the same results as the previous model. The previous
``vertical line'' model can be obtained by assuming a delta-function
prior distribution which ensures the line is vertical. Similarly,
given a fixed data set, any model ``B'' for the distribution of $M-R$
curves is equivalent to any other model ``A'', if one modifies the
prior distribution for model B so that
\begin{equation}
P[{\cal M}_{B}]=P[{\cal D}|{\cal
    M}_{A}] P[{\cal M}_{A}]/P[{\cal D}|{\cal M}_{B}] \, .
\label{eq:alt_prior}
\end{equation}
This shows that choosing an alternate EOS parameterizations is
equivalent to choosing a different prior distribution for the original
EOS parameters. 

\begin{figure}
\includegraphics[width=9.0cm]{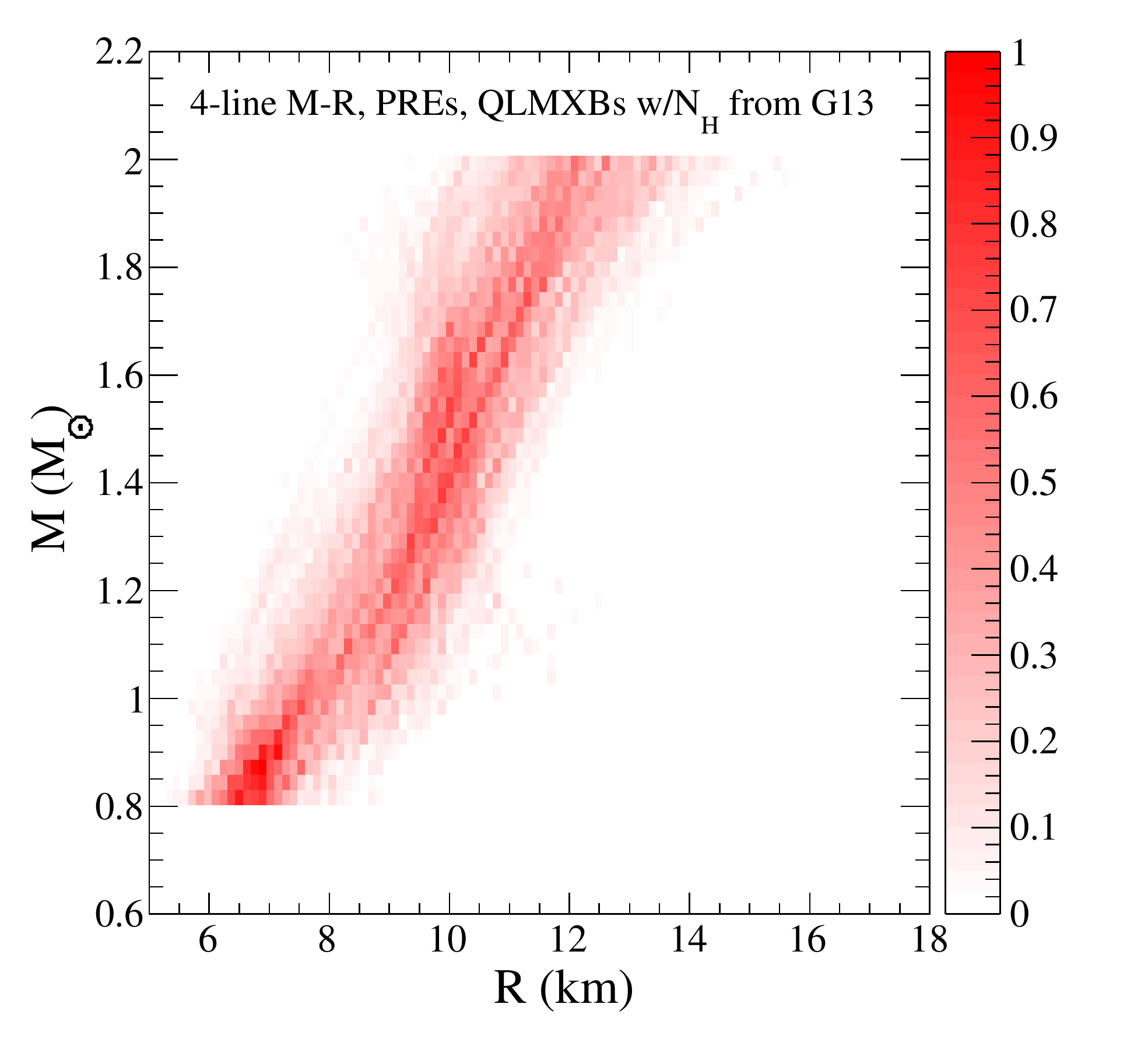}
\includegraphics[width=8.5cm]{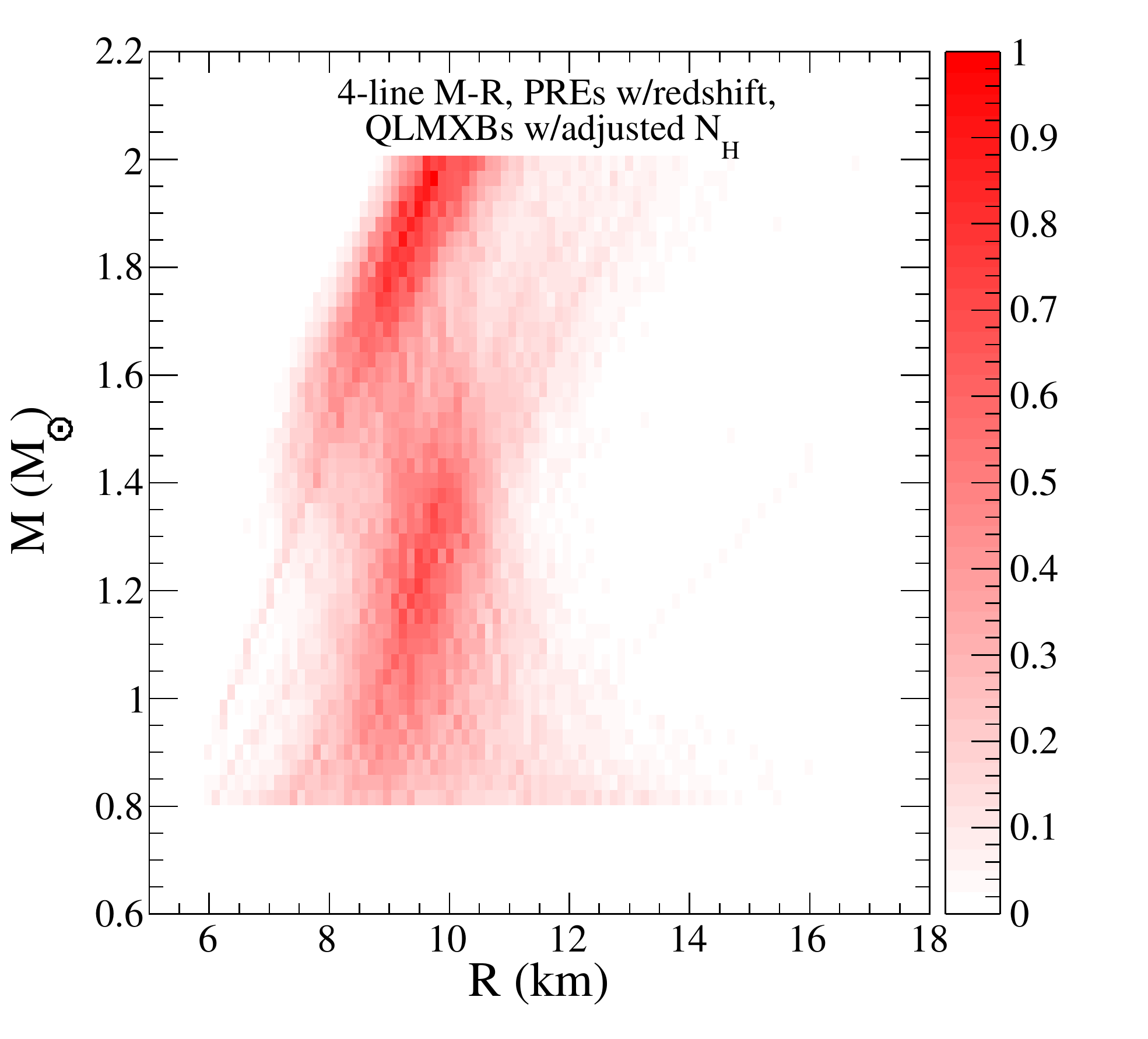}
\caption{ $M-R$ curves obtained from the PRE and QLMXB data for a
  generic $M-R$ curve made from four line segments. The top panel
  assumes that $z_{\mathrm{ph}}=0$ for the PRE sources and the G13
  values for $N_H$ for the QLMXBs. The bottom panel assumes that the
  photosphere is redshifted for the PRE sources and uses the 
  values of $N_H$ from Ref. \cite{Dickey90} for the QLMXBs.\label{mrd1}}
\end{figure}

\begin{figure}
\includegraphics[width=9.0cm]{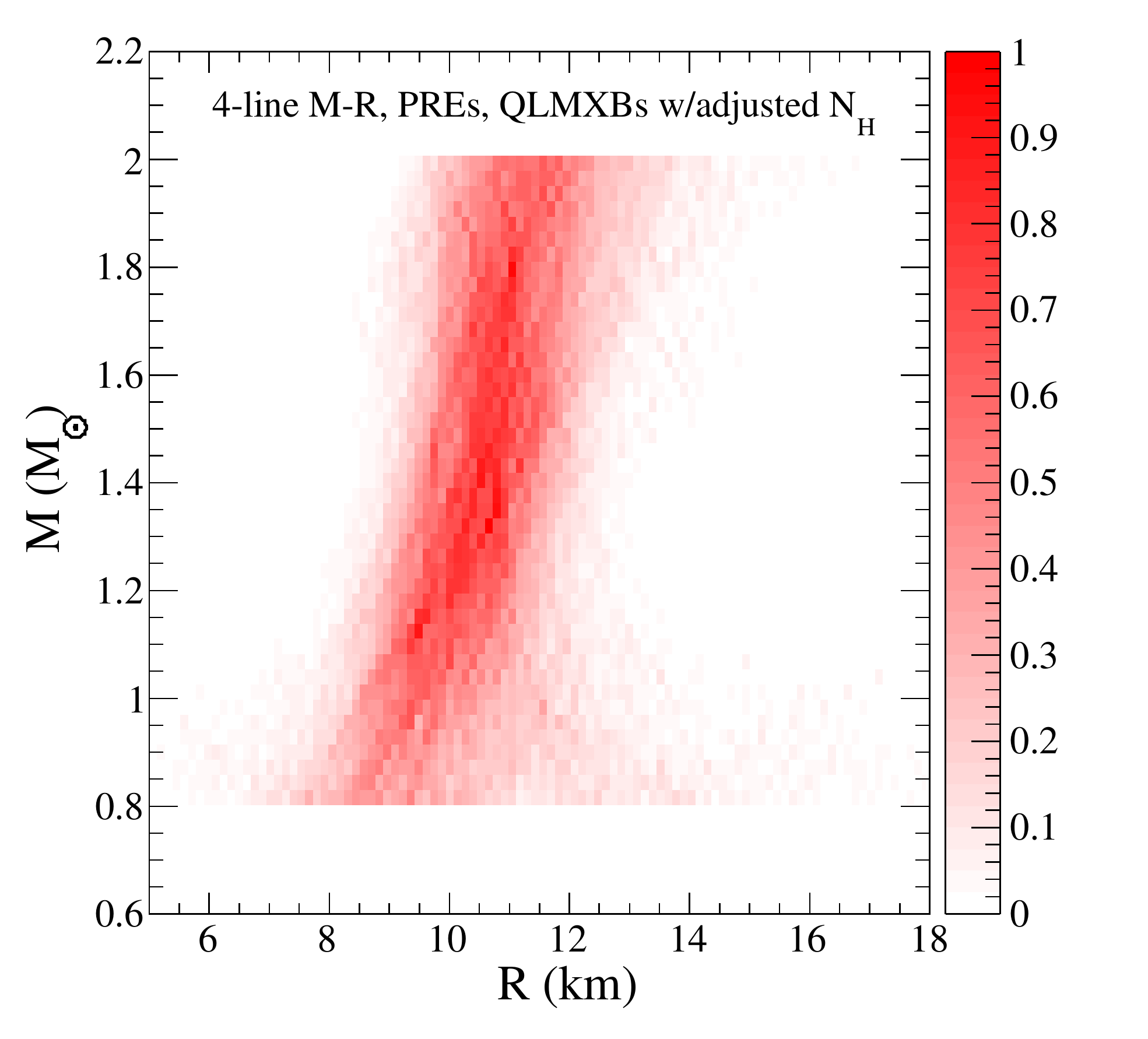}
\includegraphics[width=8.5cm]{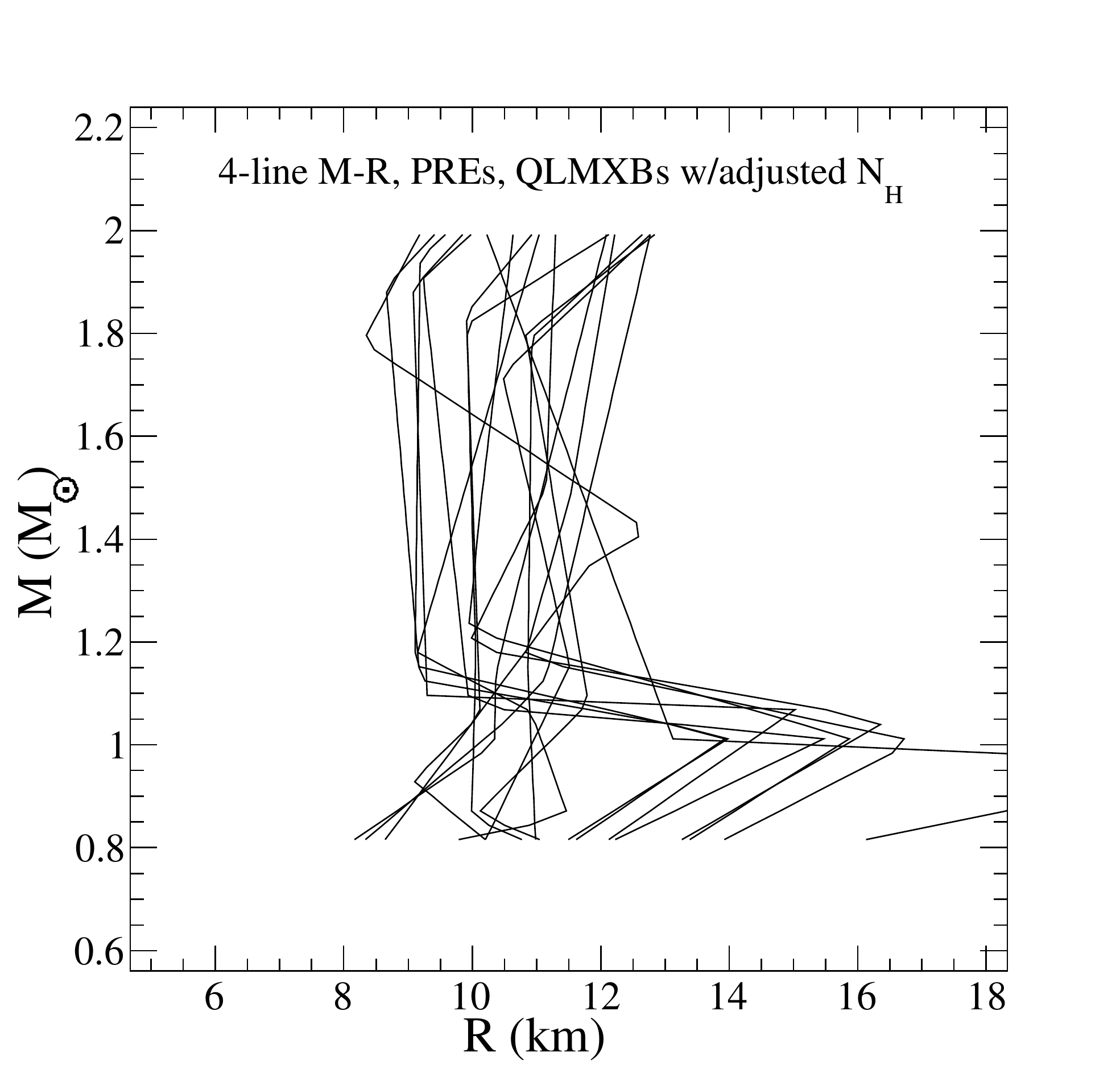}
\caption{The top panel is the same as in Figure \ref{mrd1} except that
  the values for $N_H$ for the QLMXBs are taken from Ref.
  \cite{Dickey90}. The bottom panel gives a small sampling of $M-R$
  curves used to construct the distributions in the top
  panel. \label{mrd2}}
\end{figure}

Finally, we consider more generic $M-R$ curves, made up of four line
segments with masses between 1 and 2$M_{\odot}$ with a total of 8
parameters. The results are given in Figs.~\ref{mrd1} and \ref{mrd2}.
The top panel of Fig.~\ref{mrd1} and the top panel of Fig.~\ref{mrd2}
show different assumptions for the hydrogen column densities and
atmospheric composition as before. The bottom panel of Fig.~\ref{mrd1}
shows results with the values of $N_H$ from Ref. \cite{Dickey90} for
the QLMXBs and with $z_{\mathrm{ph}}=z$ for the PREs. Small radii are
still preferred for low mass neutron stars. This will be the most
significant difference between models with no assumptions about
low-density matter and the results given below.

It is important to note here that Figs.~\ref{line_mvsr}, \ref{mrd1},
and the top panel of Fig.~\ref{mrd2} represent a set of
one-dimensional histograms: one histogram for each fixed mass
$\hat{M}$ as in Equation~(\ref{eq:margs}), each of them separately
normalized, and then plotted together. The distribution of radii for a
1$M_{\odot}$ neutron star has a smaller density (lighter color) than
the distribution of the radii for a 1.5$M_{\odot}$ because the
distribution is broader, i.e. the radius of a 1$M_{\odot}$ neutron
star is less well-constrained. This does not mean that 1$M_{\odot}$
neutron stars are less probable than 1.5$M_{\odot}$ neutron stars.
Also, while it is tempting to see the general $M-R$ curve in the top
panel of Fig.~\ref{mrd2} as nearly linear, this does not mean that the
radius of a low-mass and the radius of a high-mass neutron star are
necessarily correlated (in this model). Several $M-R$ curves from the
same simulation in the top panel of Fig.~\ref{mrd2} are given in
bottom panel of Fig.~\ref{mrd2}. These curves can contain kinks at
moderate masses which effectively decouple the low- and high-mass
properties of the typical $M-R$ curve. Note also that many of these
$M-R$ curves are incompatible with the TOV equations and physical EOSs.

\subsection{Models With Equations of State}

\begin{figure}
\includegraphics[width=9.5cm]{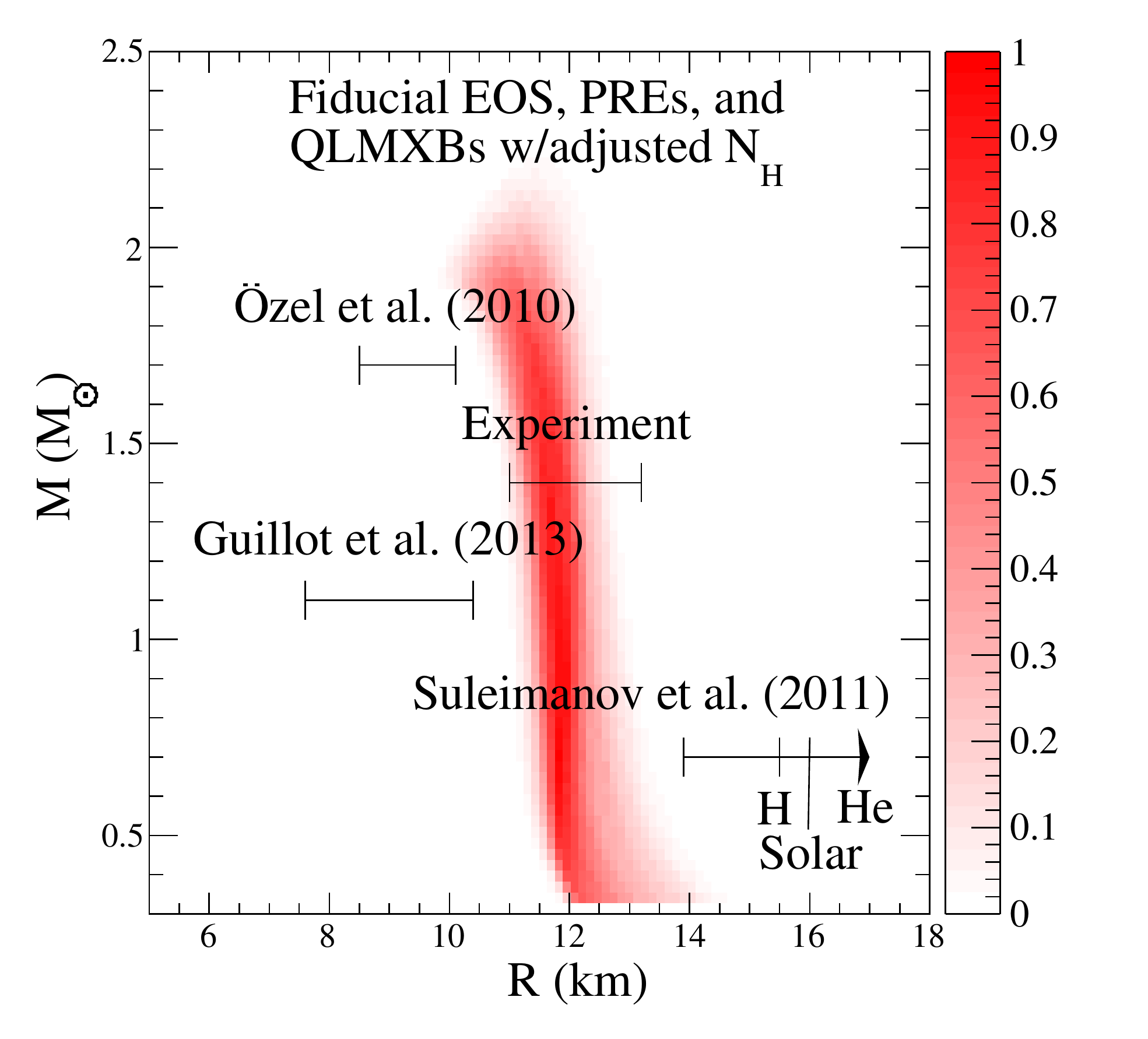}
\caption{Probability distributions for $M$ and $R$ for the five PRE
  burst sources, assuming $z_{\mathrm{ph}}=0$, and five QLMXB sources,
  assuming adjusted values of $N_H$ from Ref. \cite{Dickey90} and the
  possibility of either H or He atmospheres for four of the
  sources. The EOS is parametrized with two polytropes at high
  density. The radius ranges implied by the analyses of Ozel et
  al. (95\% confidence weighted average
  \cite{Ozel:2010,Ozel12,Guver13}), Guillot et al. (joint analysis
  assuming a fixed radius for all masses \cite{Guillot13}), Suleimanov
  et al. (90\% confidence intervals for pure H, solar with $Z=0.02$,
  and pure He atmosphers \cite{Suleimanov:2011}), and nuclear
  experiments (90\% confidence interval for $1.4M_\odot$ stars, \S
  \ref{sec:comp}) are also shown for comparison (with arbitrary 
  vertical locations). \label{fid}}
\end{figure}

\begin{figure}
\includegraphics[width=9.5cm]{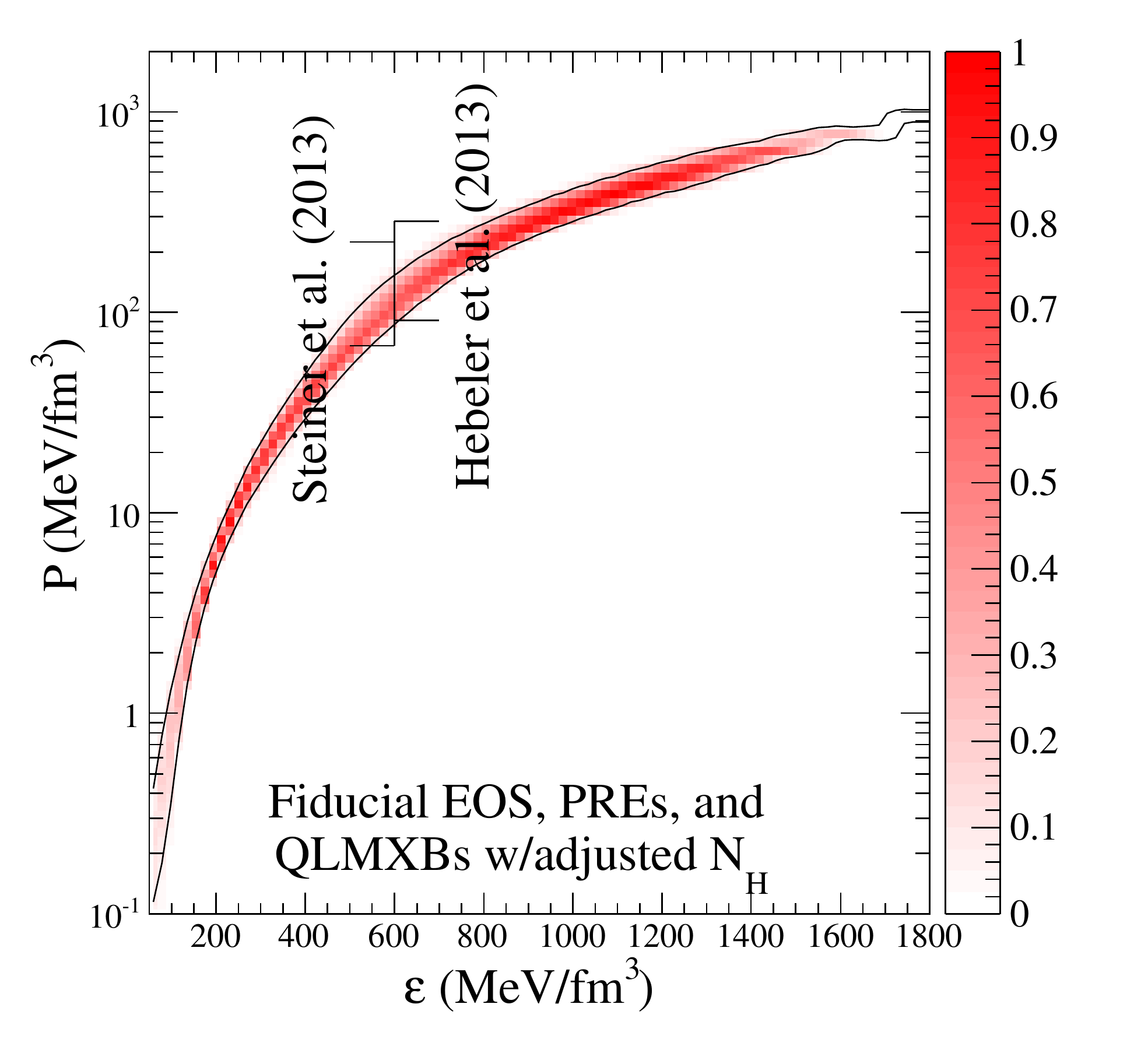}
\caption{
  Probability distributions for pressure and energy density from the
  Bayesian analysis of 5 PRE and 5 QLMXB sources under the same
  conditions as in Fig. \ref{fid}. Also shown are two
  representative ranges for the pressure from Refs.~\cite{SLB13} and
  \cite{Hebeler13}. \label{fid_eos}}
\end{figure}

We now employ the model of Ref.~\cite{SLB10}, using a neutron star
crust, a phenomenological EOS near the saturation density, and the
implicit assumption that the TOV equations relate the EOS to $M$ and
$R$. At higher densities, we use two polytropes. (This is also
referred to as ``Model A'' in Ref.~\cite{SLB13}.) For now, we keep the
same uniform initial mass function for each neutron star in the
sample. In addition to causality, we now ensure that the maximum mass
is above 2 solar masses. We use the alternative hydrogen column
densities from Ref. \cite{Dickey90} and allow for both hydrogen and
helium atmospheres (except for the source in $\omega$ Cen) for the
QLMXBs, and take $z_{\mathrm{ph}}=0$ for the PRE sources. The
predicted $M-R$ distribution for the full set of 5 PRE and 5 QLMXB
sources is shown in Figure \ref{fid} and observed to be relatively
vertical. This is a natural consequence of (i) causality, (ii) the
requirement of generating a 2 solar mass neutron star, (iii) the
existence of a neutron star hadronic crust, and (iv) the observation
of neutron stars with inferred small values of $R_{\infty}$. The $M-R$
curve also predicts larger radii for low-mass neutron stars. The range
of radii for a 1.4$M_{\odot}$ neutron star, 11.3$-$12.1 km (68\%
confidence) is also slightly larger than that suggested by the same
model (model A) in Ref.~\cite{SLB13}, a consequence of the larger
radii implied by the possibility of helium atmospheres in four of the
QLMXB sources. The 90\% confidence range for the radius of
$1.4M_\odot$ stars is tabulated in table~\ref{tab:bf2}, and the
corresponding EOS is given in Figure~\ref{fid_eos}, along with the
pressure ranges at an energy density of 600 MeV/fm$^{3}$ from
Refs.~\cite{SLB13} and \cite{Hebeler13}.

\begin{figure}
\includegraphics[width=9.5cm]{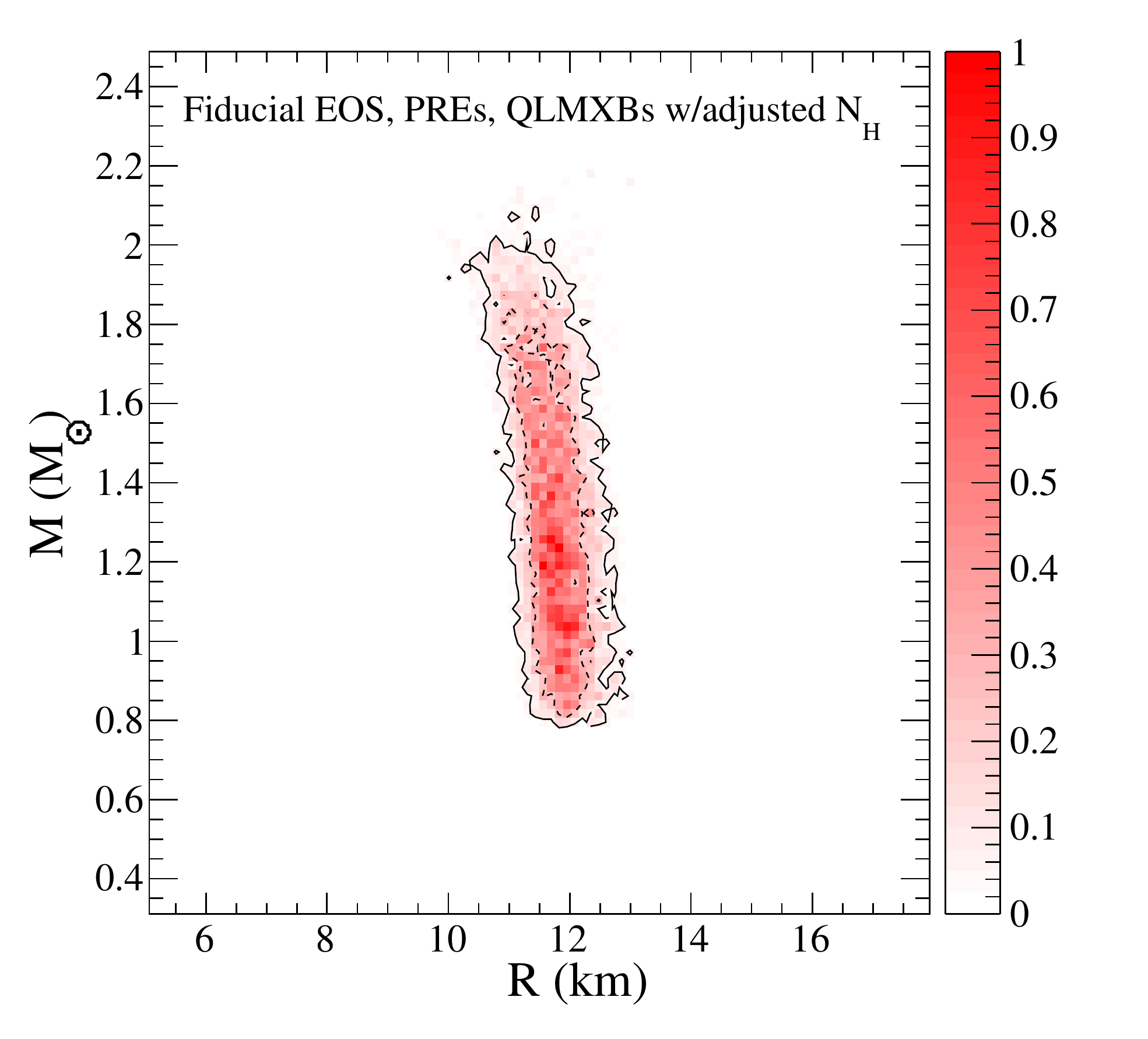}
\caption{Probability distributions for $M$ and $R$ of the neutron star
  in M13 with the same assumptions as given for Figure
  \ref{fid}.\label{fidM13}}
\end{figure}

Fig.~\ref{fidM13} gives the posterior $(R,M)$ distribution for just
one of the 10 neutron stars, the QLMXB in M13. (This posterior
distribution is a true two-dimensional histogram, unlike that in Fig.
\ref{fid}.)  Note that this posterior distribution for the neutron
star in M13 is much more strongly-peaked than the input $(R,M)$
distribution for M13 which was used in the simulation (which is
similar to that shown in Figure \ref{alt} assuming an H
atmosphere). The posterior $(R,M)$ distribution lies along the
predicted $M-R$ curve (as it must) and implies that this neutron star
is likely to have $M<1.8 M_{\odot}$.

It is natural to inquire how sensitive our results are to the nature
of the astronomical sources utilized. Table~\ref{tab:bf2} shows that
the 90\% confidence intervals for the radii of $1.4M_\odot$ stars is
increased by 0.3--0.4 km if only PRE burst sources are considered,
while the radii are increased by 0.1--0.2 km if only QLMXB sources are
included. These small differentials imply that our results are not
very sensitive to the type of source included, and that the
constraints of maximum mass, causality, a hadronic crust EOS and the
TOV equation are powerful. The lower limit on the radius from QLMXB
sources in the last row is 11.2 km, and adding the PRE sources (which
have significant probability at low radius) implies a only a slightly
smaller lower limit of 11.1 km. In short, limits to $L$ are not
much affected by the type of source included in our analyses.

The neutron star observations constrain the nuclear symmetry energy,
as expected from the correlation described in Eq.~\ref{radius}
above. The corresponding constraint, from our preferred model, on $L$
is between 37.0 and 55.3 MeV, to 68\% confidence. This range is
similar to that found for Model A in Ref.~\cite{SLB13} except the
lower range for $L$ has been decreased because of the very small radii
of the neutron stars in globular clusters NGC 6304 and M28. Using a
different EOS parameterization which allows for stronger phase
transitions (e.g., model C in Ref.~\cite{SLB13}) increases the upper
68\% confidence limit on $L$ to about 65 MeV. This happens because a
phase transition partially decouples the low- and high-density
behavior of the EOS, allowing small radii even if $L$ is relatively
large.

\subsection{Alternative Mass Distributions}

\begin{figure}
\includegraphics[width=8.0cm]{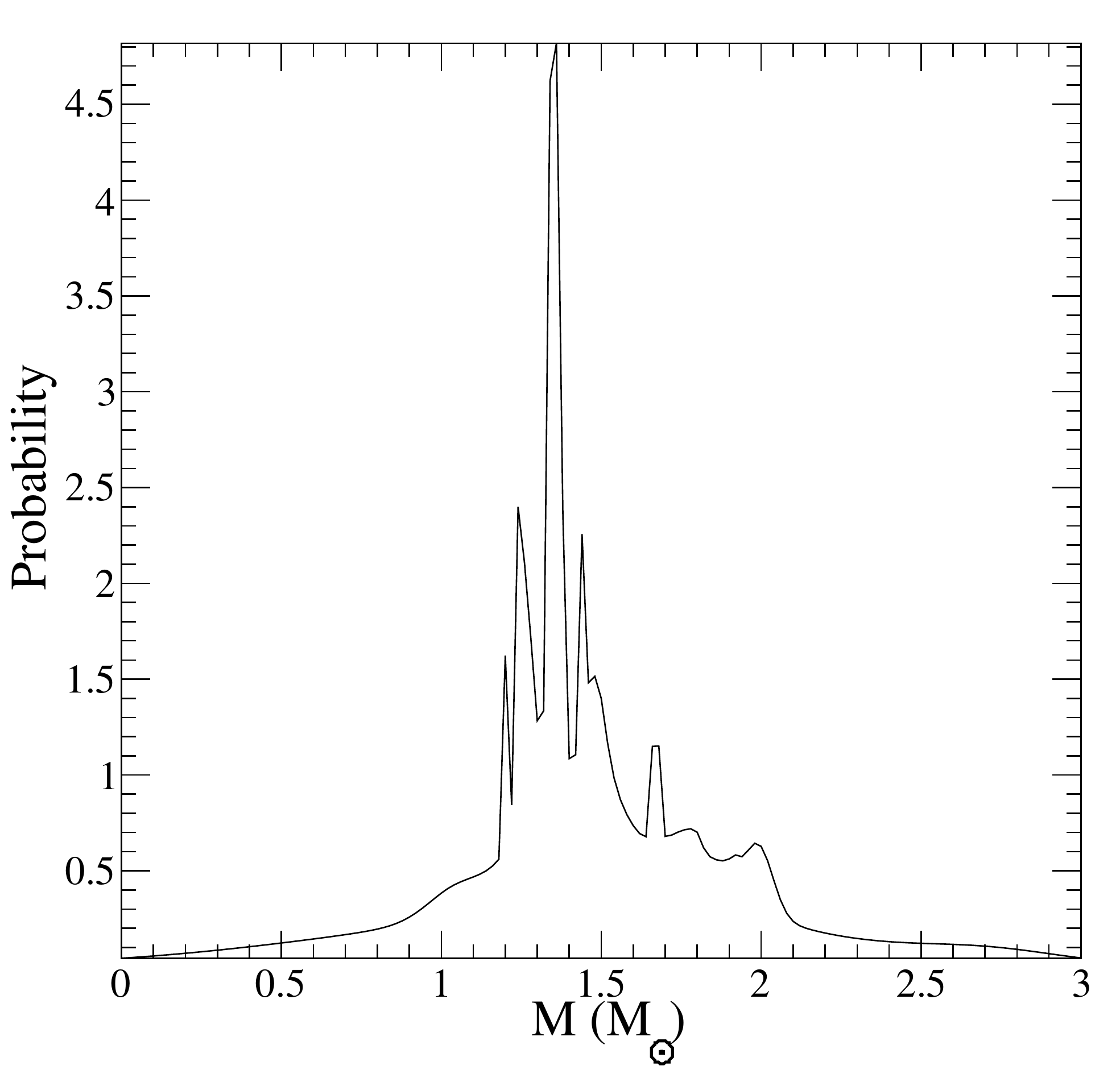}
\includegraphics[width=9.5cm]{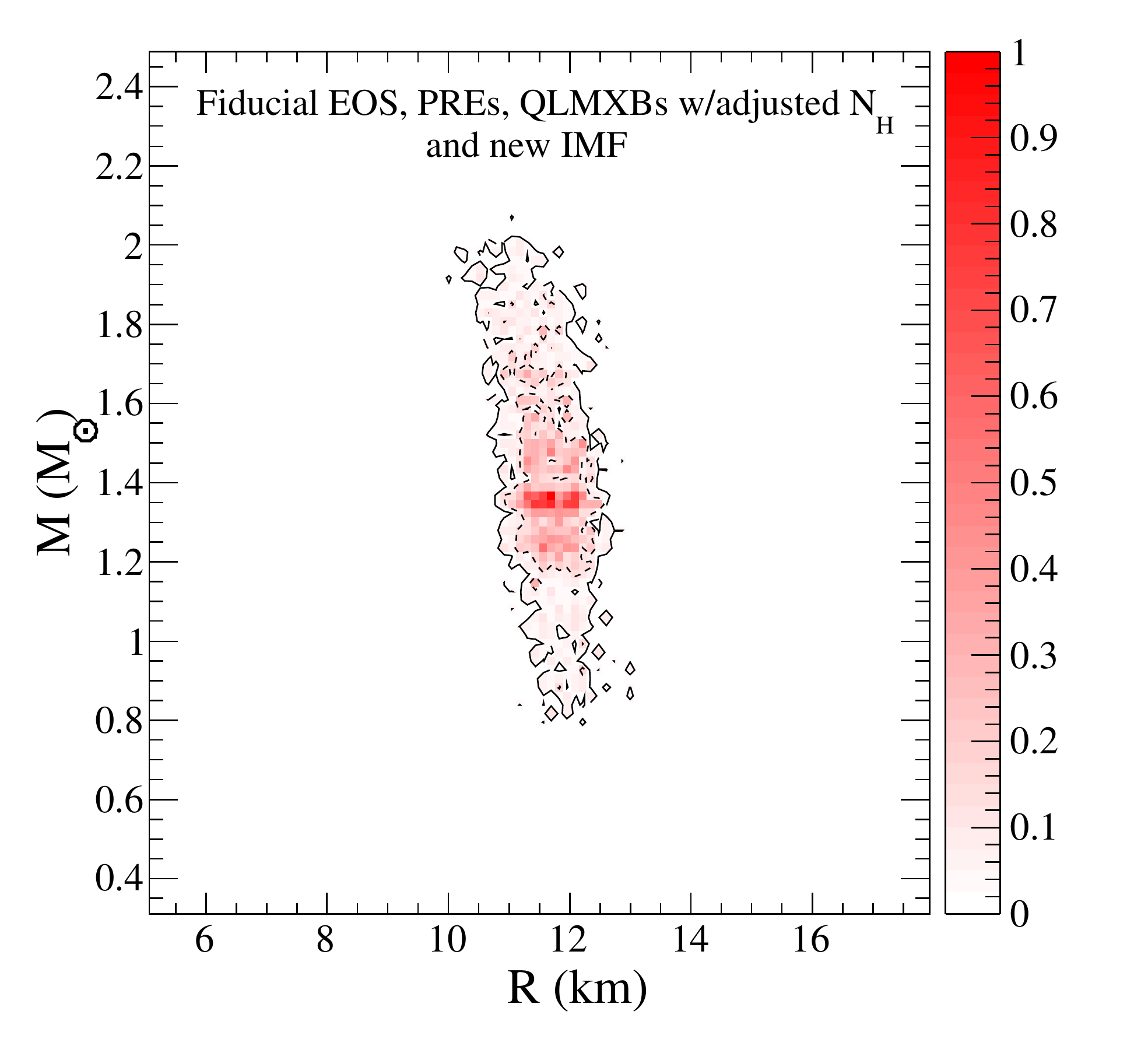}
\caption{The top panel shows the mass distribution inferred 
from the mass data in Ref.~\cite{Lattimer12} and the bottom
panel shows the posterior mass and radius distribution of
the neutron star in M13 having used the PRE and QLMXB data
as in Fig.~\ref{fid} but assuming the neutron star mass 
function is as given in the top panel.\label{mdist_res}}
\end{figure}

One can assess the effects of a different neutron star initial mass
function by modifying the prior distribution for the neutron star
masses. We assume the same mass distribution as in that Ref.
\cite{Lattimer12}, which is obtained summing the mass probability
distributions for each star, weighting each of them equally. The
individual probability distributions are assumed to be Gaussians
centered on the tabulated masses, with their $1\sigma$ error widths.
The top panel of Fig.~\ref{mdist_res} shows this prior mass
distribution and the bottom panel shows the resulting posterior
$(R,M)$ distribution for the neutron star in M13. The sharp peaks in
the initial mass function naturally lead to a stronger mass and radius
constraint for this neutron star.

\subsection{Bayes Factors}

To compare two models, ${\cal M}_1$ and ${\cal M}_2$, one 
uses the Bayes factor
\begin{equation}
B_{12} = 
\frac{\int_{{\cal M}_1} P[{\cal D}|{\cal M}_1=m_1] P[{\cal M}_1=m_1]~dm_1}
{\int_{{\cal M}_2} P[{\cal D}|{\cal M}_2=m_2] P[{\cal M}_2=m_2]~dm_2}
\label{eq:bfactor}
\end{equation}
If $B_{12}>1$, then model 1 is preferred, and if $B_{12}<1$, then
model 2 is preferred. A typical phrasing is that $B_{12}>3$ implies
the evidence for model 1 is substantial, $B_{12}>10$ implies the
evidence is strong, $B_{12}>30$ implies the evidence is very strong,
$B_{12}>100$ implies the evidence is decisive. In contrast to the
integrals in Equation~(\ref{eq:margs}), the normalization of the integrals in
the Bayes factor is important. In addition, the same Markov chain
cannot be trivially used to compute the Bayes factor, and so it
typically requires separate Monte Carlo integrations. Because the
integrands in Equations~(\ref{eq:margs}) and (\ref{eq:bfactor}) demand a
solution of the TOV equations at every point, they are computationally
expensive. However, we have found that the Markov chains created to
compute the integrals in Equation~(\ref{eq:margs}) can be re-used with a
simple interpolation scheme to avoid having to solve the TOV equations
again for the integrations in Equation~(\ref{eq:bfactor}).

Note that, in this Bayesian formalism, the actual number of parameters
plays very little role. This is in contrast to the frequentist
approach where one must divide $\chi^2$ by the number of ``degrees of
freedom'' in order to determine the goodness-of-fit. Nevertheless, the
Bayes factor can act as an Occam's razor, disfavoring models which
have ``extra'' parameters. Thus it is often not necessary to
choose models with the minimum number of parameters, as the Bayes
factor comparison will select the preferred model automatically. 

The Bayes factor provides a quantitative statistical method for
determining whether or not the Eddington flux at
touchdown ought is redshifted in PRE X-ray bursts. When Monte Carlo
trial points for $z_{\mathrm{ph}}=z$ are rejected due to the fact that
they lead to unphysical masses and radii, they cannot contribute to
the integral in Equation~(\ref{eq:bfactor}). Ref.~\cite{SLB10} found that, in
each case, at least 90\% of the trials are rejected for a typical PRE
X-ray burst source. For one neutron star, this would lead to a Bayes
factor of 10 at least in support of $z_{\mathrm{ph}}=0$. Because of
the product appearing in Equation~(\ref{eq:pdm}), one must count a factor of
at least 10 for each neutron star, leading to overwhelming support of
the model that the photosphere is not fully redshifted at touchdown.
\begin{table*}[]
\caption{90\% confidence ranges for radii of $1.4M_\odot$ stars and integrals $I$ for computing the Bayes
  factor for various models (see the text). \label{tab:bf2}}
\begin{tabular}{lcc}
Model & $R_{1.4}$ 90\% confidence range & $I$ \\
\hline
Alt/H+He QLMXB; $z=0$ PRE & 11.13 -- 12.33 & \\
$z=0$ PRE only & 11.56 -- 12.64 & \\
Base, QLMXB only     & 11.01 -- 11.94 & $(1.77\pm0.09)\times 10^{-8}$ \\
Alt, QLMXB only      & 10.62 -- 11.50 & $(4.65\pm0.48)\times 10^{-3}$ \\
H+He, QLMXB only     & 11.29 -- 12.83 & $(4.50\pm0.21)\times 10^{-3}$ \\
Alt/H+He, QLMXB only & 11.24 -- 12.59 & $(2.14\pm0.19)\times 10^{+2}$ \\
\hline
\end{tabular}
\end{table*}

The Bayes factor can also be used to test the alternative values for
the hydrogen column densities decribed in section~\ref{qlmxb}.
Table~\ref{tab:bf2}, adapted from LS13, contains the computed 90\%
confidence limits for the radius of $1.4M_\odot$ stars for all QLMXBs
under different scenarios.  "Base" assumes $N_H$ values from G13 and H
atmospheres for all sources, "Alt" uses the alternative values of
$N_H$ from Ref. \cite{Dickey90} and H atmospheres for all sources,
"H+He" uses $N_H$ values from G13 but allows for either He or H
atmospheres for all sources except the one in $\omega$ Cen, and
"Alt/H+He" uses $N_H$ values from Ref. \cite{Dickey90} and allows for
either He or H atmospheres. The integral for computing the Bayes
factor, sometimes called the ``evidence'', is given in the last column
for the QLMXB only cases. The Bayes factor for comparing the
alternative hydrogen column densities to those given in G13 is very
large~$\sim 10^6$ (from the ratio of the evidence in the first two
rows of the table), showing the alternative model is strongly
preferred.  Almost as preferred is the model in which G13 hydrogen
column densities are used, but the possibility of either H or He
atmospheres for four of the sources is entertained.  Finally, the
combination of the alternative column densities from
Ref. \cite{Dickey90} plus the possibility of either H or He
atmospheres is greatly preferred to the other scenarios.


\section{Discussion}
\label{sec:disc}

A plethora of nuclear experimental data indicates that the symmetry
energy parameters $S_v$ and $L$ are constrained to a greater degree
than just a few years ago. Although these constraints have varying
degrees of model dependence that need to be further explored, they are
well-supported by studies of pure neutron matter, which can determine
these parameters assuming that higher-than-quadratic terms in the
symmetry energy expansion in neutron excess are ignored. It is
expected that future theoretical studies of neutron matter with small
proton concentrations will allow the validity of the quadratic
expansion to be ascertained. From studies of solutions to the
hydrostatic structure equations in general relativity \cite{LP01},
these symmetry energy restrictions and the quadratic approximation
allow the radii of neutron stars to be determined to about 10\%
accuracy \cite{Lattimer13}.
For the experimental constraints studied here, the deduced radius of
$1.4M_\odot$ neutron stars is $R_{1.4}=12.1\pm1.1$ km. Neutron matter
studies suggest slightly smaller values by about 0.2 km.

In comparison, the astrophysical
determination of individual neutron star radii have much less
precision. Nevertheless, Bayesian studies (cf., \cite{SLB10,SLB13}) of
the ensemble of individual sources for which both mass and radius
information is available, imply typical radii (i.e., for $1.2-1.8
M_\odot$ stars) in the range 11.2 -- 12.8 km.
 There is emerging an important interplay between
the nuclear physics and the astronomical observations: {\em we find a
  concordance between the observations and the nuclear experiments.}
With almost
any reasonable assumptions regarding the nature of the EOS at high
densities and the parameters of models for shorter PRE X-ray bursts
and QLMXBs, the powerful constraints of causality, observations of
 2$M_{\odot}$ neutron stars, and the existence of a nuclear neutron
star crust, lead to $M-R$ curves which are nearly vertical and radii
for moderate-mass neutron stars that are compatible with nuclear data
and theoretical studies of neutron matter.

Thus, neutron star mass and radius observations are clearly beginning to
make quantitative constraints on both the EOS and the parameter $L$
which describes the density dependence of the symmetry energy. Two
major classes of neutron star observations have provided important
constraints: PRE X-ray bursts and the surface emission of QLMXBs. In
both of these classes of neutron star observations, the theoretical
models which interpret X-ray photons and produce the inferred neutron
star mass and radius are an important source of uncertainty. 

PRE X-ray bursts are interpreted as resulting from the vertical motion
of the photosphere. Assumptions about the position of the photosphere
at touchdown can change radius estimates by about 2 km.  If the
photosphere of PRE X-ray burst neutron stars is redshifted at
touchdown, we find that the observed fluxes and normalizations tend to
be inconsistent with the model, judging from the small number of Monte
Carlo trials over the observed uncertainty ranges of touchdown fluxes,
distances, and normalizations that result.  In addition, the 95\%
confidence radius range from Ref.~\cite{Ozel12,Guver13,Ozel:2010},
which comes from PRE sources alone and assumes the photosphere at
touchdown is at the stellar surface, i.e., $z_{\mathrm{ph}}=z$, is
also incompatible with nuclear experiment, as seen in Figure
\ref{fid}.  However, these are not the only difficulties surrounding
the interpretation of PRE X-ray bursts, and color correction factors
and composition are also important uncertain parameters. For example,
Suleimanov et al. have argued \cite{Suleimanov:2011} that the short
PRE bursts studied by Ozel et al. and in this contribution might have
significant disk absorption and $f_c$ evolution during the burst that
would dramatically increase the inferred radii.  Ref.
\cite{Suleimanov:2011} instead studied longer PRE bursts and found
radii in excess of 13.9 km to 90\% confidence (Figure \ref{fid}),
assuming stellar masses less than $2.3M_\odot$.  Importantly, both the
ranges suggested by Ozel et al. and Suleimanov et al. are inconsistent
with nuclear systematics.

In the case of QLMXBs, there is no photospheric dynamics to complicate
the interpretation of the neutron star atmosphere, but the composition
of the atmosphere and the magnitude of X-ray absorption between us and
the source are both major uncertainties. Differences of assumed X-ray
absorption magnitudes result in both larger and smaller radii. If the
hydrogen column densities are assumed to be those obtained from
self-consistent fitting of X-ray spectra \cite{Guillot13}, in some
cases the observed neutron stars are (i) too small to satisfy
causality limits, and (ii) too large to be consistent with the
available nuclear data and any reasonable neutron star model. On the
basis of our Bayesian model, however, we conclude that, on average,
$N_H$ values from Ref.~\cite{Dickey90} are statistically favored in
comparison to those obtained from self-consistently fitting
\cite{Guillot13} the X-ray spectra. The alternative $N_H$ values also
lead to a more uniform distribution of masses and radii among the
sources. The radius range deduced by G13 in their joint study in which
it is assumed that all neutron stars have the same radius is
inconsistent with both our results from the joint study of PRE bursts
and QLMXBs and with inferences from nuclear experiments to 90\%
confidence.

The large degree of model-dependence in interpreting astronomical
observations suggests more sophisticated modeling is in order.  It
will be necessary to model PRE bursts using hydrodynamical radiation
transport simulations to fit the overall light-curve behavior to fully
resolve the discrepancies and to provide reliable $M$ and $R$
estimates. Similarly, for the QLMXBs, there is a clear
necessity of obtaining further observations for fixing the
interstellar X-ray absorption for QLMXBs.  Moreover, there is evidence
that models of QLMXBs allowing for the possibility of He as well as H
atmospheres are favored, a question which further observations may also
be able to decide.

\section*{Acknowledgements}
\begin{acknowledgement}
J. M. L. is supported by the U.S. DOE grant DE-AC02-87ER40317 and A. W. S. is supported by U.S. DOE Grant DE-FG02-00ER41132. 
\end{acknowledgement}

\bibliographystyle{prsty}
\bibliography{ms}

\begin{thebibliography}{10}

\bibitem{LP01}
J.~M. Lattimer and M. Prakash, Astrophys. J. {\bf 550},  426  (2001).

\bibitem{Lattimer13}
J.~M. Lattimer and Y. Lim, Astrophys. J. {\bf 771},  51  (2013).

\bibitem{Demorest:2010}
P.~B. Demorest, T. Pennucci, S.~M. Ransom, M.~S.~E. Roberts, and J.~W.~T.
  Hessels, Nature {\bf 467},  1081  (2010).

\bibitem{Antoniadis13}
J. Antoniadis, P.~C.~C. Freire, N. Wex, T.~M. Tauris, R.~S. Lynch, {\it
  et~al.}, Science {\bf 340},  448  (2013).

\bibitem{Audi03}
G. Audi, A.~H. Wapstra, and C. Thibault, Nucl. Phys. A {\bf 729},  337  (2003).

\bibitem{Myers69}
W.~D. Myers and W.~J. Swiatecki, Ann. Phys. {\bf 55},  395  (1969).

\bibitem{LS89}
E. Lipparini and S. Stringari, Phys. Rep. {\bf 175},  103  (1989).

\bibitem{MS66}
W.~D. Myers and W.~J. Swiatecki, Nucl. Phys. A {\bf 81},  1  (1966).

\bibitem{Steiner05}
A.~W. {Steiner}, M. {Prakash}, J.~M. {Lattimer}, and P.~J. {Ellis}, Phys. Rep.
  {\bf 411},  325  (2005).

\bibitem{Danielewicz03}
P. Danielewicz, Nuc. Phys. A {\bf 727},  233  (2003).

\bibitem{Kortelainen10}
M. Kortelainen, T. Lesinski, J. Mor\'e, W. Nazarewicz, J. Sarich, {\it et~al.},
  Phys. Rev. C {\bf 82},  024313  (2010).

\bibitem{MS90}
W.~D. Myers and W.~J. Swiatecki, Ann. Phys. {\bf 204},  401  (1990).

\bibitem{Moller12}
P. M\"oller, W.~D. Myers, H. Sagawa, and S. Yoshida, Phys. Rev. Lett. {\bf
  108},  052501  (2012).

\bibitem{Gandolfi12}
S. Gandolfi, J. Carlson, and S. Reddy, Phys. Rev. C {\bf 85},  032801  (2012).

\bibitem{Hebeler10}
K. Hebeler, J.~M. Lattimer, C.~J. Pethick, and A. Schwenk, Phys. Rev. Lett.
  {\bf 105},  161102  (2010).

\bibitem{Chen10}
L.-W. Chen, C.~M. Ko, B.-A. Li, and J. Xu, Phys. Rev. C {\bf 82},  024321
  (2010).

\bibitem{Ray79}
L. Ray, Phys. Rev. C {\bf 19},  1855  (1979).

\bibitem{Krasznahorkay94}
A. Krasznahorkay, J.~A. Balanda, J.~A. Bordewijk, M.~N. Brandenburg, {\it
  et~al.}, Nucl. Phys. A {\bf 567},  521  (1994).

\bibitem{Krasznahorkay99}
A. Krasznahorkay, P. Fujiwara, P. van Aarlo, H. Akimune, I. Daito, {\it
  et~al.}, Phys. Rev. Lett. {\bf 82},  3216  (1999).

\bibitem{Trzeinska01}
A. Trzeinska, P. Jastrwebski, F.~J. Hartmann, R. Schmidt, T. von Egidy, and B.
  Klos, Phys. Rev. Lett. {\bf 87},  082501  (2001).

\bibitem{Klimkiewicz07}
A. Klimkiewicz, N. Paar, P. Adrich, M. Fallot, K. Boretzky, {\it et~al.}, Phys.
  Rev. C {\bf 76},  051603(R)  (2007).

\bibitem{Terashima08}
S. Terashima, H. Sakaguchi, H. Takeda, T. Ishikawa, M. Itoh, {\it et~al.},
  Phys. Rev. C {\bf 77},  024317  (2008).

\bibitem{Tamii11}
A. Tamii, I. Poltoratska, P. von Neumann-Cosel, Y. Fujita, T. Adachi, {\it
  et~al.}, Phys.Rev.Lett. {\bf 107},  062502  (2011).

\bibitem{Roca-Maza13}
X. Roca-Maza, M. Brenna, G. Col\'o, M. Centelles, X. Vi\~nas, {\it et~al.},
  Phys. Rev. C {\bf 88},  024316  (2013).

\bibitem{Reinhard10}
P.-G. Reinhard and W. Nazarewicz, Phys. Rev. C {\bf 81},  051303  (2010).

\bibitem{Trippa08}
L. Trippa, G. Col\'o, and E. Vigezzi, Phys. Rev. C {\bf 77},  061304  (2008).

\bibitem{Tsang:2009}
M.~B. Tsang, Y. Zhang, P. Danielewicz, M. Famiano, Z. Li, W.~G. Lynch, and
  A.~W. Steiner, Phys. Rev. Lett. {\bf 102},  122701  (2009).

\bibitem{Danielewicz13}
P. Danielewicz and J. Lee, Nucl. Phys. A {\bf 922},  1  (2014).

\bibitem{Shetty07}
D.~V. Shetty, S.~J. Yennello, and G.~A. Souliotis, Phys. Rev. C {\bf 76},
  024606  (2007).

\bibitem{Friedman12}
E. Friedman, Nucl. Phys. A {\bf 896},  46  (2012).

\bibitem{Clark03}
B.~C. Clark, L. Kerr, and S. Hama, Phys. Rev. C {\bf 67},  054605  (2003).

\bibitem{Zenhiro10}
J. Zenhiro {\it et~al.}, Phys. Rev. C {\bf 82},  044611  (2010).

\bibitem{Starodubsky94}
V.~E. Starodubsky and N.~M. Hintz, Phys. Rev. C {\bf 49},  2118  (1994).

\bibitem{Steiner:2012}
A.~W. Steiner and S. Gandolfi, Phys. Rev. Lett. {\bf 108},  081102  (2012).

\bibitem{Hebeler13}
K. Hebeler, J.~M. Lattimer, C.~J. Pethick, and A. Schwenk, Astrophys. J. {\bf
  773},  11  (2013).

\bibitem{Lattimer12}
J.~M. Lattimer, Annu. Rev. Nucl. Part. Sci. {\bf 62},  485  (2012).

\bibitem{SLB10}
A.~W. {Steiner}, J.~M. {Lattimer}, and E.~F. {Brown}, Astrophys. J. {\bf 722},
  33  (2010).

\bibitem{Ozel06}
F. Ozel, Nature {\bf 441},  1115  (2006).

\bibitem{Ozel09}
F. \"Ozel, T. G\"uver, and D. Psaltis, Astrophys. J. {\bf 693},  1775  (2009).

\bibitem{Guver10a}
T. G\"uver, F. \"Ozel, A. Cabrera-Lavers, and P. Wroblewski, Astrophys. J. {\bf
  712},  946  (2010).

\bibitem{Guver10b}
T. G\"uver, P. Wroblewski, L. Camarota, and F. \"Ozel, Astrophys. J. {\bf 719},
   1807  (2010).

\bibitem{Ozel12}
F. \"Ozel, A. Gould, and T. G\"uver, Astrophys. J. {\bf 748},  5  (2012).

\bibitem{Guver13}
T. G\"uver and F. \"Ozel, Astrophys. J. {\bf 765},  1  (2013).

\bibitem{Ozel:2010}
F. \"Ozel, G. Baym, and T. G\"uver, Phys. Rev. D {\bf 82},  101301  (2010).

\bibitem{Wilms00}
J. Wilms, A. Allen, and R. McCray, Astrophys. J. {\bf 542},  914  (2000).

\bibitem{Guillot13}
S. Guillot, M. Servillat, N.~A. Webb, and R.~E. Rutledge, Astrophys. J. {\bf
  772},  7  (2013).

\bibitem{LS13}
J.~M. Lattimer and A.~W. Steiner, arxiv.org:1305.3242  (2013).

\bibitem{Dickey90}
J.~M. Dickey and F.~J. Lockman, Annu. Rev. Astron. Astrophys. {\bf 28},  215
  (1990).

\bibitem{Haggard04}
D. Haggard, A.~M. Cool, J. Anderson, P.~D. Edmonds, P.~J. Callanan, {\it
  et~al.}, Astrophys. J. {\bf 613},  512  (2004).

\bibitem{SLB13}
A.~W. Steiner, J.~M. Lattimer, and E.~F. Brown, Astrophys. J. Lett. {\bf 765},
  5  (2013).

\bibitem{Suleimanov:2011}
V. Suleimanov, J. Poutanen, M. Revnivtsev, and K. Werner, Ap. J {\bf 742},  122
   (2011).

\end{thebibliography}

\end{document}